\documentclass[a4,11pt]{article}
\usepackage{graphicx} 
\usepackage{float}
\usepackage{amsmath}
\usepackage{amsfonts}
\usepackage{latexsym}
\usepackage{amssymb}
\usepackage{setspace}
\usepackage{comment}

\makeatletter
 
  \@addtoreset{equation}{section}
 \makeatother

\begin{document}
\title{A Polynomial Scheme of Asymptotic Expansion \\
for Backward SDEs and Option pricing \footnote{To appear in {\it Quantitative Finance}.
All the contents expressed in this research are solely those of the authors and do not represent any views or 
opinions of any institutions. The authors are not responsible or liable in any manner for any losses and/or damages caused by the use of any contents in this research.
}
}

\author{Masaaki Fujii\footnote{Graduate School of Economics, The University of Tokyo. e-mail: mfujii@e.u-tokyo.ac.jp}
}
\date{
First version: May 2, 2014\\
This version: December 22, 2014
}
\maketitle



\newtheorem{definition}{Definition}
\newtheorem{assumption}{$[$ A}
\newtheorem{condition}{$[$ C}
\newtheorem{lemma}{Lemma}
\newtheorem{proposition}{Proposition}
\newtheorem{theorem}{Theorem}
\newtheorem{remark}{Remark}
\newtheorem{example}{Example}
\newtheorem{corollary}{Corollary}
\def\n{{\bf n}}
\def\A{{\bf A}}
\def\B{{\bf B}}
\def\C{{\bf C}}
\def\D{{\bf D}}
\def\E{{\bf E}}
\def\F{{\bf F}}
\def\G{{\bf G}}
\def\H{{\bf H}}
\def\I{{\bf I}}
\def\J{{\bf J}}
\def\K{{\bf K}}
\def\L{{\bf L}}
\def\M{{\bf M}}
\def\N{{\bf N}}
\def\O{{\bf O}}
\def\P{{\bf P}}
\def\Q{{\bf Q}}
\def\R{{\bf R}}
\def\S{{\bf S}}
\def\T{{\bf T}}
\def\U{{\bf U}}
\def\V{{\bf V}}
\def\W{{\bf W}}
\def\X{{\bf X}}
\def\Y{{\bf Y}}
\def\Z{{\bf Z}}
\def\cala{{\cal A}}
\def\calb{{\cal B}}
\def\calc{{\cal C}}
\def\cald{{\cal D}}
\def\cale{{\cal E}}
\def\calf{{\cal F}}
\def\calg{{\cal G}}
\def\calh{{\cal H}}
\def\cali{{\cal I}}
\def\calj{{\cal J}}
\def\calk{{\cal K}}
\def\call{{\cal L}}
\def\calm{{\cal M}}
\def\caln{{\cal N}}
\def\calo{{\cal O}}
\def\calp{{\cal P}}
\def\calq{{\cal Q}}
\def\calr{{\cal R}}
\def\cals{{\cal S}}
\def\calt{{\cal T}}
\def\calu{{\cal U}}
\def\calv{{\cal V}}
\def\calw{{\cal W}}
\def\calx{{\cal X}}
\def\caly{{\cal Y}}
\def\calz{{\cal Z}}
%
\def\sskip{\hspace{0.5cm}}
\def\simleq{ \raisebox{-.7ex}{\em $\stackrel{{\textstyle <}}{\sim}$} }
\def\leqsim{ \raisebox{-.7ex}{\em $\stackrel{{\textstyle <}}{\sim}$} }
\def\ep{\epsilon}
\def\half{\frac{1}{2}}
\def\iku{\rightarrow}
\def\Iku{\Rightarrow}
\def\ikup{\rightarrow^{p}}
\def\inclusion{\hookrightarrow}
\def\cadlag{c\`adl\`ag\ }
\def\up{\uparrow}
\def\down{\downarrow}
\def\doti{\Leftrightarrow}
\def\douti{\Leftrightarrow}
\def\dochi{\Leftrightarrow}
\def\douchi{\Leftrightarrow}%
\def\yy{\\ && \nonumber \\}
\def\y{\vspace*{3mm}\\}
\def\nn{\nonumber}
\def\be{\begin{equation}}
\def\ee{\end{equation}}
\def\bea{\begin{eqnarray}}
\def\eea{\end{eqnarray}}
\def\beas{\begin{eqnarray*}}
\def\eeas{\end{eqnarray*}}
%
\def\hd{\hat{D}}
\def\hv{\hat{V}}
\def\hsd{{\hat{d}}}
\def\hx{\hat{X}}
\def\hsx{\hat{x}}
\def\bsx{\bar{x}}
\def\bsd{{\bar{d}}}
\def\bx{\bar{X}}
\def\ba{\bar{A}}
\def\bb{\bar{B}}
\def\bc{\bar{C}}
\def\bv{\bar{V}}
\def\balpha{\bar{\alpha}}
\def\bbalpha{\bar{\bar{\alpha}}}
\def\combi{\l(\begin{array}{c}\alpha\\ \beta \end{array}\r)}
\def\f{^{(1)}}
\def\s{^{(2)}}
\def\ss{^{(2)*}}
\def\l{\left}
\def\r{\right}
\def\a{\alpha}
\def\b{\beta}
\def\L{\Lambda}

\def\E{{\bf E}}
\def\P{{\bf P}}
\def\Q{{\bf Q}}
\def\R{{\bf R}}

\def\cadlag{{c\`adl\`ag~}}

\def\calf{{\cal F}}
\def\calp{{\cal P}}
\def\calq{{\cal Q}}
\def\wtW{\widetilde{W}}
\def\wtB{\widetilde{B}}
\def\wtPsi{\widetilde{\Psi}}
\def\wt{\widetilde}
\def\mbb{\mathbb}
\def\ep{\epsilon}
\def\del{\delta}
\def\part{\partial}
\def\wh{\widehat}
\def\bsigma{\bar{\sigma}}
\def\yy{\\ && \nonumber \\}
\def\y{\vspace*{3mm}\\}
\def\nn{\nonumber}
\def\be{\begin{equation}}
\def\ee{\end{equation}}
\def\bea{\begin{eqnarray}}
\def\eea{\end{eqnarray}}
\def\beas{\begin{eqnarray*}}
\def\eeas{\end{eqnarray*}}
\def\l{\left}
\def\r{\right}

\newcommand{\Slash}[1]{{\ooalign{\hfil/\hfil\crcr$#1$}}}
\vspace{15mm}

\begin{abstract}
A new asymptotic expansion scheme for backward SDEs (BSDEs) is proposed.
The perturbation parameter ``$\ep$" is introduced just to scale the 
forward stochastic variables within a BSDE. In contrast to the standard 
{\it small-diffusion} asymptotic expansion method, the dynamics of variables given 
by the forward SDEs is treated exactly. 
Although it requires a special form of the quadratic covariation terms of the continuous part,
it allows rather generic drift as well as jump components to exist.
The resultant approximation is given by a polynomial function in terms of the {\it unperturbed} forward 
variables whose coefficients are uniquely specified by the solution of the recursive system of 
linear ODEs. Applications to a jump-extended Heston 
and $\lambda$-SABR models for European contingent claims, as well as  the utility-optimization problem
in the presence of a terminal liability are discussed.

\end{abstract}
\vspace{20mm}
{\bf Keywords :}
Stochastic Control, Asymptotic Expansion, BSDE, random measure, Heston, SABR,
Utility optimization
\newpage
\section{Introduction}
The backward stochastic differential equation (BSDE) was introduced by Bismut (1973)~\cite{Bismut}
under a linear setup and was later extended by Pardoux \& Peng (1990)~\cite{P-Peng} to general non-linear situations. 
Although the research activity had been contained in a relatively small mathematical community, 
it has been rapidly gaining traction 
with financial researchers and practitioners, in particular, since the last 
financial crisis. This is because that one
almost inevitably encounters BSDEs when he/she tries to handle various non-linear effects 
arising from credit risk, collateralization, funding and regulatory 
costs,  and various other sources of incompleteness arising from the new market realities. 
See, for example, Duffie \& Huang (1996)~\cite{Duffie-Huang}, 
Fujii \& Takahashi (2013)~\cite{FT-asymmetric}, Cr\'epey (2013)~\cite{Crepey-TVA} and a summary of 
recent practical topics in the financial industry \cite{RBook, Brigo}.
Various  interesting financial applications, such as for insurance, utility indifference pricing and optimal contract theory 
can be found in books \cite{Delong, Carmona, Cvitanic, CrepeyB}. One can consult with \cite{ElKaroui, MY} as a text 
for general mathematical treatments of BSDEs.

There now exists vast literature on their numerical treatments, ranging from the famous 
four-step scheme proposed by Ma et al. (1994)~\cite{Ma4}, its discretized implementation by Douglas et al. (1996)~\cite{Douglas},
various Monte-Carlo techniques making use of the least-square regression method
(See, for example, Bouchard \& Touzi (2004)~\cite{Bouchard}, Bender \& Denk (2007)~\cite{Bender},
Gobet et al. (2005)~\cite{Gobet1} and Gobet \& Lemor (2010)~\cite{Gobet2}),
a branching diffusion method by Henry-Labordere (2012)~\cite{Lab},
and a particle method by Fujii \& Takahashi (2012a)~\cite{FTP}. Unfortunately though, many of them require 
a good amount of experience and deep expertise to achieve stable results, such as an appropriate choice of basis functions,
 the order of regressions, and of course,  a good programming technique.

It is obvious that a simple analytical approximation method is deeply wanted.
In Fujii \& Takahashi (2012b)~\cite{FT-analytical},  we have developed a driver perturbation 
method combined with a standard asymptotic expansion technique for the forward SDEs.
Its error estimate was recently provided by Takahashi \& Yamada (2013)~\cite{TY}. It is systematic and straightforward,
but one still needs to endure long tough calculations especially for 
higher order corrections,  which is stemming from the needs of evaluation of conditional expectations 
at each order of expansion.  An interesting exceptional case arises if a so-called quadratic-growth
BSDE (qgBSDE) is associated with linear Gaussian forward SDEs, and at the same time, if its terminal value
is given by, at most, quadratic form of the Gaussian variables. In this case, 
the value function is given by a quadratic function of the Gaussian variables whose coefficients
are completely determined by the ordinary differential equations (ODEs) involving ones with
Riccati form. See, for example, Schroder \& Skiadas (1999)~\cite{SS} as an early research.
Recently, this property was applied to the mean-variance (quadratic) hedging problem by Fujii \& Takahashi 
(2013,~2014)~\cite{FT-MVH1, FT-MVH2},
making  use of the beautiful BSDE expression derived by Mania \& Tevzadze (2003)~\cite{Mania}.
Notice that the Riccati equation may possibly diverge in a finite time-interval in a general setup.
In such a case, one needs to shorten the maturity of the corresponding problem.

In this paper, we propose a new scheme which approximates a solution of a BSDE by a polynomial 
function of the underlying variables. In a Markovian setup, it is well-known that the solution of a BSDE
is given by a Markovian function of the underlying variables~\cite{Ma4}. Therefore, it is intuitively
clear that the solution should be  well approximated by a polynomial function for short maturities 
within which the size of the underlying variables (after appropriate rescaling and shift of their means) remains
relatively small. Despite the apparent similarity to the usual asymptotic expansion, 
the new scheme yields a recursive system of {\it linear} ODEs which can be obtained by simply matching 
the coefficients of the assumed polynomial solution to those of the BSDE's driver.
Although we have to assume that the forward processes have a special form of 
quadratic covariation of the continuous part, they can have rather general drifts and random jump components.
In that sense, the method can be interpreted as a generalizations of  {\it exact} but {\it exceptional} quadratic-solution 
example to an {\it approximate} polynomial-solution technique with wider applications.
\\ 

The organization of the paper is as follows: In Section 2, the main idea 
of the polynomial expansion scheme is explained. In order to show its
usefulness and accuracy, we apply the method to several well-known problems 
in the remaining part of the paper.
In Section 3 and 4,
the proposed scheme is applied to European contingent claims using
a jump-extended Heston model and the $\lambda$-SABR model. We provide
the closed expression for the recursive system of linear ODEs which 
specifies the approximate solution at an arbitrary order. 
In Section 5, the optimization problem for the exponential utility in the presence of
a terminal liability is analyzed. The closed-form system of the linear 
ODEs is derived for this setup, too.
Each model is associated with several illustrative numerical examples.
Finally in Appendix, some details  omitted in the main text are provided.

\section{Polynomial Expansion Scheme}
\subsection{Problem Setup}
Let us consider the following system of forward and backward SDEs 
in a probability space $(\Omega, \calf, \mbb{P})$:
\bea
\label{BSDE-Vorig}
V_t&=&H(X_T)-\int_t^T \bar{f}\Bigl(s,X_s,V_s,\bar{Z}_s,\int_K U(s,z)Q(s,dz)\Bigr)ds\nn \\
&&-\int_t^T \bar{Z}_s dW_s-\int_t^T \int_K U(s,z)\wt{\caln}(ds,dz) \\
X_t&=&x+\int_0^t b(s,X_s)ds+\int_0^t \sigma(s,X_s)dW_s+\int_0^t \int_K z\caln(ds,dz)
\label{SDE-X}
\eea
where $x\in\mbb{R}$ is a constant, $W$ one-dimensional standard Brownian motion and $\caln$ a 
random measure whose deterministic jump distribution is given by
$Q(t,\cdot)$ with some (compact) space $K$ for its support. 
$\wt{\caln}(dt,dz)$ is the corresponding $\mbb{P}$-compensated random measure 
\be
\wt{\caln}(dt,dz):=\caln(dt,dz)-\lambda(t,X_t)Q(t,dz)dt
\ee
where $\lambda(t,X_t)$ denotes the jump intensity.
We assume that $H: \mbb{R}\rightarrow \mbb{R}$, $\bar{f}: [0,T]\times \mbb{R}^4\rightarrow \mbb{R}$,
$b:[0,T]\times \mbb{R}\rightarrow \mbb{R}$
and $\lambda:[0,T]\times \mbb{R}\rightarrow \mbb{R}_+$
are all smooth functions.
In addition, we assume that the quadratic covariation of $X$ from its continuous part is
given by, at most, quadratic form of $X$ itself
\bea
d\langle X^c\rangle_t = \Bigl(\sigma_2(t) X_t^2+ \sigma_1(t) X_t+\sigma_0(t)\Bigr)dt~
\label{qcov}
\eea
where the superscript ``$c$" denotes the continuous part of $X$.
Here, $(\sigma_i(t))_{i=1,2,3}$ is the set of deterministic functions in such a way that
it guarantees the right-hand side of (\ref{qcov}) is non-negative for 
every possible value taken by $X$. We assume that the forward-backward SDE system 
of (\ref{BSDE-Vorig}) and (\ref{SDE-X}) has a well-posed solution.

Since the stochasticity of $(V_t)_{t\geq 0}$ is provided solely by $(X_t)_{t\geq 0}$, we can rewrite the BSDE as
\bea
V_t&=&H(X_T)-\int_t^T f\Bigl(s,X_s,V_s,Z_s,\int_K U(s,z)Q(s,dz)\Bigr)ds\nn \\
&&-\int_t^T Z_s dX_s^c -\int_t^T \int_K U(s,z)\caln(ds,dz)
\label{BSDE-X}
\eea
with appropriate redefinition of $f(\cdot)$ and $Z$~\footnote{For example, $Z$ and $\bar{Z}$ 
is connected by the relation $\bar{Z}_s=Z_s\sigma(s,X_s)$. }. Here,
$dX_t^c:=b(t,X_t)dt+\sigma(t,X_t)dW_t$ denotes the continuous part of the $X$'s change. 
Thus, in the following, we consider the equivalent system given by $(\ref{BSDE-X})$ and $(\ref{SDE-X})$.

\subsection{Asymptotic Expansion}
\label{sec-asympgeneral}
\subsubsection{General Idea}
\label{sec-general}
In order to obtain a polynomial expansion, we introduce $\ep$ so that we can count 
the order of the underlying $X$. Let consider the following perturbed BSDE:
\bea
V_t^\ep &=&H(\ep X_T)-\int_t^T f\Bigl(s,\ep X_s,V^\ep_s,Z^\ep_s,\int_K U^\ep(s,z)Q(s,dz)\Bigr)ds\nn \\
&&-\int_t^T Z^\ep_s dX_s^c -\int_t^T \int_K U^\ep(s,z)\caln(ds,dz)~.
\label{BSDE-epX}
\eea
Here, the superscripts $\ep$ in $V,Z,U$ emphasizes that these variables are now
dependent on the parameter $\ep$. An important difference from the usual
{\it small diffusion} asymptotic expansion method proposed by 
(Yoshida (1992a)~\cite{YoshidaE}, Takahashi (1999)~\cite{TA1}, Kunitomo \& Takahashi (2003)~\cite{KT2} for the 
pricing of contingent claims, Yoshida (1992b)~\cite{Yoshida} for statistical applications ) 
based on Watanabe (1987)~\cite{Watanabe} theory is that the underlying process $(X_t)_{t\geq 0}$ itself 
is not perturbed and only its size is scaled by $\ep$ {\it within} the BSDE. 

We  assume that that the expansion
\bea
\label{Vep}
V_t^\ep&=&\sum_{n=0}^\infty \ep^n V^{[n]}_t  \\
V_t^{[n]}&=&\sum_{m=0}^n \frac{X_t^m}{m!}v^{[n]}_m(t)
\label{Vtn}
\eea
is well defined, where $(v^{[n]}_m(t), 0\leq m \leq n)$ are all deterministic bounded functions 
in a given time interval $t\in[0,T]$.  In particular, the It\^o-formula 
should be applicable to (\ref{Vtn}) so that one obtains the 
well-defined forward SDE for $(V^{[n]}_t)_{0\leq t\leq T}$.
It leads to the corresponding expansions of the control variables
\bea
&&Z_t^\ep=\sum_{n=0}^\infty \ep^n Z_t^{[n]} \\
&&U^{\ep}(t,z)=\sum_{n=0}^\infty \ep^n U^{[n]}(t,z)
\eea
whose expressions can be easily derived once (\ref{Vtn}) is obtained.

If the maturity is short enough so that size of $X$ remains small,
truncating the expansion in (\ref{Vep}) at a certain order $n$ and putting $(\ep=1)$
are expected to give an approximation of the original problem.
Note that, since $\ep$ is introduced as a combination $(\ep X)$, 
discussing the size of $\ep$ separately from $X$ is not useful.
Let us denote the truncated $n$-th order approximation as (using the superscript $(n)$
instead of $[n]$)
\bea
\label{Vn-truncated}
&&V_t^{(n)}=\sum_{j=0}^n V_t^{[j]} \\
&&Z_t^{(n)}=\sum_{j=0}^n Z_t^{[j]},
\quad U^{(n)}(t,z)=\sum_{j=0}^n U^{[j]}(t,z)~.
\label{Cn-truncated}
\eea
Note that all of them are given by the polynomial functions of $X$, at most, of the order of $n$.

One can check the accuracy of approximation by comparing
\bea
\label{Vtilde}
\wt{V}^{(n)}(T)&=&V_0^{(n)}+\int_0^T f\Bigl(s,X_s,V_s^{(n)},Z_s^{(n)},\int_K U^{(n)}(s,z)Q(s,dz)\Bigr)ds\nn \\
&&+\int_0^T Z_s^{(n)}dX_s^c +\int_0^T \int_K U^{(n)}(s,z)\caln(ds,dz)
\eea
and $H(X_T)$ in a ``{\it path-wise}" fashion. In numerical examples given in later sections,
we shall observe that the polynomial approximation gives a good {\it path-wise} approximation, at least
for the paths along which $(|X_t(\omega)|)_{t\geq 0}$ does not significantly grow to a big value.
In practical applications, the capability of checking $(H(X_T)-\wt{V}^{(n)}(T))$
directly should be a great help for setting aside an appropriate amount of risk-reserve 
for the hedging program to be implemented.
By the very nature of polynomial approximation, one can imagine that 
a higher order expansion may yield an unstable result in a very volatile market,
or for a problem with long maturity.
The above comparison gives useful information for an appropriate order of expansion for a
given situation.

As we shall see below, all the functions $(v^{[n]}_m(t))_{m,n}$ except $v^{[0]}_0(t)$ are specified  by {\it linear} ODEs.
In later sections which deal with specific models, we provide a closed form recursive system of 
linear ODEs which fixes the coefficients of the polynomials up to an arbitrary order. 
However, in this section, let us adopt a slightly tedious {\it step-by-step} 
explanation, which we hope to give a clearer image to the readers.

\subsubsection{Zero-th order}
It is obvious that
\bea
V^{[0]}_t=H(0)-\int_t^T f(s)ds
\eea
where $f(s):=f(s,0,V_s^{[0]},0,0)$. Hence, the coefficient should be determined by
\bea
\dot{v}^{[0]}_0(t)=f(t,0,v_0^{[0]}(t),0,0), \qquad v^{[0]}_0(T)=H(0).
\eea
This is the only non-linear ODE we encounter.
We assume that the finite solution exist to the relevant time interval $t\in[0,T]$.
This should be the case for most of the natural applications, since the $0$-th order
problem corresponds to the market where all the underlyings are constant.

\subsubsection{First order}
Thanks to the smoothness assumption, one has
\bea
&&V_t^{[1]}=\part_xH(0)X_T-\int_t^T Z_s^{[1]}dX_s^c-\int_t^T \int_K U^{[1]}(s,z)\caln(ds,dz)\nn \\
&&\quad -\int_t^T \left\{\frac{\bigl.}{\bigr.}
\part_x f(s)X_s+\part_v f(s) V_s^{[1]}+\part_z f(s) Z_s^{[1]}
+\part_uf(s) \int_K U^{[1]}(s,z)Q(s,dz)\right\}ds~.\nn\\
\label{BSDE-V1}
\eea
On the other hand, let us suppose the above solution is given by
\bea
V_t^{[1]}= v^{[1]}_1(t)X_t+v^{[1]}_0(t)~.
\label{v1-candidate}
\eea
Then, it yields the dynamics
\bea
&&dV_t^{[1]}=\Bigl\{\dot{v}^{[1]}_1(t)X_t+\dot{v}^{[1]}_0(t)\Bigr\}dt\nn \\
&&\quad +v^{[1]}_1(t)dX_t^c+v^{[1]}_1(t)\int_K z\caln(dt,dz)~.
\label{SDE-V1}
\eea
By comparing (\ref{BSDE-V1}) and (\ref{SDE-V1}), we should have
\bea
&&Z_t^{[1]}=v^{[1]}_1(t) \\
&&U^{[1]}(t,z)=v^{[1]}_1(t)z~.
\eea

Substituting the expressions of $V^{[1]},Z^{[1]}$ and $U^{[1]}$ into (\ref{BSDE-V1})
and matching its driver to the drift part of (\ref{SDE-V1}), one obtains
\bea
&&\dot{v}^{[1]}_1(t)=\part_x f(t)+\part_v f(t) v^{[1]}_1(t) \\
&&\dot{v}^{[1]}_0(t)=\part_v f(t)v^{[1]}_0(t)+\Bigl(\part_z f(t) +\part_u f(t)q(t,1)\Bigr) v^{[1]}_1(t)
\eea
with terminal conditions $v^{[1]}_1(T)=\part_x H(0)$ and $v^{[1]}_0(T)=0$.
Here, we have used the notation
\bea
q(t,n)=\int_K z^n Q(t,dz)
\eea
to denote the n-th jump moment. In the reminder of the paper, we assume the 
existence of the moments  relevant for the approximation scheme.
It is clear that the assumed solution (\ref{v1-candidate}) and the corresponding 
control variables with the coefficients satisfying the 
above ODEs is in fact one solution of the BSDE (\ref{BSDE-V1}) as long as the forward SDE (\ref{SDE-V1})
is well-defined.
Due to the linearity of the ODEs,  the solution should be unique
among the assumed polynomial forms.

\subsubsection{Second order}
For the 2nd and higher order corrections, the assumption on the 
quadratic covariation term plays an important role.
As before, let us suppose that the solution takes the polynomial form: 
\bea
V_t^{[2]}=v^{[2]}_2(t)\frac{X_t^2}{2!}+v^{[2]}_1(t)X_t+v^{[2]}_0(t)~.
\label{v2-candidate}
\eea
Then, a simple application of It\^o-formula yields
\bea
&&dV_t^{[2]}=\left(\dot{v}^{[2]}_2(t)\frac{X_t^2}{2!}+\dot{v}^{[2]}_1(t)X_t+\dot{v}^{[2]}_0(t)\right)dt
+\frac{1}{2}v^{[2]}_2(t)d\langle X^c\rangle_t \nn \\
&&\qquad+\Bigl(v^{[2]}_2(t)X_t+v^{[2]}_1(t)\Bigr)dX_t^c\nn \\
&&\qquad+\int_K\left(v^{[2]}_2(t)\frac{(X_{t-}+z)^2-X_{t-}^2}{2}+v^{[2]}_1(t)z\right)
\caln(dt,dz)~.
\label{SDE-V2}
\eea

It should be clear that the assumption made in $(\ref{qcov})$ is necessary to 
guarantee that the {\it highest} polynomial order assumed in $V_t^{[n]}$ remains $n$
under the dynamics of $(X_t)_{t\geq 0}$.
The expression in (\ref{SDE-V2}) now implies
\bea
&&Z_t^{[2]}=v^{[2]}_2(t)X_t+v^{[2]}_1(t) \\
&&U^{[2]}(t,z)=v^{[2]}_2(t)\Bigl(X_{t-}z+\frac{z^2}{2}\Bigr)+v^{[2]}_1(t) z~.
\eea

On the other hand, the 2nd order part of (\ref{BSDE-epX}) leads to a BSDE
\bea
&&V^{[2]}_t=\frac{X_T^2}{2!}\part_x^2 H(0)-\int_t^T Z^{[2]}_s dX_s^c -\int_t^T \int_K U^{[2]}(s,z)\caln(ds,dz)\nn \\
&&-\int_t^T \left\{\part_v f(s)V_s^{[2]}+\part_z f(s)Z_s^{[2]}
+\part_u f(s)\Bigl(\int_K U^{[2]}(s,z)Q(s,dz)\Bigr) \frac{\bigl.}{\bigr.}
\right. \nn \\
&&\qquad +\frac{1}{2}\part_x^2 f(s)X_s^2 +\frac{1}{2}\part_v^2 f(s) [V_s^{[1]}]^2+
\frac{1}{2}\part_z^2 f(s)[Z_s^{[1]}]^2+\frac{1}{2}\part_u^2 f(s)
\Bigl(\int_K U^{[1]}(s,z)Q(s,dz)\Bigr)^2\nn \\
&&\qquad +X_s\left(
\part_{x,v}f(s)V_s^{[1]}+\part_{x,z}f(s)Z_s^{[1]}+\part_{x,u}f(s)\Bigl(\int_K U^{[1]}(s,z)Q(s,dz)\Bigr)\right)\nn \\
&&\qquad+V_s^{[1]}\left(\part_{v,z}f(s)Z_s^{[1]}+\part_{v,u}f(s)\Bigl(\int_K U^{[1]}(s,z)Q(s,dz)\Bigr)\right)\nn \\
&&\qquad\left. +\part_{z,u}f(s)Z_s^{[1]}\int_K U^{[1]}(s,z)Q(s,dz) \frac{\bigl.}{\bigr.} \right\}
\label{BSDE-V2}
\eea
Although there appear many cross terms,
the same procedures as in the first-order case of matching the coefficients of $X$ in the driver
of (\ref{BSDE-V2}) to those in (\ref{SDE-V2}) give us a set of linear ODEs.

After a simple calculation, one can confirm that the relevant ODEs are 
given by
\be
\dot{v}^{[2]}_2(t)=\Bigl(\part_v f(t)-\sigma_2(t)\Bigr)v^{[2]}_2(t)+
\part_v^2 f(t)\bigl[v^{[1]}_1(t)\bigr]^2+2\part_{x,y}f(t)v^{[1]}_1(t)+\part_x^2 f(t) 
\ee
\bea
\dot{v}^{[2]}_1(t)&=&\part_v f(t) v^{[2]}_1(t)+
\Bigl(\part_z f(t)+\part_u f(t) q(t,1)-\frac{\sigma_1(t)}{2}\Bigr)v^{[2]}_2(t)\nn \\
&&+\Bigl(\part_{v,z}f(t)+\part_{v,u}f(t)q(t,1)\Bigr)[v^{[1]}_1(t)]^2
+\part_v^2 f(t) v^{[1]}_1(t)v^{[1]}_0(t)\nn \\
&&+\Bigl(\part_{x,z}f(t)+\part_{x,u}f(t)q(t,1)\Bigr)v^{[1]}_1(t)+
\part_{x,v}f(t) v^{[1]}_0(t)
\eea
\bea
\dot{v}^{[2]}_0(t)&=&\part_v f(t) v^{[2]}_0(t)+\frac{1}{2}\Bigl(
\part_u f(t) q(t,2)-\sigma_0(t)\Bigr)v^{[2]}_2(t)\nn \\
&&+\Bigl(\part_z f(t)+\part_u f(t) q(t,1)\Bigr)v^{[2]}_1(t)
+\frac{1}{2}\part_v^2 f(t) [v^{[1]}_0(t)]^2\nn \\
&&+\frac{1}{2}\Bigl(\part_z^2 f(t)+\part_u^2 f(t) q(t,1)^2+2\part_{z,u}f(t)q(t,1)\Bigr)
[v^{[1]}_1(t)]^2\nn\\
&&+\Bigl(\part_{v,z}f(t)+\part_{v,u}f(t)q(t,1)\Bigr)v^{[1]}_1(t)v^{[1]}_0(t)
\eea
with terminal conditions
\bea
v^{[2]}_2(T)=\part_x^2 H(0),\quad v^{[2]}_1(T)= v^{[2]}_0(T)=0~.
\eea
Given the solution for the $1$st-order expansion $(v^{[1]}_1(t), v^{[1]}_0(t))$,
the above ODEs can be solved {\it one-by-one} following the order of 
$v^{[2]}_2\rightarrow v^{[2]}_1 \rightarrow v^{[2]}_0$.
If the forward dynamics (\ref{SDE-V2}) of the hypothesized solution is well-defined, 
then it is clear that it actually gives one solution for the BSDE of the 
2nd order (\ref{BSDE-V2}).
Due to the linearity of the ODE, the solution should be unique among the assumed forms.
\\

It is clear that one can repeat the procedures up to an arbitrary order.
At any order $n~(\geq 1)$, the relevant ODEs 
specifying the coefficients of polynomial solution are linear and they 
give the unique solution. If the hypothesized polynomial solution 
is well-defined in the given interval, it at least provides one solution for 
the BSDE of the $n$-th order.

\subsubsection*{Remark:}
In the above example, the distribution $(Q(t,\cdot))_{t\geq 0}$ is not 
necessary be deterministic. $q(t,n)$ can be a polynomial function of $X$ 
at most of the order of $n$. However, it is important to note that this point
is dependent on how the jump component is introduced in the model:~If $X$ has a 
proportional jump, then $q(t,n)$ specifying the 
$n$-th moment of the proportional jump factor should be independent of $X$.
For example, one can consider the conditions to keep $(\ref{BSDE-V2})$
as a 2nd-order polynomial of $X$.

\subsubsection{Mathematical justification for convergence and error estimate}
Unfortunately, we have not yet obtained a good understanding of the mathematical properties 
of the proposed expansion. Despite the similarity to Takahashi~\cite{TA1} and Kunitomo \& Takahashi~\cite{KT2} 
in the way that the parameter $\ep$ is introduced and its similar application to 
BSDEs in Fujii \& Takahashi~\cite{FT-analytical},
it is not yet clear if we can simply borrow the arguments in Takahashi \& Yamada~\cite{TY} for justification 
to the current polynomial scheme. A rigorous proof is left for 
an important topic for the future research.

However, we would like to emphasize that the above limitation is not a significant 
drawback for practical applications. The great advantage to have an explicit 
form of an approximate solution is allowing one to test its accuracy directly for a given setup (See the discussion 
in Section~\ref{sec-general}.).
The test like this is necessary for any methods since the convenient assumptions needed for
the mathematical justification will be violated in  realistic situations anyway.
In contrast to the proposed scheme (and the one in \cite{FT-analytical}), one can see that
carrying out this check is not a simple task for purely simulation-based 
techniques.

In addition, it is interesting to notice that we have not used any special properties of 
Wiener integral after expressing the BSDE in term of $dX$ as in (\ref{BSDE-X}).
If one can loosen the conditions necessary for the quadratic covariation terms,
one may possibly obtain an unified way of approximation for rather general semimartingales.

\section{Pricing European Options \\ with a Jump-extended Heston Model}
\label{sec-Heston}
\subsection{Problem Setup}
We assume that the asset price $S$ and its stochastic {\it variance-factor} $Y$ have the 
following dynamics under a probability space $(\Omega,\calf,\mbb{Q})$:
\bea
\label{S-hes}
&&S_t=S_0+\int_0^t S_s\left(\sigma(s)\sqrt{\bar{Y}_s}dW_s+\int_K (e^z-1)\wt{\caln}(ds,dz)\right) \\
&&\bar{Y}_t=1+\int_0^t\left(\alpha(s)\sqrt{\bar{Y}_s}dB_s+\kappa(s)(1-\bar{Y}_s)ds\right)
\label{Y-hes}
\eea
where $\mbb{Q}$ is supposed to be a certain equivalent martingale measure chosen by market participants.
$W$ and $B$ denote one-dimensional $\mbb{Q}$-Brownian motions with $d\langle W, B\rangle_t=\rho(t)dt$.
$\wt{\caln}$ denotes $\mbb{Q}$-compensated random  measure specified by
\bea
\wt{\caln}(dt,dz)=\caln(dt,dz)-\bar{\lambda}(t,\bar{Y}_t)Q(t,dz)dt
\eea
with the jump intensity $\bar{\lambda}$ and its deterministic distribution function $Q(t,\cdot)$.
$\sigma(\cdot),\alpha(\cdot), \rho(\cdot)$ and $\kappa(\cdot)$ are appropriate deterministic functions.
We allow $\bar{\lambda}:[0,T]\times \mbb{R}_+\rightarrow \mbb{R}_+$ to be a smooth 
generic function of $\bar{Y}$, and hence (\ref{S-hes}) and (\ref{Y-hes}) {\it do not} 
consist of the analytically solvable affine system.

In order to make the expansion around the origin a good approximation, we perform 
a change of variables
\bea
X_t:=\ln\left(\frac{S_t}{S_0}\right)\nn \\
Y_t:=\bar{Y}_t-1~.
\eea
Then, they follow the dynamics
\bea
\label{SDE-Xh}
&&X_t=\int_0^t \left(\sigma(s)\sqrt{Y_s+1}dW_s-\Bigl[\frac{\sigma(s)^2}{2}(Y_s+1)+
\lambda(s,Y_s)\beta(s)\Bigr]ds\right)\nn \\
&&\hspace{10mm} +\int_0^t \int_K z\caln(ds,dz)\\
&&Y_t=\int_0^t \left(\alpha(s)\sqrt{Y_s+1}dB_s-\kappa(s)Y_sds\right)
\label{SDE-Yh}
\eea
where $\beta(\cdot)$ is a deterministic function defined by
\be
\beta(t)=\int_K (e^z-1)Q(t,dz)~
\ee
and  $\lambda(t,Y_t):=\bar{\lambda}(t,Y_t+1)$.

Let us consider the valuation problem for a European option in a BSDE form:
\bea
V_t=H(X_T)-\int_t^T \bar{Z}_sdW_s-\int_t^T \bar{\Gamma}_sdB_s-\int_t^T \int_K U(s,z)\wt{\caln}(ds,dz)
\eea
where $H(X_T)$ denotes the terminal payoff at the maturity $T$ in terms of $X$, 
and $V_t$ denotes its present value at time $t~(<T)$.
Simple redefinition of the control variables $(\bar{Z}, \bar{\Gamma})$, one obtains
\bea
&&V_t=H(X_T)-\int_t^T Z_s dX_s^c-\int_t^T \Gamma_s dY_s -\int_t^T \int_K U(s,z)\caln(ds,dz) \nn \\
&&-\int_t^T\left\{Z_s\Bigl[\frac{\sigma(s)^2}{2}(Y_s+1)+\lambda(s,Y_s)\beta(s)\Bigr]
+\kappa(s)\Gamma_s Y_s-\lambda(s,Y_s)\int_KU(s,z)Q(s,dz)\right\}ds~.\nn
\eea
One can now apply the proposed polynomial expansion scheme to the BSDE
if $H(\cdot)$ is a smooth function. Although we can directly approximate the option payoff by 
a polynomial function, we shall take an alternative road that does not 
involve such approximation. We are going to consider $H(x)=x^m$
for $m=1,2,3,\cdots$. Then the corresponding value function $V_t$ gives the moments of $X_T$.
We finally use the Edgeworth expansion to get an estimate of the probability density function 
of $X_T$ (and hence $S_T$) to calculate 
the standard Call and Put options.

\subsection{Polynomial Expansion}
We consider the system of a perturbed BSDE
\bea
&&V_t^\ep=H(\ep X_T)-\int_t^T Z_s^\ep dX_s^c-\int_t^T \Gamma_s^\ep dY_s-\int_t^T \int_K U^\ep(s,z)\caln(ds,dz)\nn\\
&&-\int_t^T\left\{Z_s^\ep \Bigl[\frac{\sigma(s)^2}{2}(\ep Y_s+1)+\lambda(s,\ep Y_s)\beta(s)\Bigr]
+\ep \kappa(s)\Gamma_s^\ep Y_s-\lambda(s,\ep Y_s)\int_KU^\ep (s,z)Q(s,dz)\right\}ds\nn\\
\label{epV-Heston}
\eea
and the forward SDEs (\ref{SDE-Xh}) and (\ref{SDE-Yh}).
We expand the solution in term of $\ep$ as
\bea
\label{asymphes1}
&&V_t^\ep=\sum_{n=0}^\infty \ep^n V_t^{[n]} \\
&&Z_t^\ep =\sum_{n=0}^\infty \ep^n Z_t^{[n]},\quad \Gamma_t^\ep=\sum_{n=0}^\infty \ep^n \Gamma_t^{[n]},
\quad U^\ep(t,z)=\sum_{n=0}^\infty \ep^n U^{[n]}(t,z)~.
\label{asymphes2}
\eea

In the next lemma, we  give the solution of the above expansion in terms of a recursive system of linear ODEs.
We denote the number of choices selecting $m$ out of $n~(\geq m)$ by $C_{(n,m)}=\displaystyle{\frac{n!}{(n-m)!m!}}$.
We also use the convention for the summation symbol that $\sum_{i}^j\equiv 0$ when $(j<i)$.

{\lemma{If it exists, the polynomial solution for the expansion in (\ref{asymphes1}) and (\ref{asymphes2}) is 
uniquely given by 
\bea
\label{hypoVn}
&&V_t^{[n]}=\sum_{m=0}^n\sum_{k=0}^m \frac{X_t^{m-k}Y_t^k}{(m-k)!k!}v^{[n]}_{m-k,k}(t) \\
\label{hypoZn}
&&Z_t^{[n]}=\sum_{m=1}^n \sum_{k=0}^{m-1} \frac{X_t^{m-k-1}Y_t^k}{(m-k-1)!k!}v^{[n]}_{m-k,k}(t)  \\
\label{hypoGn}
&&\Gamma_t^{[n]}=\sum_{m=1}^n \sum_{k=1}^{m}\frac{X_t^{m-k}Y_t^{k-1}}{(m-k)!(k-1)!}v^{[n]}_{m-k,k}(t) \\
\label{hypoUn}
&&U^{[n]}(t,z)=\sum_{m=0}^{n-1}\sum_{k=0}^m \frac{X_t^{m-k}Y_t^k}{(m-k)!k!}
\Bigl(\sum_{l=m+1}^n \frac{z^{l-m}}{(l-m)!}v^{[n]}_{l-k,k}(t)\Bigr)
\eea
with the set of deterministic functions $v^{[n]}_{m-k,k}(t)$ of $(0\leq k\leq m\leq n)$
satisfying the following recursive system of linear ODEs
\bea
&&\hspace{-10mm}\dot{v}^{[n]}_{m-k,k}(t)=-\mbb{I}_{(m\leq n-1,1\leq k)} k
\left( \frac{\sigma(t)^2}{2}v^{[n]}_{m-k+2,k-1}(t)+\rho(t)\sigma(t)\alpha(t)v^{[n]}_{m-k+1,k}(t)+
\frac{\alpha(t)^2}{2}v^{[n]}_{m-k,k+1}(t)\right)\nn \\
&&-\mbb{I}_{(m\leq n-2)}\left(
\frac{\sigma^2(t)}{2}v^{[n]}_{m-k+2,k}(t)+\rho(t)\sigma(t)\alpha(t)v^{[n]}_{m-k+1,k+1}(t)
+\frac{\alpha(t)^2}{2}v^{[n]}_{m-k,k+2}(t)\right)\nn \\
&&+\mbb{I}_{(m\leq n-1,1\leq k)}k\left(\frac{\sigma(t)^2}{2}v^{[n-1]}_{m-k+1,k-1}(t)
+\kappa(t)v^{[n-1]}_{m-k,k}(t)\right)+\mbb{I}_{(m\leq n-1)}\frac{\sigma(t)^2}{2}v^{[n]}_{m-k+1,k}(t)\nn\\
&&+\mbb{I}_{(m\leq n-1)}\sum_{l=0}^kC_{(k,l)}\part^l_y\lambda(t,0)
\left(\beta(t)v^{[n-l]}_{m-k+1,k-l}(t)-\sum_{j=1}^{n-m}v^{[n-l]}_{j+m-k,k-l}(t)
\frac{q(t,j)}{j!}\right)
\label{H-ODE}
\eea
having the terminal conditions $v^{[n]}_{n,0}(T)=\part_x^n H(0)$ with all the other components zero.
\label{lemma-1}
}}\\ \\
{\it Proof:}
Let us suppose that the polynomial-form solution given in (\ref{hypoVn}) exists.
Then, the application of It\^o-formula and simple rearrangements of summation yield
\bea
&&dV_t^{[n]}=\sum_{m=0}^n \sum_{k=0}^m \frac{X_t^{m-k}Y_t^k}{(m-k)!k!}\left\{ \frac{\bigl.}{\bigr.}
\dot{v}^{[n]}_{m-k,k}(t) \right. \nn \\
&&\quad+\mbb{I}_{(m\leq n-1,1\leq k)}k\left(
\frac{\sigma_t^2}{2}v^{[n]}_{m-k+2,k-1}(t)+\rho_t\sigma_t\alpha_tv^{[n]}_{m-k+1,k}(t)+
\frac{\alpha_t^2}{2}v^{[n]}_{m-k,k+1}(t)\right)\nn \\
&&\quad\left.+\mbb{I}_{(m\leq n-2)}\left(
\frac{\sigma_t^2}{2}v^{[n]}_{m-k+2,k}(t)+\rho_t\sigma_t\alpha_t v^{[n]}_{m-k+1,k+1}(t)+
\frac{\alpha_t^2}{2}v^{[n]}_{m-k,k+2}(t)\right)\frac{\bigl.}{\bigr.}\right\}dt\nn\\
&&\quad+\sum_{m=1}^n\sum_{k=0}^{m-1}v^{[n]}_{m-k,k}(t)\frac{X_t^{m-k-1}Y_t^k}{(m-k-1)!k!}dX_t^c
+\sum_{m=1}^n\sum_{k=1}^m v^{[n]}_{m-k,k}(t)\frac{X_t^{m-k}Y_t^{k-1}}{(m-k)!(k-1)!}dY_t\nn \\
&&\quad+\sum_{m=0}^{n-1}\sum_{k=0}^m\frac{X_t^{m-k}Y_t^k}{(m-k)!k!}
\Bigl(\sum_{l=m+1}^n v^{[n]}_{l-k,k}(t)\int_K \frac{z^{l-m}}{(l-m)!}\caln(dt,dz)\Bigr)
\label{Vn-H-hypo}
\eea
which then implies (\ref{hypoZn}), (\ref{hypoGn}) and (\ref{hypoUn}).

On the other hand, extracting the $n$-th order part from the BSDE (\ref{epV-Heston}), one obtains
\bea
&&V_t^{[n]}=\frac{X_T^n}{n!}\part_x^n H(0)-\int_t^T Z_s^{[n]}dX_s^c-\int_t^T \Gamma_s^{[n]}dY_s
-\int_t^T \int_K U^{[n]}(s,z)\caln(ds,dz)\nn \\
&&-\int_t^T \left\{
\frac{\sigma_s^2}{2}\Bigl(Y_s Z_s^{[n-1]}+Z_s^{[n]}\Bigr)
+\beta(s)\sum_{l=0}^{n-1}\frac{\part_y^l\lambda(s,0)}{l!}Y_s^lZ_s^{[n-l]}
+\kappa_s Y_s \Gamma_s^{[n-1]}
\right. \nn \\
&&\qquad\qquad \left. -\sum_{l=0}^{n-1}\frac{\part_y^l\lambda(s,0)}{l!}Y_s^l
\int_K U^{[n-l]}(s,z)Q(s,dz)\right\}ds
\label{BSDE-Vn-Heston}
\eea
Substituting the control variables $Z,\Gamma$ and $U$ with assumed form in
(\ref{hypoZn}), (\ref{hypoGn}) and (\ref{hypoUn}), and reordering the 
summation, one can confirm that (\ref{BSDE-Vn-Heston}) becomes
\bea
&&V_t^{[n]}=\frac{X_T^n}{n!}\part_x^n H(0)-\int_t^T Z_s^{[n]}dX_s^c-\int_t^T \Gamma_s^{[n]}dY_s
-\int_t^T \int_K U^{[n]}(s,z)\caln(ds,dz)\nn \\
&&- \sum_{m=0}^n\sum_{k=0}^m\int_t^T\frac{X_s^{m-k}Y_s^k}{(m-k)!k!}\mbb{I}_{(m\leq n-1)} \left\{
\frac{\sigma_s^2}{2}\Bigl(\mbb{I}_{(1\leq k)}k v^{[n-1]}_{m-k+1,k-1}(s)+
v^{[n]}_{m-k+1,k}(s)\Bigr) \frac{\Bigl.}{\bigr.}
\right.\nn \\
&&\qquad+\sum_{l=0}^kC_{(k,l)}\beta(s)\part^l_y\lambda(s,0)v^{[n-l]}_{m-k+1,k-l}(s)+
\mbb{I}_{(1\leq k)}k\kappa(s)v^{[n-1]}_{m-k,k}(s)\nn \\
&&\qquad\left. -\sum_{l=0}^k C_{(k,l)}\part^l_y \lambda(s,0)\Bigl(
\sum_{j=1}^{n-m}v^{[n-l]}_{j+m-k,k-l}(s)\frac{q(s,j)}{j!}\Bigr)\right\}ds~.
\label{Vn-H-hypoB}
\eea
Then, matching the coefficients in the drift term of (\ref{Vn-H-hypo}) to those of 
(\ref{Vn-H-hypoB}) yields the system of the linear ODEs (\ref{H-ODE}). The terminal conditions
should be clear from the expression (\ref{Vn-H-hypoB}).

As long as the forward SDE (\ref{Vn-H-hypo}) is well-defined when using the solution of the 
ODEs (\ref{H-ODE}), it actually gives one possible solution for the $n$-th order BSDE (\ref{BSDE-Vn-Heston}).
Due to the linearity of the ODEs, the uniqueness of the solution within the assumed form
should be clear. $~\blacksquare$
\\

Note that the above system of ODEs can be easily solved {\it one-by-one}
by evaluating in the following order:
\bea
&&n:0\longrightarrow n_{\rm max}\\
&&m:n\longrightarrow 0 \\
&&k:0\longrightarrow m ~.
\eea

\subsection{Pricing formula for a European Option}
Suppose that we have obtained the good estimate of moments of $\gamma_m =\mbb{E}[X_T^m]$
for $m=1,2,\cdots$ from the truncated approximation of the BSDE (\ref{epV-Heston}) with $H(x)=x^m$.
The $n$-th order cumulant $\chi_n$ is given, in terms of these moments, by
\bea
\chi_n=n! \sum_{\{k_m\}}(-1)^{r-1}(r-1)!\sum_{m=1}^n \frac{1}{k_m!}\left(
\frac{\gamma_m}{m!}\right)^{k_m}
\eea
where the summation $\sum_{\{k_m\}}$ is taken for all the n-uplets of non-negative integers $\{k_1,\cdots, k_n\}$
satisfying the  Diophantine equation
\bea
k_1+2k_2+\cdots+n k_n=n~.
\eea
$r$ is defined by $r:=k_1+k_2+\cdots+k_n$.

Then, the Edgeworth expansion of the $X_T$'s density using up to the $n$-th order cumulant
is given by
\bea
p_n(x)=\phi(x;\mu,\Sigma^2)\left\{
1+\sum_{s=1}^{n-2}\sum_{\{k_m\}}\frac{1}{\Sigma^{s+2r}}H_{s+2r}\Bigl(
\frac{x-\mu}{\Sigma}\Bigr)\prod_{m=1}^s \frac{1}{k_m!}\Bigl(
\frac{\chi_{m+2}}{(m+2)!}\Bigr)^{k_m}\right\}
\label{xdensity}
\eea
where $\mu:=\chi_1$, $\Sigma:=\sqrt{\chi_2}$ and
\bea
\phi(x;\mu,\Sigma^2)=\frac{1}{\sqrt{2\pi}\Sigma}\exp\left(-\frac{1}{2}\Bigl(
\frac{x-\mu}{\Sigma}\Bigr)^2\right)~.
\eea
Here, the summation $\sum_{\{k_m\}} $is taken for all the s-uplets of non-negative integers
satisfying 
\be
k_1+2k_2+\cdots+s k_s=s
\ee
and 
\be
r:=k_1+k_2+\cdots+k_s~
\ee
in (\ref{xdensity}).
$H_n()$ denotes the Hermite polynomial defined by
\bea
H_n(x):=(-1)^n e^{\frac{x^2}{2}}\frac{d^n}{dx^n}e^{-\frac{x^2}{2}}~.
\eea
See, for example, Blinnikov and Moessner (1998)~\cite{Blinnikov} for a simple
derivation of the formulas and informative numerical examples of the density approximation
from the moments.

Then an approximated price of a Call option on $S_T$ with strike $K$ 
based on the $n$-th order $(n\geq 2)$ Edgeworth expansion~\footnote{We mean that 
the expansion using the cumulants $(\chi_i),~i=1,2, \cdots,n$.} is given by
\bea
&&C_n^K=\int_{-\infty}^{\infty} (S_0 e^x-K)^+ p_n(x)dx\nn \\
&&=\int_d^{\infty}(S_0 e^{\Sigma y+\mu}-K)\phi(y)
\left\{1+\sum_{s=1}^{n-2}\sum_{\{k_m\}}\frac{1}{\Sigma^{s+2r}}H_{s+2r}(y)
\prod_{m=1}^s \frac{1}{k_m!}\Bigl(\frac{\chi_{m+2}}{(m+2)!}\Bigr)^{k_m}\right\}dy \nn \\
\label{Call-Heston}
\eea
where
\bea
d:=\frac{\ln(K/S_0)-\mu}{\Sigma}, \quad \phi(y)=\frac{1}{\sqrt{2\pi}}e^{-\frac{y^2}{2}}~.
\eea
All the necessary integrations in (\ref{Call-Heston}) can be performed analytically 
thanks to the following properties of the Hermite polynomials:
\bea
\label{Hermite-1}
&&\int_d^{\infty}\phi(y)H_n(y)dy=\phi(d)H_{n-1}(d) \\
&&\int_d^{\infty}e^{\Sigma y}\phi(y)H_n(y)dy=e^{\Sigma d}\phi(d)H_{n-1}(d)
+\Sigma \int_d^{\infty}e^{\Sigma y}\phi(y)H_{n-1}(y)dy.
\label{Hermite-2}
\eea
Put options can be evaluated similarly.

\subsection{Numerical Examples}
\label{sec-Heston-num}
\begin{figure}[H]
\begin{center}	
\includegraphics[width=76mm]{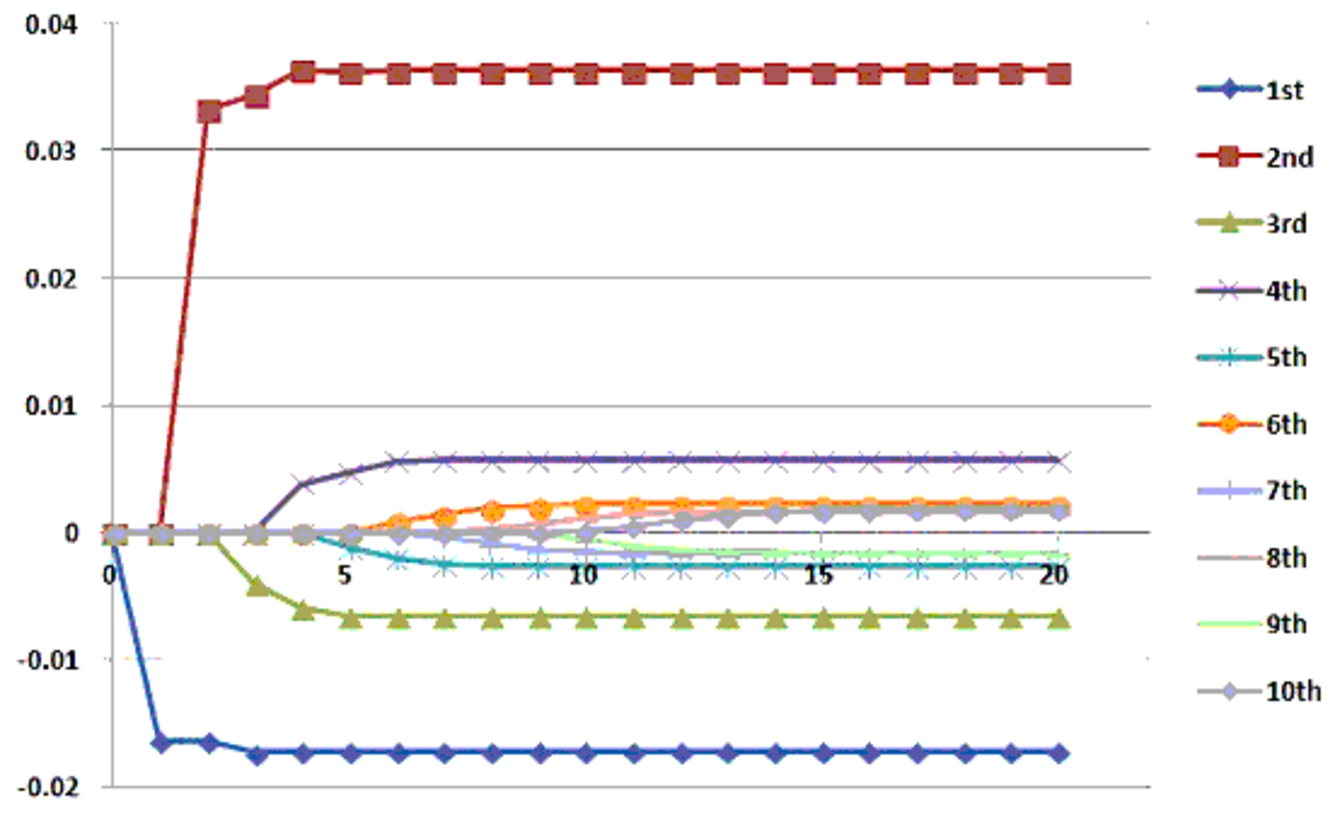}
\includegraphics[width=76mm]{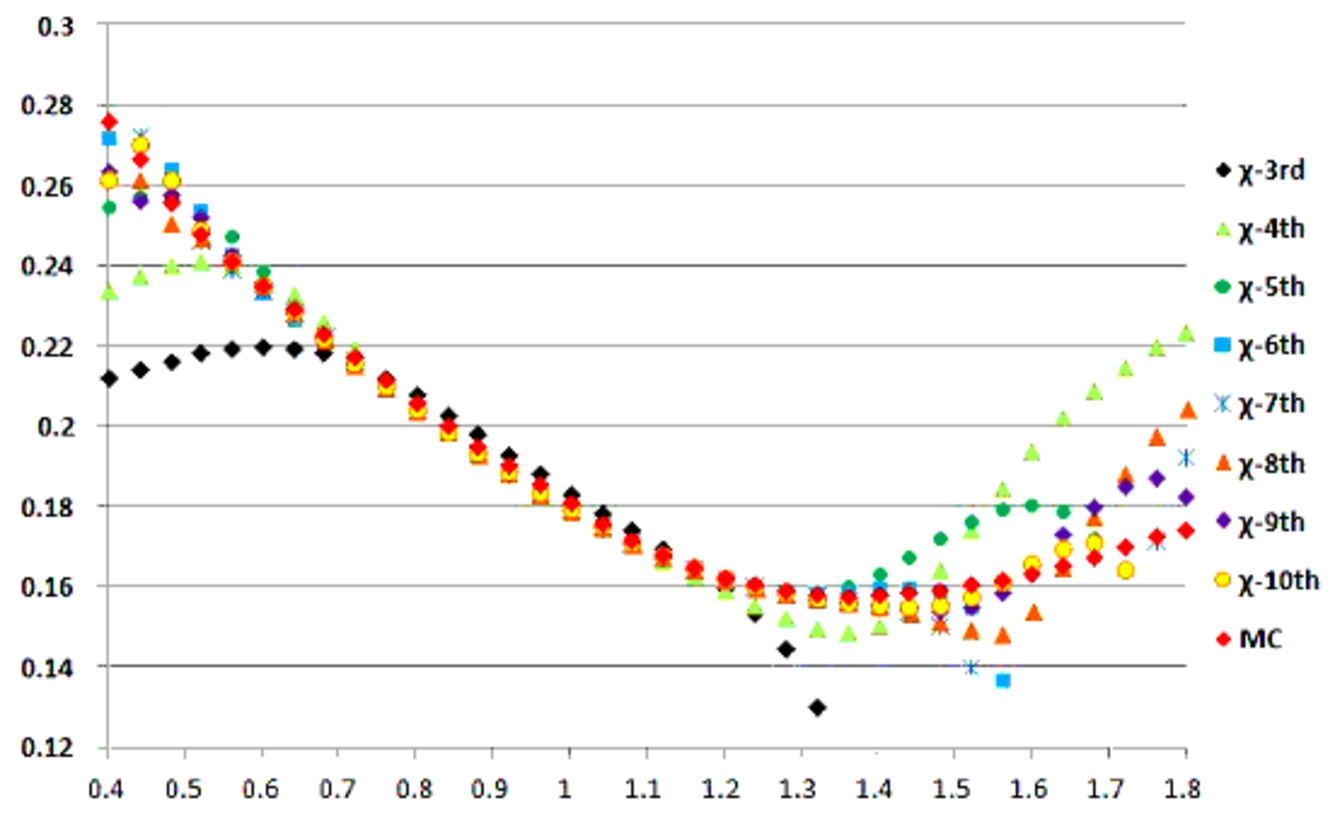}
\end{center}
\vspace{-2mm}
\caption{\footnotesize
Estimation of moments and implied volatilities. $T=1, \sigma=0.15, \alpha=0.6, \rho=-0.6, \kappa=0.1, \mu_J=-0.02, \sigma_J=0.03$ and 
$\lambda(t,Y_t)=8(Y_t+1)^2$.}
\label{Fig-1}
\end{figure}

\begin{figure}[H]
\begin{center}	
\includegraphics[width=76mm]{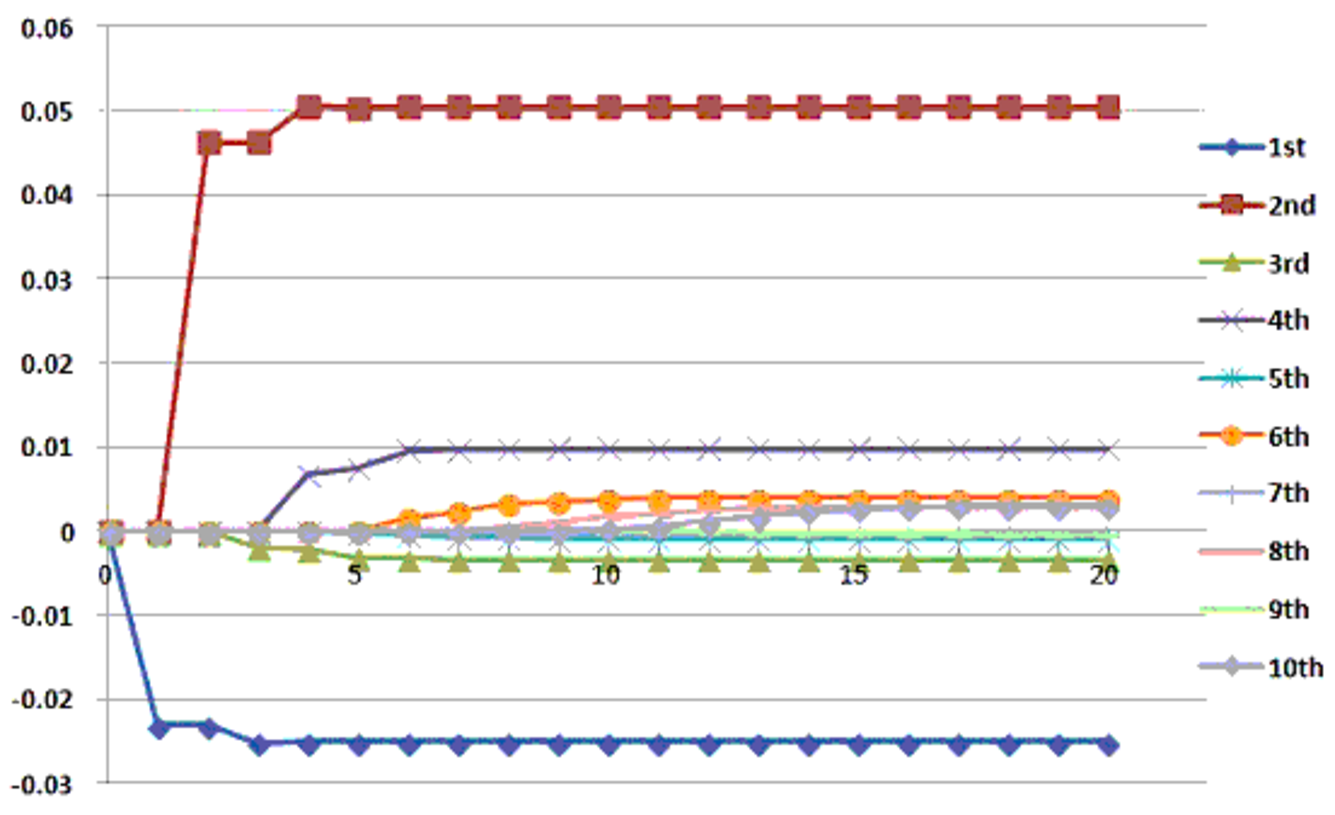}
\includegraphics[width=76mm]{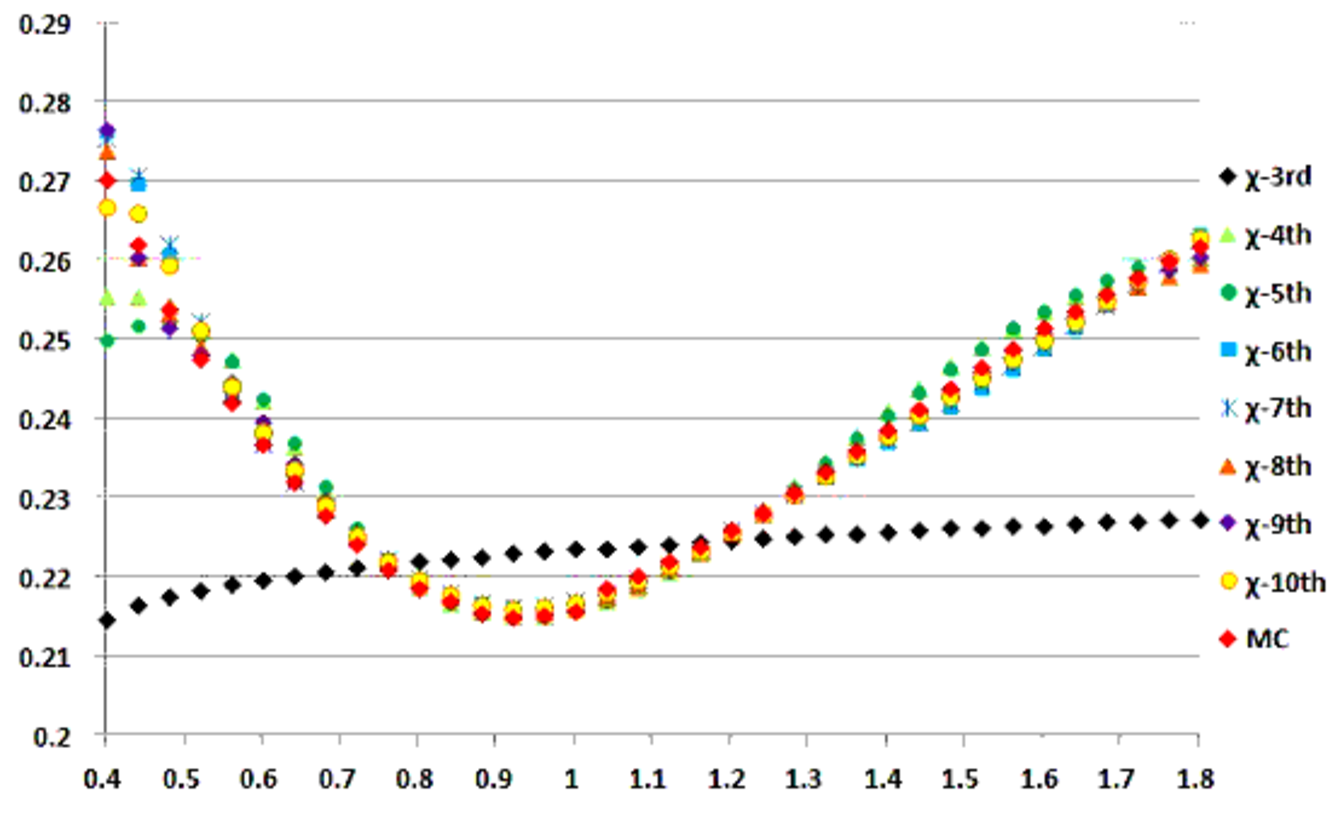}
\end{center}
\vspace{-2mm}
\caption{ \footnotesize
Estimation of moments and implied volatilities. $T=1, \sigma=0.15, \alpha=0.6, \rho=0, \kappa=0.1, \mu_J=-0.02, \sigma_J=0.03$ and 
$\lambda(t,Y_t)=8(Y_t+1)^2$.}
\label{Fig-2}
\end{figure}

\begin{figure}[H]
\begin{center}	
\includegraphics[width=76mm]{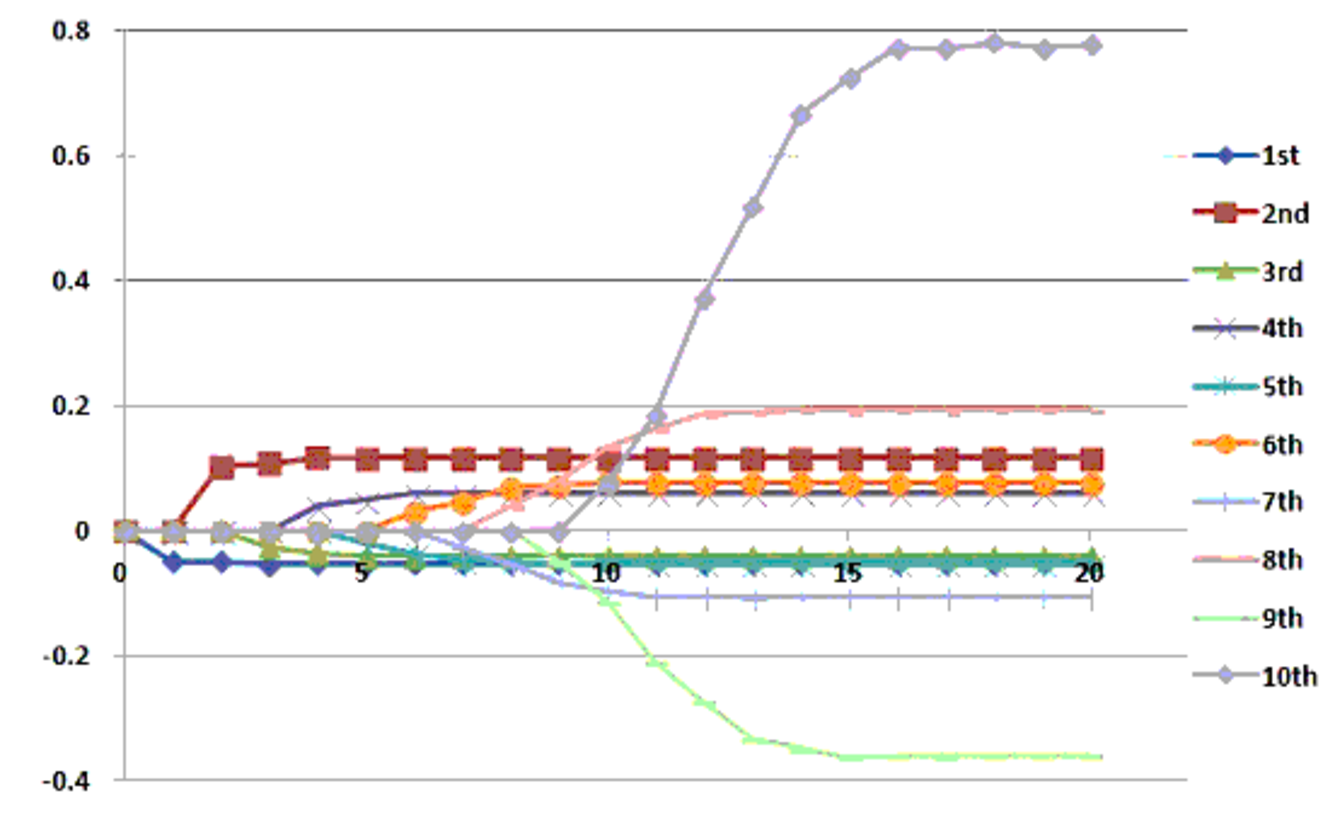}
\includegraphics[width=76mm]{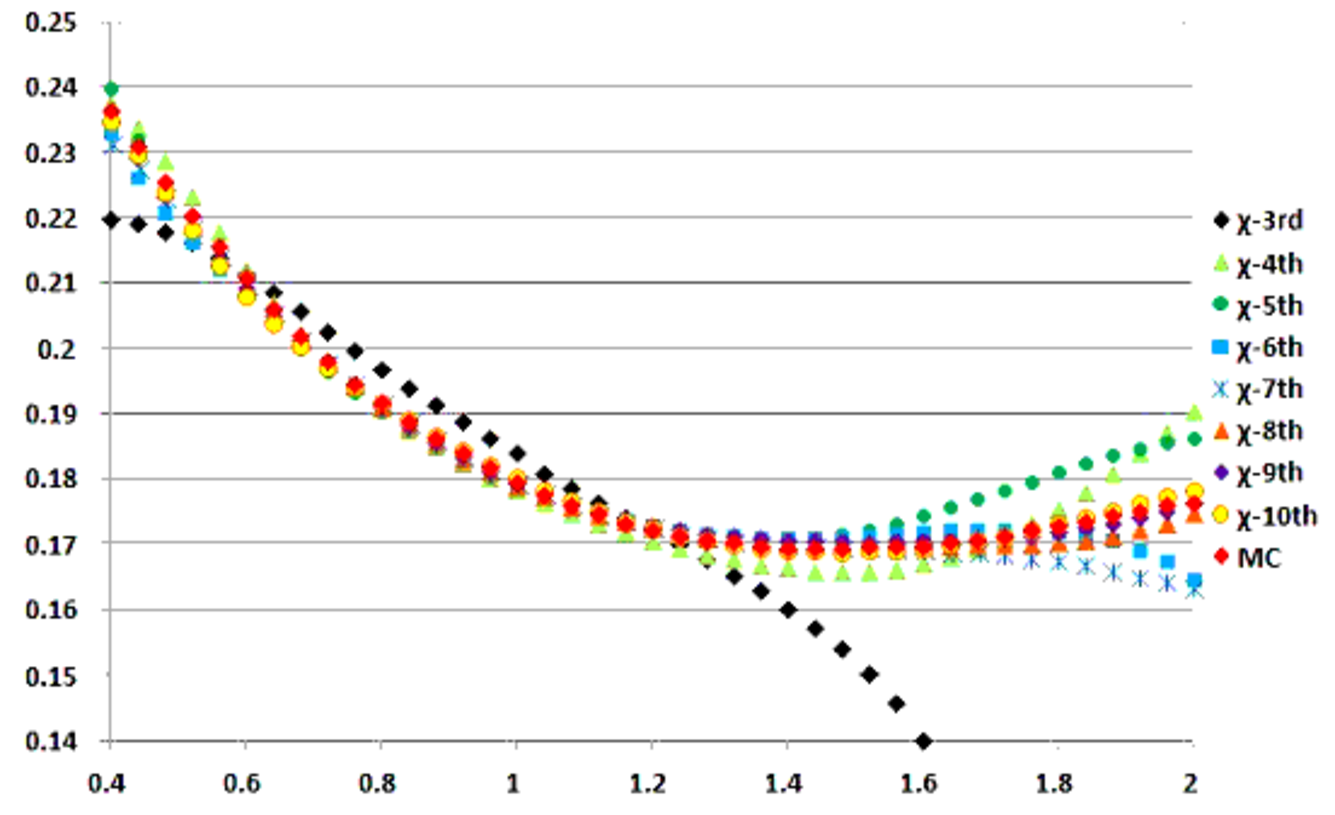}
\end{center}
\vspace{-2mm}
\caption{\footnotesize
Estimation of moments and implied volatilities. $T=3, \sigma=0.15, \alpha=0.5, \rho=-0.5, \kappa=0.1, \mu_J=0.01, \sigma_J=0.035$ and 
$\lambda(t,Y_t)=5Y_t^2+10Y_t+8$.}
\label{Fig-3}
\end{figure}

\begin{figure}[H]
\begin{center}	
\includegraphics[width=76mm]{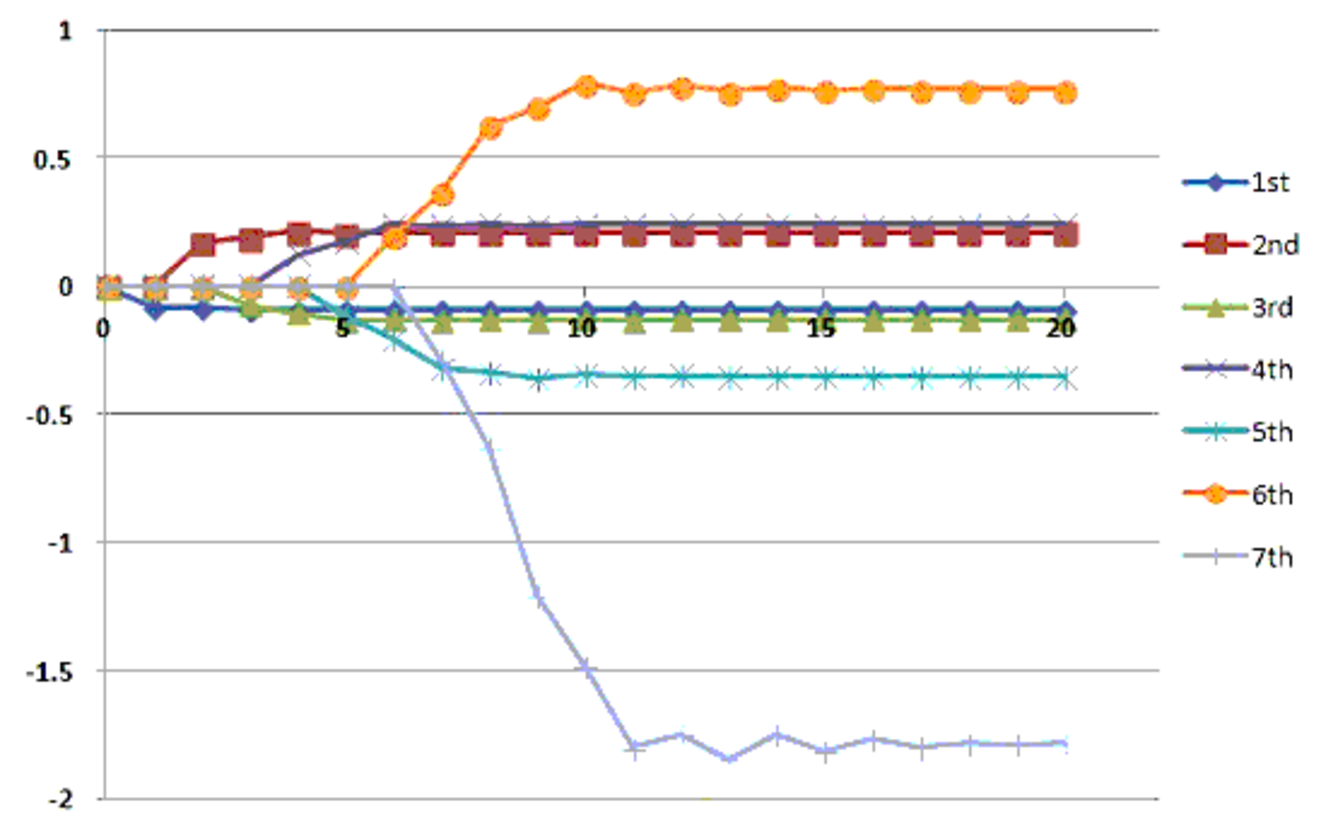}
\includegraphics[width=76mm]{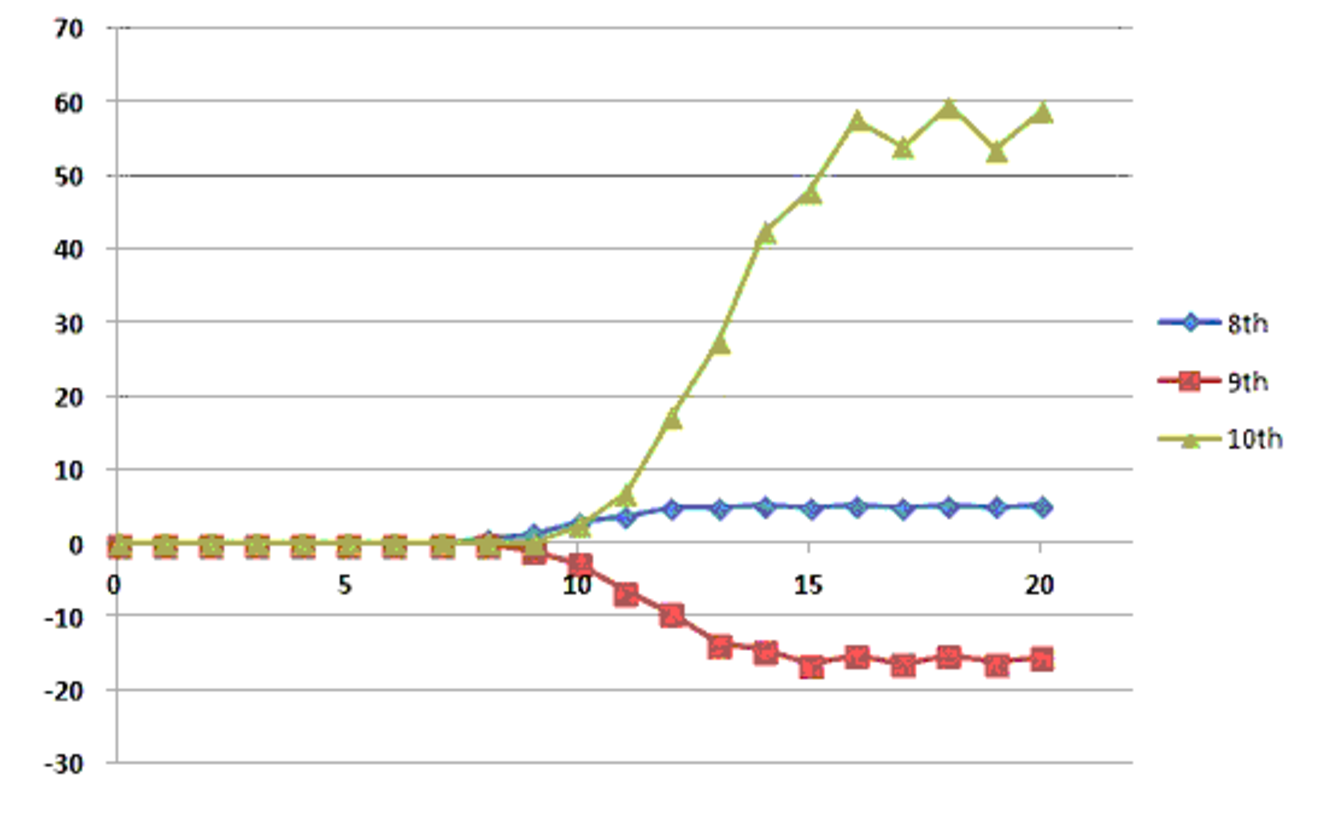}
\end{center}
\vspace{-2mm}
\caption{\footnotesize
Estimation of moments. $T=5, \sigma=0.15, \alpha=0.5, \rho=-0.5, \kappa=0.1, \mu_J=0.01, \sigma_J=0.035$ and 
$\lambda(t,Y_t)=5Y_t^2+10Y_t+8$.}
\label{Fig-4}
\end{figure}

\begin{figure}[H]
\begin{center}
\includegraphics[width=85mm]{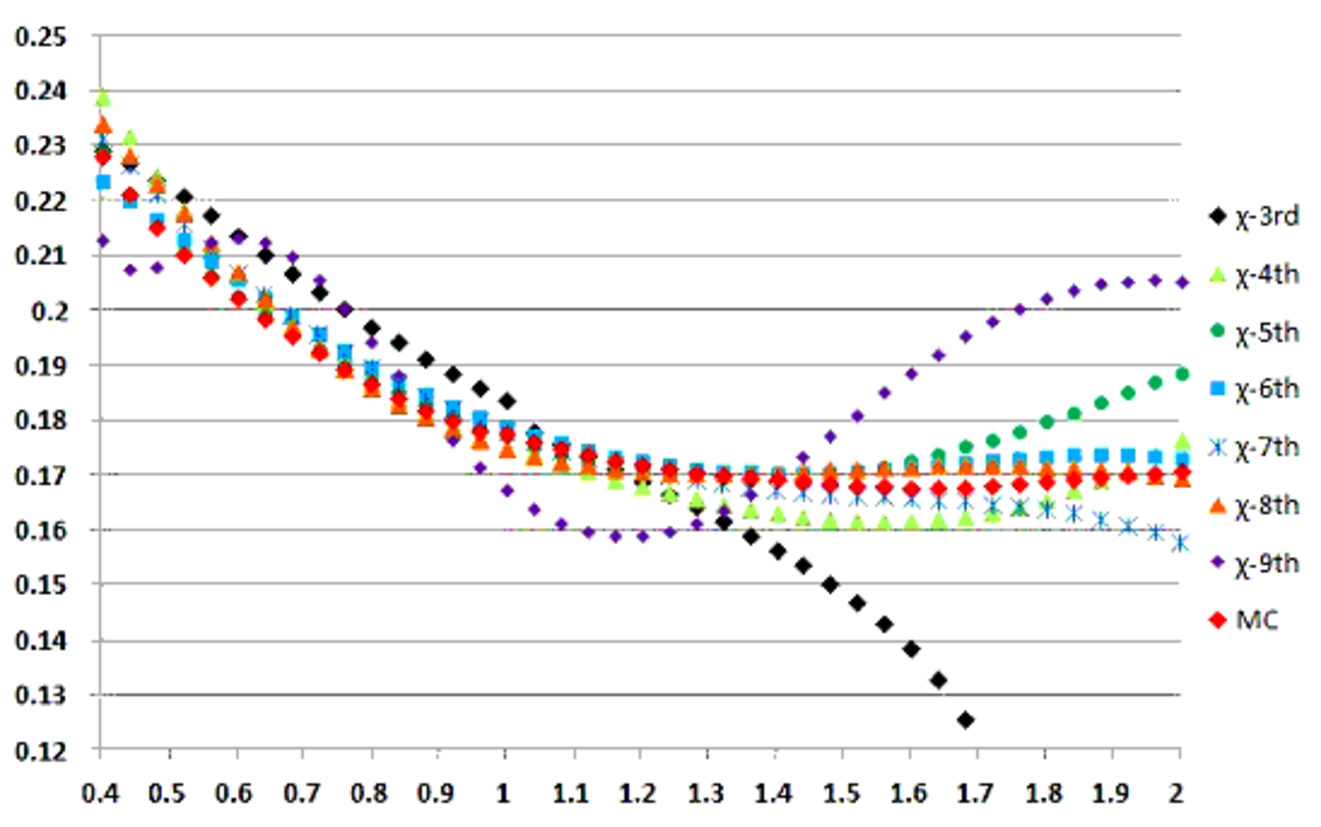}
\end{center}
\vspace{-2mm}
\caption{\footnotesize
Estimation of implied volatilities. $T=5, \sigma=0.15, \alpha=0.5, \rho=-0.5, \kappa=0.1, \mu_J=0.01, \sigma_J=0.035$ and  $\lambda(t,Y_t)=5Y_t^2+10Y_t+8$.}
\label{Fig-5}
\end{figure}

For numerical examples, we choose a set of constant parameters and a Gaussian jump density given 
as~\footnote{It does not have a compact support but the scheme still seems to work well in 
this example.}
\be
Q(t,dz)=\frac{1}{\sqrt{2\pi}\sigma_J}\exp\left(-\frac{1}{2}\Bigl(\frac{z-\mu_J}{\sigma_J}\Bigr)^2\right)dz~.
\ee
In each of Figure~\ref{Fig-1} to \ref{Fig-3}, the approximation of the moments $\gamma_m=\mbb{E}[X_T^m]~(m=1,\cdots, 10)$
with the expansion order up to $n=20$ based on the result of
Lemma~\ref{lemma-1} is given in the left-hand panel. 
Each line is connecting one of the  $\{\gamma_m\}$ estimated by 
the polynomial expansion up to the order specified by the horizontal axis.
Note that the approximation of $\gamma_m$ becomes non-zero
only for $n\geq m$. One can see that the lower-order moments  converge rather quickly.
The right-hand panel gives the comparison of the implied volatilities
approximated by the Edgeworth expansion using the corresponding order of cumulants  and the result from 
the Monte-Carlo simulation with 500,000 paths.  We have used Put options
for lower strikes by directly applying the corresponding formula without relying on the 
Put-Call parity.
The horizontal axis denotes the size of 
the strikes scaled by $S_0$, i.e.  $K/S_0$.
In Figure~\ref{Fig-4}, the higher moments $(\gamma_8, \gamma_9,\gamma_{10})$ 
are shown separately and the results for the implied volatilities are given in Figure~\ref{Fig-5}~\footnote{
The estimation based on $\chi_{10}$ is omitted since it seems to give a totally useless result.}.

As one can see from Figures~\ref{Fig-3} and \ref{Fig-4}, higher moments
grow rapidly for longer maturities and also the rate of convergence slows down.
As for  higher moments there is no guarantee
that the Edgeworth expansion converges even if the moments are accurately estimated~\footnote{Although it is similar, Gram-Charlier series generally gives much worse approximation.}.
In addition, by the very nature of polynomial expansion, when $|\gamma_m| \gg 1$
the expansion can be divergent.
As can be seen in Figure~\ref{Fig-5}, one may be better off by focusing on the 
lower moments (and cumulants) to get a stable approximation for a problem with long maturity.

\begin{figure}[H]
\begin{center}	
\includegraphics[width=76mm]{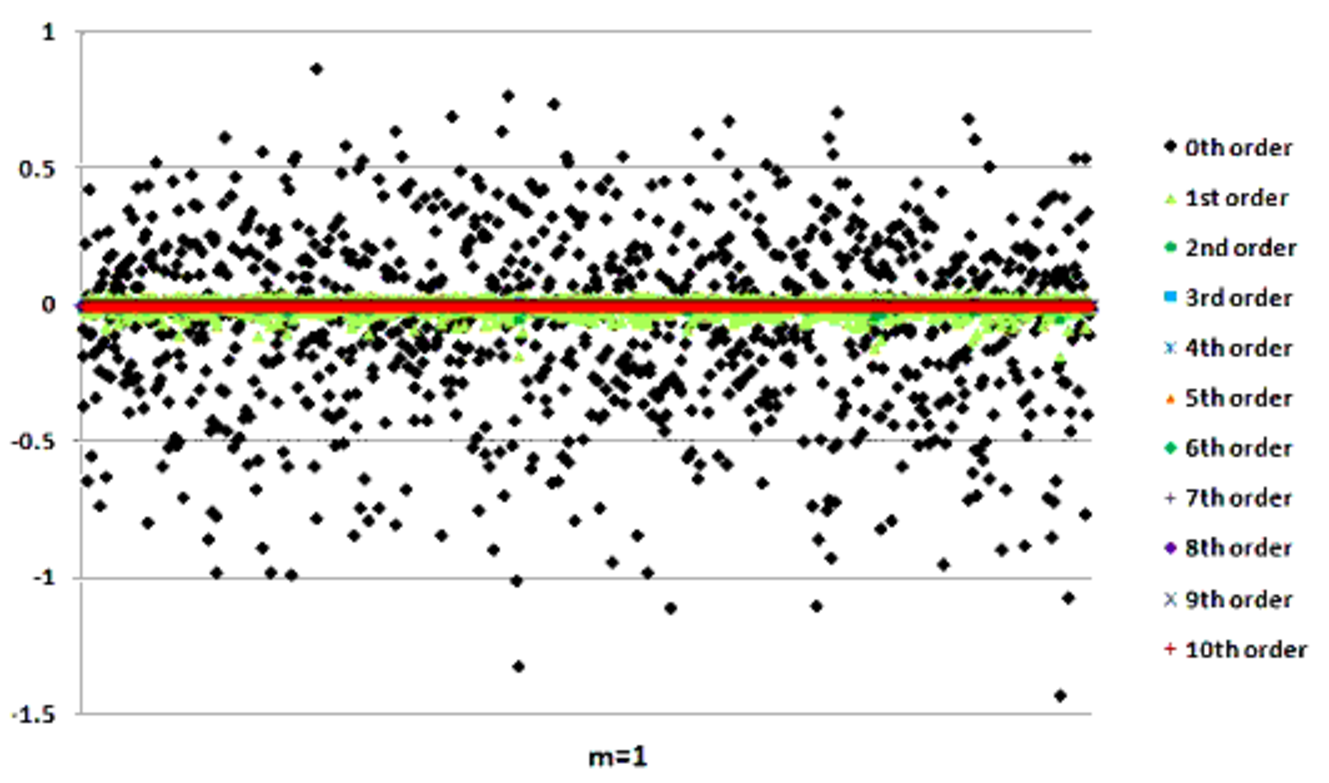}
\includegraphics[width=76mm]{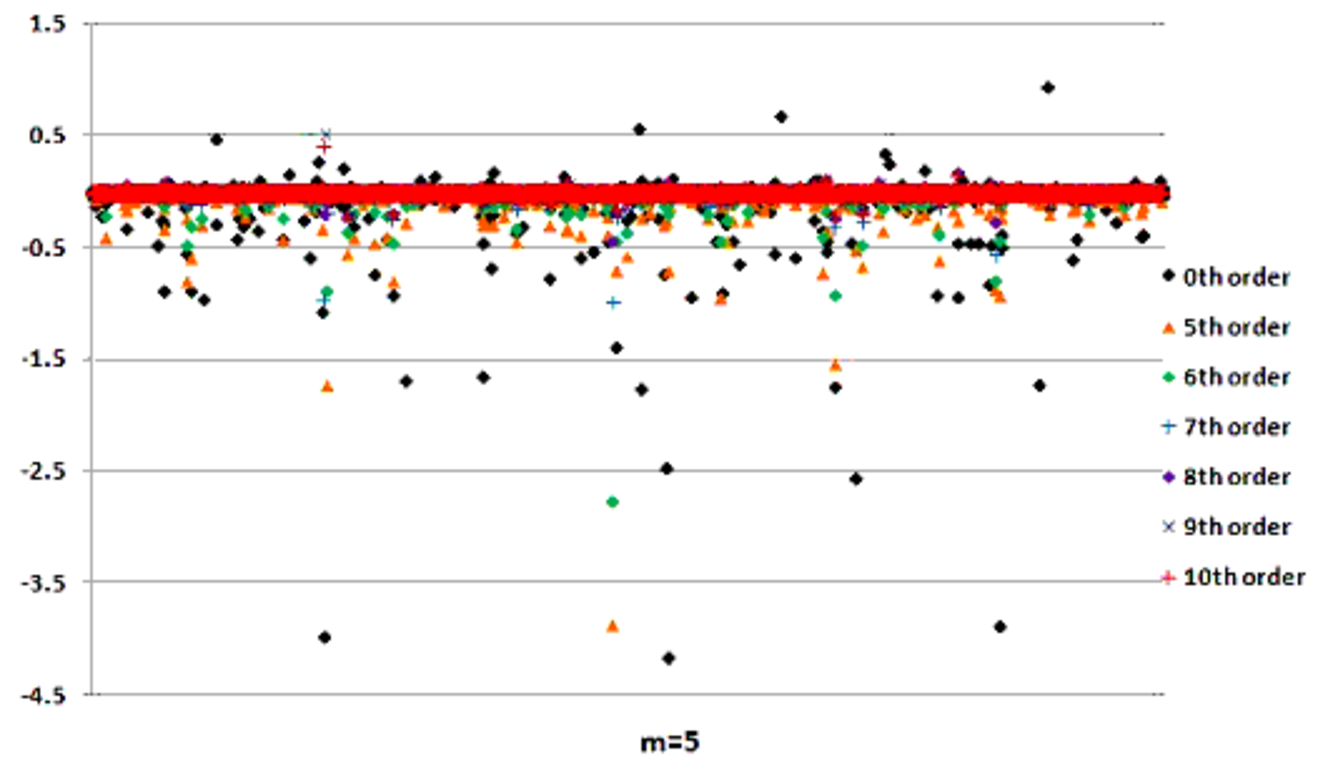}
\end{center}
\vspace{-2mm}
\caption{\footnotesize
Scattered plot of $[X_T^m-\wt{V}^{(n)}_T]$ ($m=1$ in the left and $m=5$
in the right) with various expansion orders $n$. The setup is the same as in Figure~\ref{Fig-3}. }
\label{Fig-moment}
\end{figure}

Before closing this section, let us study the path-wise nature of the current approximation scheme
for the terminal condition $H(X_T)=X_T^m$.
For each order of moment $m$ and expansion $n$, one can calculate the path-wise  truncated 
approximation error $[X_T^m-\wt{V}^{(n)}_T]$, where $\wt{V}^{(n)}$ 
is given by (\ref{Vtilde}) appropriately specified for the current model. 
In Figure~\ref{Fig-moment}, we have shown the scattered plot of this quantity for ($m=1$ and $5$)
with various orders of expansion $n$ using the same setup as in Figure~\ref{Fig-3},  i.e. $\{T=3, \sigma=0.15, \alpha=0.5, \rho=-0.5, \kappa=0.1, \mu_J=0.01, \sigma_J=0.035$ and
$\lambda(t,Y_t)=5Y_t^2+10Y_t+8\}$.
In Table~\ref{path-wise-moment}, the mean and standard deviation of  $[X_T^m-\wt{V}^{(n)}_T]$
are given for $m=\{1,2,\cdots,5\}$ in the same setup.
For ease of comparison, $\mbb{E}[X_T^m]$ estimated by simulation is also given in the lower table 
for each moment. Note that the non-trivial approximation 
exists only for $n\geq m$.
Improvement of approximation stops effectively at just a few higher order expansion $n\geq m$,
which means that the contributions of polynomial expansion 
for the target of $X_T^m$  is dominated by $m$-th and just a couple of higher order polynomials. 
This is rather natural and also consistent with the left panel of Figure~\ref{Fig-3}  
showing the convergence of approximation series for each moment.

One can observe that our scheme can provide accurate path-wise approximation of
$X^m$ but its error grows gradually for the higher moments. This fact can be naturally expected, 
since the contribution from the small number of realizations which reside in the tails of the distribution of $X_T$
becomes more important for higher moments. 
For the above example, the situation does not change meaningfully even if we use the pure diffusion model by putting $\lambda=0$.
We have observed a minor improvement of convergence only by a factor of few.
Since we have used the standard Euler scheme,
the corresponding simulation error may be contributing to the above result to some extent.

\begin{table}[H]
\scriptsize
\begin{tabular}{|c||c|c|c|c|c|c|c|c|c|c|c|c|}																			
\hline																			
$m=1$    & $n=0$ & $n=1$  & $n=2$ & $n=3$   & $n=5$  & $n=7$  & $n=10$  \\
\hline 
mean &  -0.054 & $-3.4\times 10^{-3}$&$-3.2\times 10^{-3} $&$6.7 \times 10^{-4}$ & $1.4\times 10^{-5}$ & $1.6\times 10^{-7}$ & $1.5\times 10^{-6} $\\
\hline
stdev & 0.34 &  0.028 & $4.6\times 10^{-3}$ & $9.5\times 10^{-4}$ & $1.3\times 10^{-4}$ & $1.2\times 10^{-4}$ &$1.2\times 10^{-4}$  \\ 
\hline  \hline 
$m=2$ & $n=0$ & $n=2$ & $n=3$ & $n=4$ & $n=5$ & $n=7$ & $n=10$ \\ 
\hline
mean & 0.12& 0.014 & $7.7\times 10^{-3}$ & $-1.6 \times 10^{-3}$ & $1.9\times 10^{-4}$ & $-4.7\times 10^{-5}$ & $-5.4 \times 10^{-5}$  \\  
\hline
stdev & 0.22 & 0.044 & 0.017 & $4.7\times 10^{-3}$ & $3.9\times 10^{-3}$ &  $3.9\times 10^{-3}$ & $3.9\times 10^{-3}$ \\
\hline \hline
$m=3$ & $n=0$ & $n=3$ & $n=4$ & $n=5$ & $n=6$ & $n=7$ & $n=10$ \\
\hline
mean & -0.041 & -0.017 & $-6.6\times 10^{-3}$ & $4.7\times 10^{-4}$ & $-1.8\times 10^{-4}$ & $1.5\times 10^{-4}$ & $3.9\times 10^{-5}$ \\ 
\hline
stdev & 0.27 & 0.084 & 0.043 &$5.6\times 10^{-3}$& $5.3\times 10^{-3}$& $5.3\times 10^{-3}$& $5.2\times 10^{-3}$ \\
\hline \hline
$m=4$ & $n=0$ & $n=4$ & $n=5$ & $n=6$ & $n=7$ & $n=8$ & $n=10$ \\
\hline
mean & 0.060 & 0.025 & 0.014 & $8.3\times 10^{-4}$ & $6.1\times 10^{-4}$ & $-4.4\times 10^{-4}$& $ -1.2\times 10^{-4}$ \\
\hline
stdev & 0.39 & 0.18 & 0.11 & 0.019 & 0.012 & 0.011 & 0.010 \\ 
\hline \hline
$m=5$ & $n=0$	& $n=5$ & $n=6$ & $n=7$ & $n=8$ & $n=9$ & $n=10$  \\ 
\hline
mean & -0.050 & -0.037 & -0.021 & $-5.2\times 10^{-3}$ & $-1.9\times 10^{-3}$ & $6.8\times 10^{-4}$ & $-9.0\times 10^{-6}$ \\
\hline
stdev & 0.65 & 0.40 & 0.28 & 0.084 & 0.042 & 0.025 & 0.024\\ 
\hline  
\end{tabular}

\scriptsize
\vspace{5mm}
\begin{tabular}{|c||c|c|c|c|c|}
\hline
  & $m=1$ & $m=2$& $m=3$ & $m=4$ & $m=5$ \\
  \hline
$\mbb{E}[X_T^m]$ & $-5.60\times 10^{-2}$ & $1.16\times 10^{-1}$ & $-4.30\times 10^{-2}$ & $6.33\times 10^{-2}$ & $-5.69\times 10^{-2}$ \\
\hline
\end{tabular}

\caption{\footnotesize
Mean and standard deviation of the path-wise realizations of $[X_T^{m}-\wt{V}^{(n)}_T]$ of $m=\{1,\cdots,5\}$ 
with the same setup as in Figure~\ref{Fig-3} by  simulation.
The second table gives the figures of $\mbb{E}[X_T^m]$ with $m=\{1,\cdots,5\}$ estimated by MC simulation for clarity.}
\label{path-wise-moment}
\end{table}

\section{An Application to $\lambda$-SABR Model}
\subsection{Problem Setup}
By using an appropriate change of variables, the proposed polynomial expansion scheme
can be applied to a wider choice of models than what may be naively imagined.
Let us consider (rescaled) $\lambda$-SABR model (SABR model with mean-reverting volatility)
under an equivalent martingale measure $\mbb{Q}$.
\bea
&&S_t=S_0+\int_0^t (S_0^{1-\beta})\sigma(s)\bar{Y}_s S_s^\beta dW_s \\
&&\bar{Y}_t=1+\int_0^t\Bigl( \alpha(s)\bar{Y}_s dB_s+\kappa(s)(1-\bar{Y}_s)ds\Bigr)
\eea
where $W$, $B$ are one-dimensional $\mbb{Q}$-Brownian motions with $d\langle W, B\rangle_t=\rho(t)dt$.
$(\sigma, \rho, \kappa)$ are all deterministic functions, and $\beta\in[0,1)$ is a constant.
Here, a factor of $S_0^{1-\beta}$ is included to make $\sigma$ roughly equal 
to the at-the-money implied volatility of $S$.

The change of variables 
\bea
&&X_t:=\frac{1}{1-\beta}\left(\Bigl(\frac{S_t}{S_0}\Bigr)^{1-\beta}-1\right) \\
&&Y_t:=\bar{Y}_t-1
\eea
leads to the dynamics
\bea
\label{SABR-X}
&&X_t=\int_0^t \left(\sigma(s)(1+Y_s)dW_s-\frac{\beta}{2}\sigma(s)^2b(X_s)(1+Y_s)^2ds\right)\\
&&Y_t=\int_0^t \Bigl(\alpha(s)(1+Y_s)dB_s-\kappa(s)Y_s ds\Bigr)
\label{SABR-Y}
\eea
where
\be
b(x):=\frac{1}{1+(1-\beta)x}~.
\ee
The assumption on the quadratic covariation (\ref{qcov}) is now satisfied for these new variables.

The BSDE relevant for a European contingent claim with terminal payoff $H(X_T)$ at maturity $T$
is given by
\bea
V_t&=&H(X_T)-\int_t^T \left(\frac{\beta}{2}\sigma(s)^2 b(X_s)(1+Y_s)^2 Z_s+\kappa(s)Y_s\Gamma_s\right)ds\nn \\
&&-\int_t^T Z_s dX_s-\int_t^T \Gamma_s dY_s
\label{V-SABR}
\eea
As in the Heston's case, we choose $H(x)=x^m,~(m=1,2,\cdots)$ 
to obtain the moment estimate of $X_T$ and then use the Edgeworth expansion to approximate 
its probability density.  Here, we are not claiming the Edgeworth expansion is the best choice
and different basis functions (such as Laguerre polynomials)  
can be more appropriate.

\subsection{Polynomial Expansion}
We now introduce $\ep$ to the BSDE (\ref{V-SABR}) so that we can perform polynomial expansion
\bea
V_t^\ep&=&H(\ep X_T)-\int_t^T \left(\frac{\beta}{2}\sigma(s)^2b(\ep X_s)(1+\ep Y_s)^2 Z_s^\ep
+\ep \kappa(s)Y_s\Gamma_s^\ep\right)ds\nn \\
&&-\int_t^T Z_s^\ep dX_s-\int_t^T \Gamma_s^\ep dY_s 
\label{SABR-epV}
\eea
as
\bea
\label{asympSABR1}
&&V_t^\ep=\sum_{n=0}^\infty \ep^n V_t^{[n]}  \\
&&Z_t^\ep=\sum_{n=0}^\infty \ep^n Z_t^{[n]}, \quad \Gamma_t^\ep =\sum_{n=0}^\infty \ep^n \Gamma_t^{[n]}~.
\label{asympSABR2}
\eea
We have the following lemma.

{\lemma{ If it exists, the polynomial solution for the expansion in (\ref{asympSABR1}) and (\ref{asympSABR2})
is uniquely given by 
\bea
\label{SABR-Vtn}
&&V_t^{[n]}=\sum_{m=0}^n\sum_{k=0}^m \frac{X_t^{m-k}Y_t^k}{(m-k)!k!}v^{[n]}_{m-k,k}(t)\\
\label{SABR-Ztn}
&&Z_t^{[n]}=\sum_{m=1}^n \sum_{k=0}^{m-1}\frac{X_t^{m-k-1}Y_t^k}{(m-k-1)!k!}v^{[n]}_{m-k,k}(t)\\
\label{SABR-Gtn}
&&\Gamma_t^{[n]}=\sum_{m=1}^n \sum_{k=1}^m \frac{X_t^{m-k}Y_t^{k-1}}{(m-k)!(k-1)!}v^{[n]}_{m-k,k}(t)
\eea
with the set of deterministic functions $v^{[n]}_{m-k,k}(t)$ of $(0\leq k\leq m \leq n)$
satisfying the following recursive system of linear ODEs
\bea
&&\dot{v}^{[n]}_{m-k,k}(t)=-\mbb{I}_{(2\leq k)}k(k-1)\left(
\frac{\sigma_t^2}{2}v^{[n]}_{m-k+2,k-2}(t)+\rho_t\sigma_t\alpha_t v^{[n]}_{m-k+1,k-1}(t)+
\frac{\alpha_t^2}{2}v^{[n]}_{m-k,k}(t)\right) \nn \\
&&\quad -\mbb{I}_{(m\leq n-1,1\leq k)}k\Bigl(
\sigma_t^2 v^{[n]}_{m-k+2,k-1}(t)+2\rho_t\sigma_t\alpha_t v^{[n]}_{m-k+1,k}(t)+
\alpha_t^2 v^{[n]}_{m-k,k+1}(t)\Bigr)\nn \\
&&\quad-\mbb{I}_{(m\leq n-2)}\left(
\frac{\sigma_t^2}{2}v^{[n]}_{m-k+2,k}(t)+\rho_t\sigma_t\alpha_t v^{[n]}_{m-k+1,k+1}(t)
+\frac{\alpha_t^2}{2}v^{[n]}_{m-k,k+2}(t)\right)\nn \\
&&\quad+\mbb{I}_{(m\leq n-1,1\leq k)}k \kappa_t v^{[n-1]}_{m-k,k}(t)
+\mbb{I}_{(m\leq n-1)}\frac{\beta}{2}\sigma_t^2 \sum_{l=0}^{m-k} C(m-k,l)\part^l_x b(0) \times
\nn \\
&&\quad  \left(v^{[n-l]}_{m-k-l+1,k}(t)+
\mbb{I}_{(1\leq k)} 2k~v^{[n-l-1]}_{m-k-l+1,k-1}(t)+\mbb{I}_{(2\leq k)}
k(k-1)~v^{[n-l-2]}_{m-k-l+1,k-2}(t)\frac{\bigl.}{}\right)\nn \\
\label{SABR-ODE}
\eea
having the terminal conditions $v^{[n]}_{n,0}(T)=\part_x^n H(0)$ with all the other 
components zero.
}\label{lemma-2}}
\\ \\
{\it Proof:} It can be proved in exactly the same way as Lemma 1. The derivation 
is given in the Appendix~\ref{ap-lemma-2}.

\subsection{Numerical Examples}
As in Section~\ref{sec-Heston-num}, let us provide several numerical examples for
the estimated moments and the comparison of the implied volatilities.
The number of paths for Monte-Carlo simulation is $500,000$ as before.
For this model, we cannot use the special relation in (\ref{Hermite-1}) and (\ref{Hermite-2}),
and hence we have carried out numerical integration of the estimated density for the pricing.
The styles and conventions used in each figures are the same as those in Section~\ref{sec-Heston-num}.

Although the polynomial expansion gives similar accuracy for short maturities,
its applicability to long maturities is rather limited compared to the previous extended Heston model.
The main cause seems to be the factor $k(k-1)$ appearing in the first line of the ODE given in Lemma~\ref{lemma-2},
which strongly drives $\{v^{[n]}_{m,k}\}$ especially for higher moments and makes them
unable to converge. This factor stems from the terms $\propto Y^2$ in the quadratic covariations.
In addition, since the support of $X_t$ is limited to the range $X_t \geq -\displaystyle{\frac{1}{1-\beta}}$,
the model's compatibility to the Edgeworth expansion may be lower than the Heston model.
This may be one of the reasons for somewhat unstable behavior of the implied volatilities 
when  higher-order cumulants are included. 

\begin{figure}[H]
\begin{center}	
\includegraphics[width=76mm]{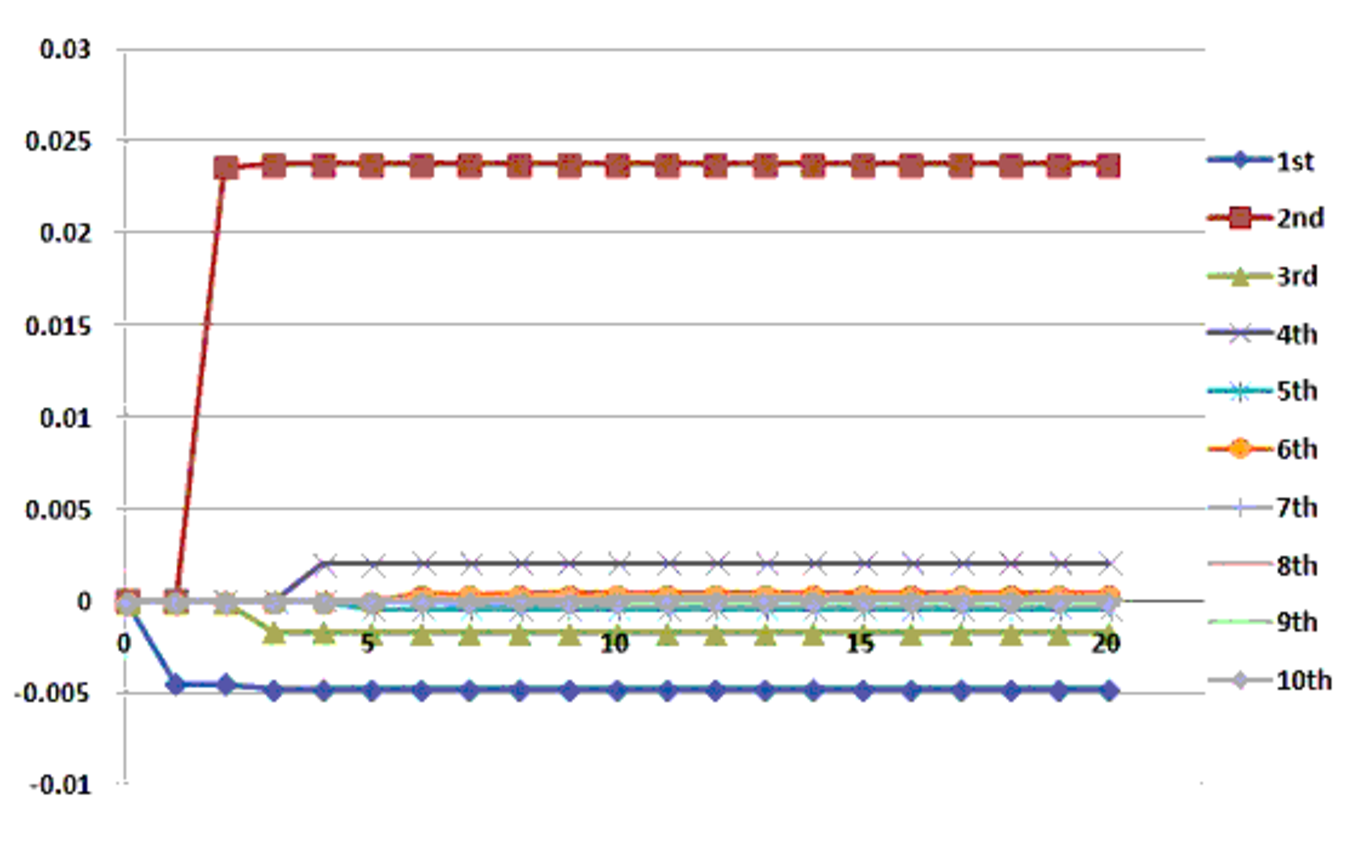}
\includegraphics[width=76mm]{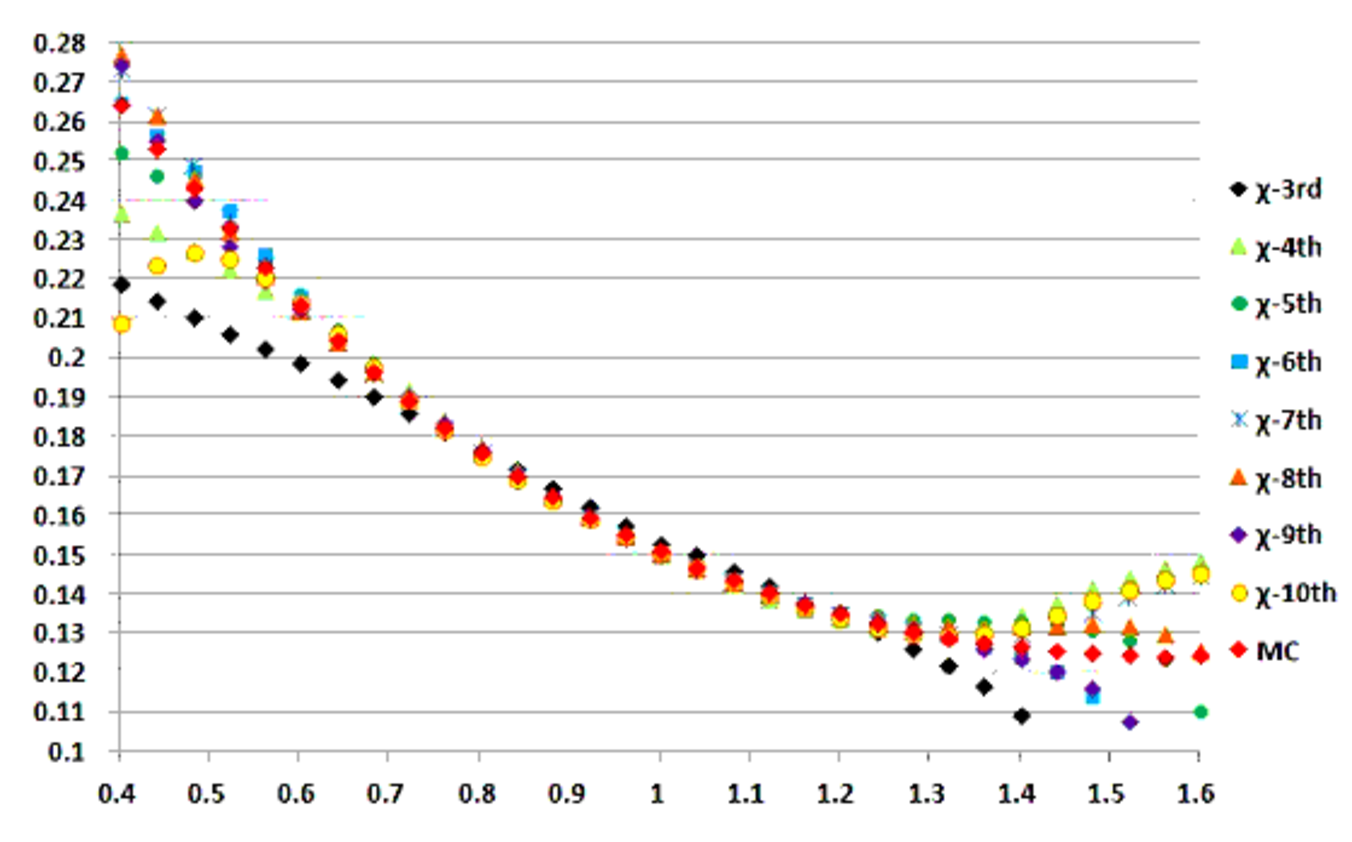}
\end{center}
\caption{ \footnotesize
Estimation of moments and implied volatilities. $T=1, \sigma=0.15, \alpha=0.3, \rho=-0.4, \kappa=0.1, \beta=0.4$.}
\label{Fig-s-2}
\end{figure}

For completeness, we give a convergence analysis for the 
path-wise realizations of the truncated approximation $[X_T^m-\wt{V}^{(n)}_T]$
as before. In Table~\ref{path-wise-moment-2}, the mean and standard deviation for $m=\{1,2,\cdots,5\}$
with various order of expansions are given under the same setup used in Figure~\ref{Fig-s-2}. 
In this model, the improvement of approximation stops more quickly than the previous Heston model case.
This is likely due to the smaller size of moments~\footnote{This 
is due to the performed  change of parameters.} and possibly other delicate model features. 
The quicker convergence of approximation series can also be seen  from the left panel of Figure~\ref{Fig-s-2}.

\vspace{5mm}
\begin{figure}[H]
\begin{center}	
\includegraphics[width=76mm]{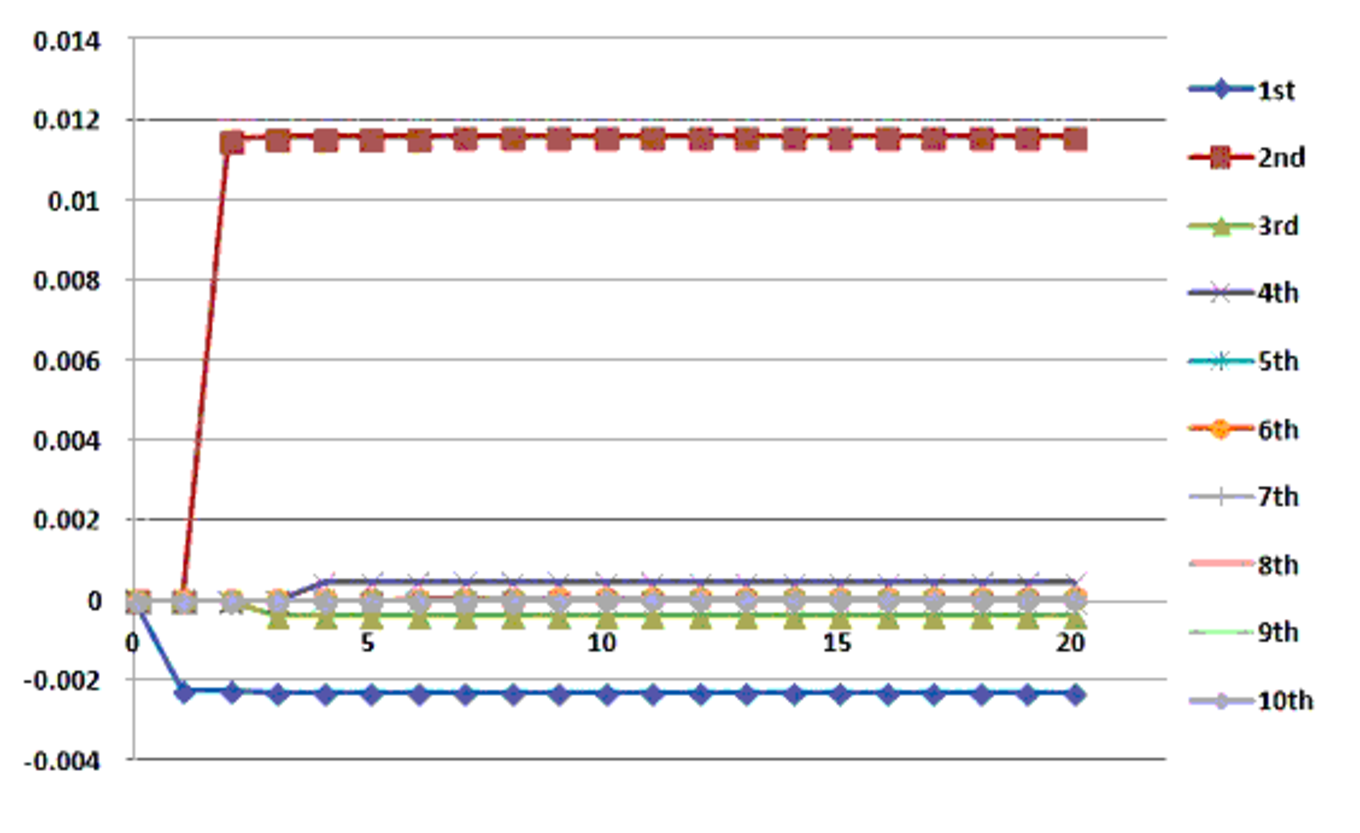}
\includegraphics[width=76mm]{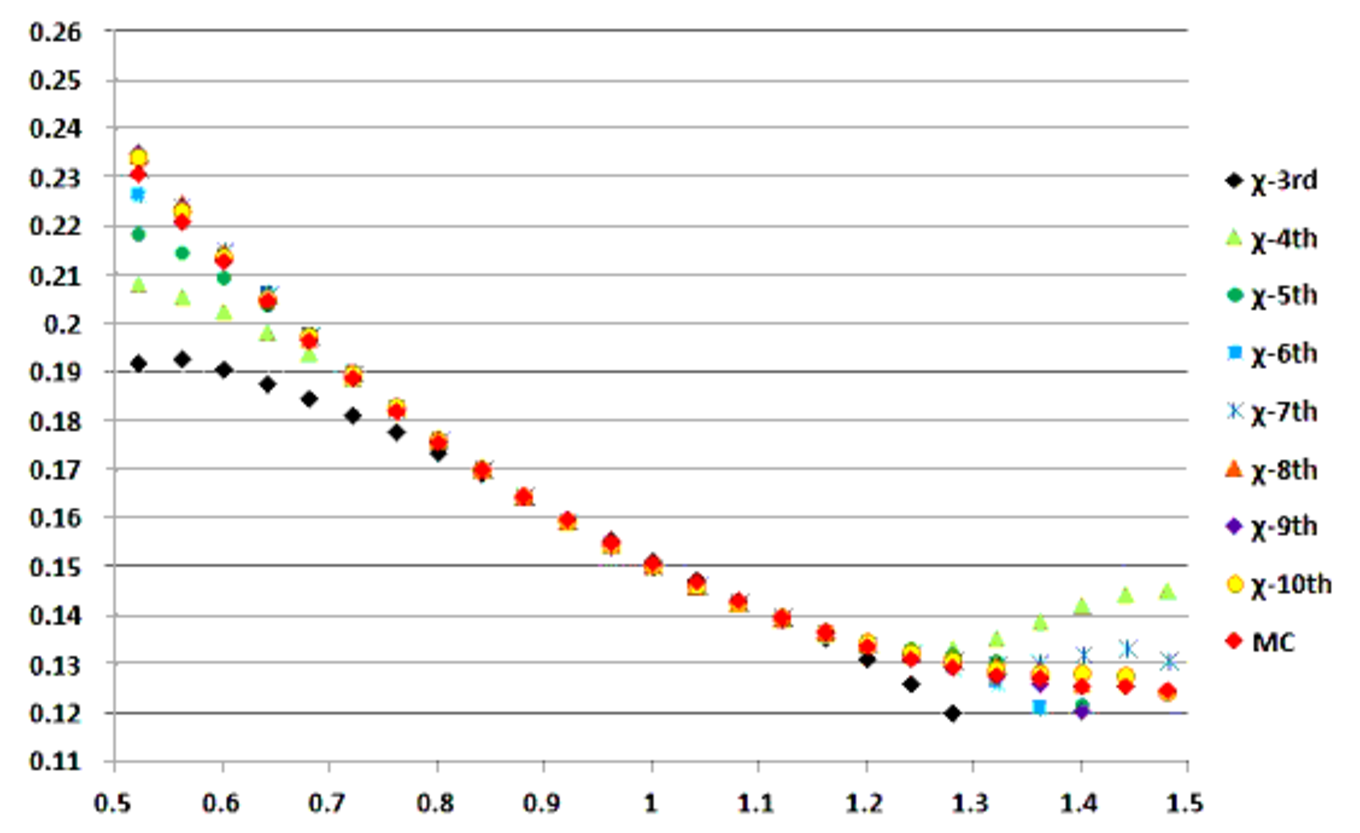}
\end{center}
\caption{ \footnotesize
Estimation of moments and implied volatilities. $T=0.5, \sigma=0.15, \alpha=0.3, \rho=-0.4, \kappa=0.1, \beta=0.4$.}
\label{Fig-s-1}
\end{figure}

\begin{figure}[H]
\begin{center}	
\includegraphics[width=76mm]{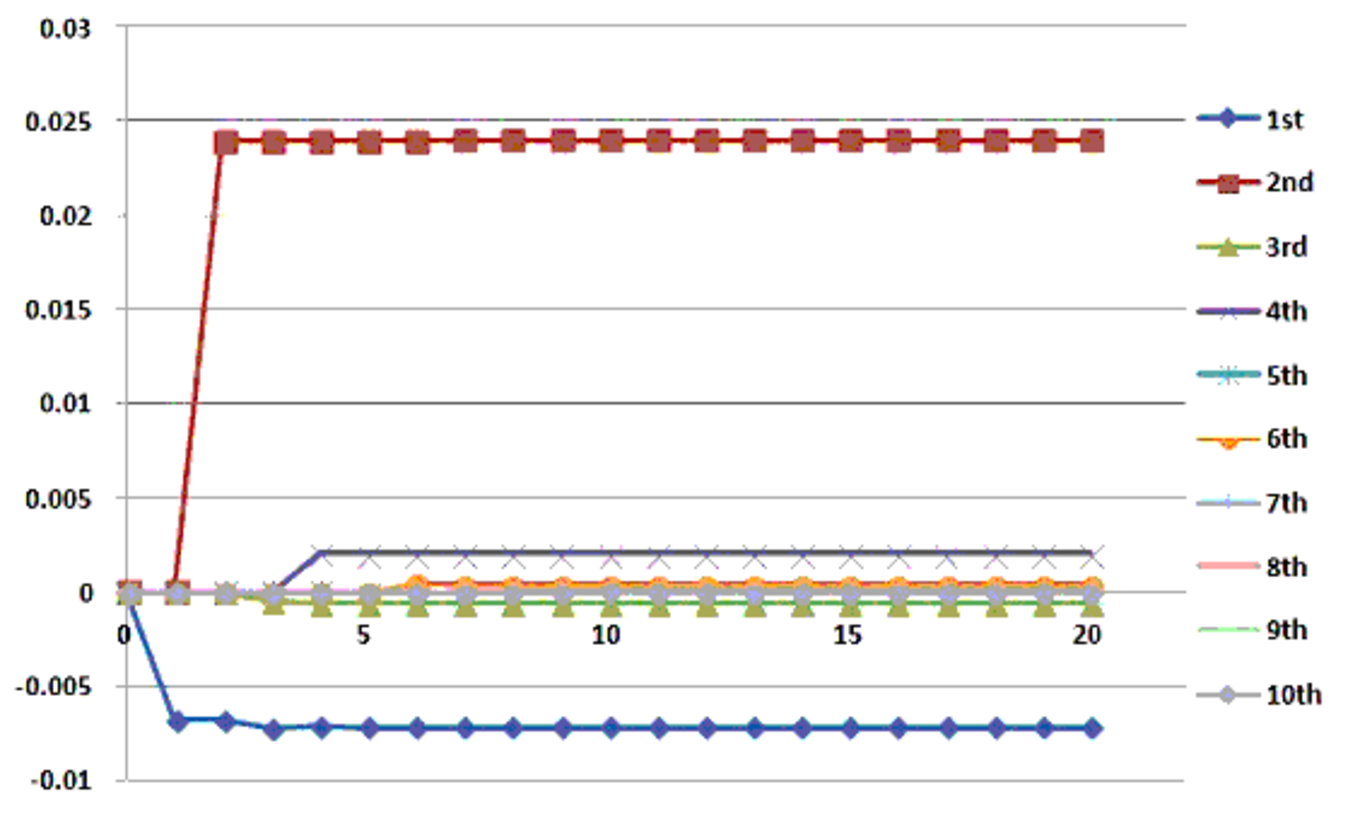}
\includegraphics[width=76mm]{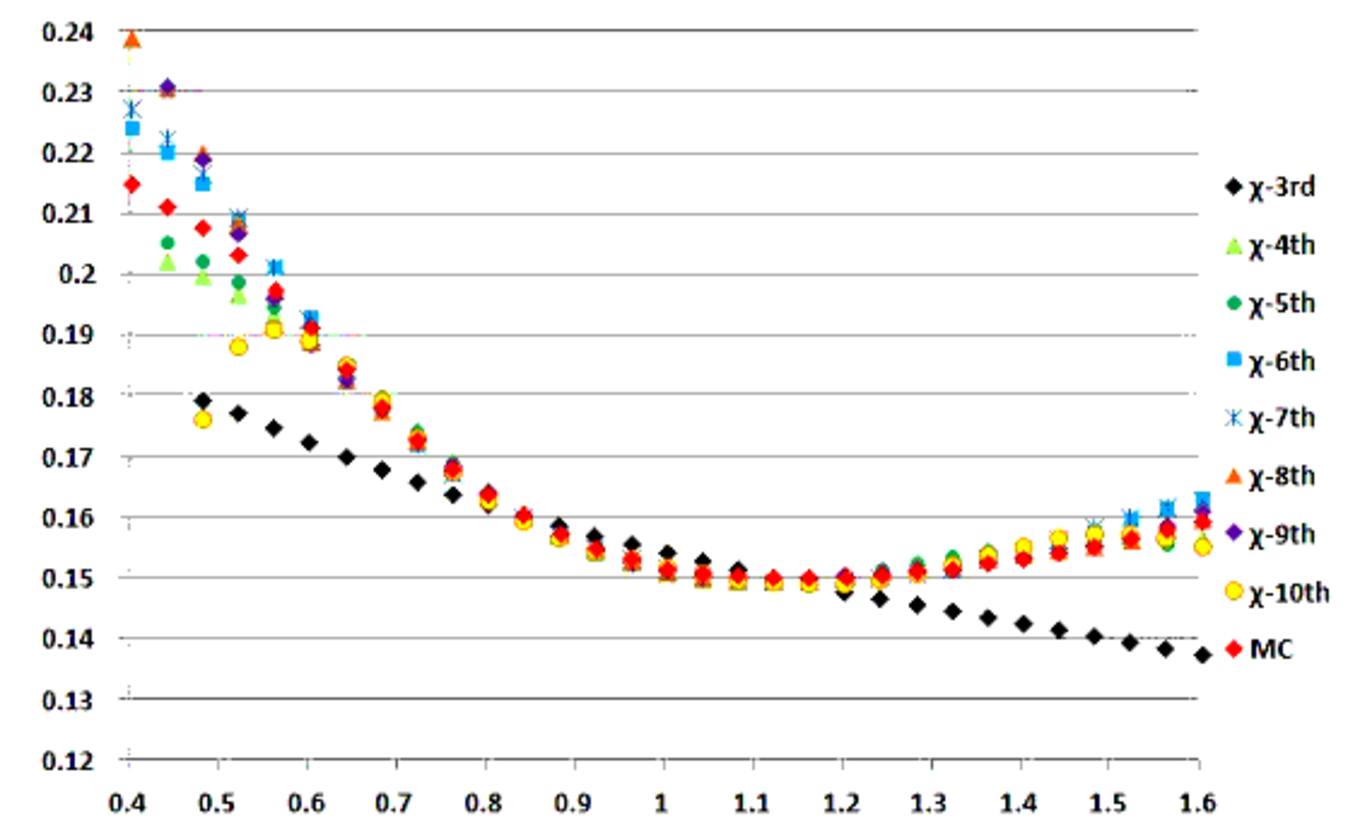}
\end{center}
\caption{\footnotesize
Estimation of moments and implied volatilities. $T=1, \sigma=0.15, \alpha=0.35, \rho=0, \kappa=0.1, \beta=0.6$.}
\label{Fig-s-3}
\end{figure}

\begin{table}[H]
\scriptsize
\begin{tabular}{|c||c|c|c|c|c|c|c|c|c|c|c|c|}																			
\hline																			
$m=1$    &$n=0$ & $n=1$  &$n=2$ & $n=3$   & $n=5$  & $n=7$  & $n=10$  \\
\hline 
mean & $-4.7\times 10^{-3}$ & $-2.8\times 10^{-4}$ & $-2.7\times 10^{-4} $ & $1.8\times 10^{-6}$ & $-4.4\times 10^{-7}$ 
& $5.5\times 10^{-8}$ & $1.0\times 10^{-7}$\\
\hline
stdev & 0.15 & $1.9\times 10^{-3}$ & $4.1\times 10^{-4}$ & $7.5\times 10^{-5}$ & $1.8\times 10^{-5}$ 
& $1.7\times 10^{-5}$ &$1.7\times 10^{-5}$  \\ 
\hline  \hline 
$m=2$ & $n=0$ & $n=2$ & $n=3$ & $n=4$ & $n=5$ & $n=7$ & $n=10$ \\ 
\hline
mean &0.024 & $2.1\times 10^{-4}$ & $5.3\times 10^{-5}$ & $2.2\times 10^{-5}$ & $1.2\times 10^{-5}$ & $1.1\times 10^{-5}$ 
& $1.0\times 10^{-5}$  \\  
\hline
stdev & 0.037 &$1.3\times 10^{-3}$ & $1.2\times 10^{-3}$ & $1.2\times 10^{-3}$ & $1.2\times 10^{-3}$ &  $1.2\times 10^{-3}$ 
& $1.2\times 10^{-3}$ \\
\hline \hline
$m=3$ & $n=0$ & $n=3$ & $n=4$ & $n=5$ & $n=6$ & $n=7$ & $n=10$ \\
\hline
mean & $-1.6\times 10^{-3}$ & $-4.0\times 10^{-5}$ & $-4.1\times 10^{-5}$ & $-5.7\times 10^{-6}$ & $-3.0\times 10^{-6}$ 
& $-1.8\times 10^{-6}$ & $-1.4\times 10^{-6}$ \\ 
\hline
stdev & 0.018 & $6.5\times 10^{-4}$ & $5.9\times 10^{-4}$  &$5.6\times 10^{-4}$& $5.6\times 10^{-4}$& $5.6\times 10^{-4}$& $5.6\times 10^{-4}$ \\
\hline \hline
$m=4$ & $n=0$ & $n=4$ & $n=5$ & $n=6$ & $n=7$ & $n=8$ & $n=10$ \\
\hline
mean & $1.9 \times 10^{-3}$ & $2.6 \times 10^{-5}$  & $2.1\times 10^{-5}$ & $7.9\times 10^{-6}$ 
& $5.4\times 10^{-6}$ & $4.7\times 10^{-6}$& $4.4\times 10^{-6}$ \\
\hline
stdev & $9.1\times 10^{-3}$ & $4.8\times 10^{-4}$ & $4.5\times 10^{-4}$ & $4.3\times 10^{-4}$ &$4.3\times 10^{-4}$ &$4.3\times 10^{-4}$ & $4.3\times 10^{-4}$\\ 
\hline \hline
$m=5$ & $n=0$	& $n=5$ & $n=6$ & $n=7$ & $n=8$ & $n=9$ & $n=10$  \\ 
\hline
mean & $-3.8\times 10^{-4}$ & $-9.7\times 10^{-6}$  & $-1.4\times 10^{-6}$ & $-4.6\times 10^{-6}$ & $-3.4\times 10^{-6}$ 
& $-2.9\times 10^{-6}$ & $-2.7\times 10^{-6}$ \\
\hline
stdev & $5.7\times 10^{-3}$  &$3.9\times 10^{-4}$ & $3.7\times 10^{-4}$  & $3.6\times 10^{-4}$ & $3.5\times 10^{-4}$ 
& $3.5\times 10^{-4}$  & $3.6\times 10^{-4}$ \\ 
\hline  
\end{tabular}

\scriptsize 
\vspace{2mm}
\begin{tabular}{|c||c|c|c|c|c|}
\hline
  & $m=1$ & $m=2$& $m=3$ & $m=4$ & $m=5$ \\
  \hline
$\mbb{E}[X_T^m]$ & $-4.78\times 10^{-3}$ & $2.39\times 10^{-2}$ & $-1.73\times 10^{-3}$ & $2.04\times 10^{-3}$ &
 $-4.48\times 10^{-4}$ \\
\hline
\end{tabular}

\caption{\footnotesize
Mean and standard deviation of the path-wise realizations of $[X_T^{m}-\wt{V}^{(n)}_T]$ of $m=\{1,\cdots,5\}$ with the setup in Figure~\ref{Fig-s-2} by  simulation. The second table gives the figures of $\mbb{E}[X_T^m]$ with $m=\{1,\cdots,5\}$ estimated by MC simulation for clarity.}
\label{path-wise-moment-2}
\end{table}

\section{Utility Optimization with  Terminal Liability}
European contingent claims, which we studied in the previous sections, can of course 
be solved without resorting to a complicated BSDE formulation.
The main motivation there was to get some insight about the performance of the 
proposed scheme by studying the two popular models.
Now, in this section, we treat a utility-optimization problem in an incomplete market where
solving a BSDE becomes crucially important.

Here, we adopt a simple Heston security market consists of 
one-risky asset with stochastic volatility. For simplicity, we
assume that the interest rate is zero. 
In the probability space of the {\it physical} measure $(\Omega,\calf, \mbb{P})$,
the dynamics of the underlying variables is assumed to be given by
\bea
&&S_t=S_0+\int_0^t S_s \sigma(s)\sqrt{\bar{Y}_s}\Bigl(dW_s+\bar{\theta}(s,S_s,\bar{Y}_s) ds\Bigr) \nn \\
&&\bar{Y}_t=1+\int_0^t \left(\frac{\bigl.}{\bigr.}
\alpha(s)\sqrt{\bar{Y}_s}\Bigl(dB_s+\rho(s)\bar{\theta}(s,S_s,\bar{Y}_s) ds\Bigr)+\kappa(s)(1-\bar{Y}_s)ds\right)
\eea
where $W,B$ are $\mbb{P}$-Brownian motions with $d\langle W, B\rangle_t=\rho(t)dt$.
$\sigma,\alpha$ and $\kappa$ are deterministic functions of time,
and $\bar{\theta}:[0,T]\times \mbb{R}_+^2 \rightarrow \mbb{R}$ gives the 
risk-premium process associating with $W$. The risk-premium for $B$ is implied by the $\rho\bar{\theta}$ as well as 
the mean-reverting term of $\bar{Y}$.

Given a portfolio strategy $(\pi_t)_{t\geq 0}$, the wealth at the terminal time $T~(>t)$ is 
given by
\bea
\calw^\pi_T(t,w)=w+\int_t^T \pi_u dS_u~.
\eea
In the reminder of this section, we are going to study the BSDE associated with the
exponential cost minimization:
\bea
V(t,w)={\rm ess}\inf_{\pi}\mbb{E}\left[
\exp\left(\frac{\bigl.}{}\gamma \Bigl(\bar{H}(S_T,Y_T)-\calw_T^\pi(t,w)\Bigr)\right)\Bigr|\calf_t \right]
\eea
where $\gamma$ is a positive constant specifying the risk averseness,
 and $\bar{H}:\mbb{R}_+^2\rightarrow\mbb{R}$ is a 
smooth function denoting the terminal liability. 
Using It\^o-Ventzell formula and the transformation
\be
V(t)=\ln\Bigl(V(t,w)e^{\gamma w}\Bigr)
\label{exT}
\ee
one can show that the following BSDE holds:
\bea
&&V_t=\gamma \bar{H}(S_T,\bar{Y}_T)-\int_t^T\left\{
\frac{1}{2}\bar{\theta}(s,S_s,\bar{Y}_s)^2-\frac{1}{2}(1-\rho(s)^2)\bar{\Gamma}_s^2\right\}ds \nn \\
&&\quad-\int_t^T \bar{Z}_s \Bigl[dW_s+\bar{\theta}(s,S_s,\bar{Y}_s)ds\Bigr] 
-\int_t^T \bar{\Gamma}_s \Bigl[dB_s+\rho(s)\bar{\theta}(s,S_s,\bar{Y}_s)ds\Bigr]~.
\label{BSDE-util}
\eea
It is well-known that the transformation (\ref{exT}) makes  $V(t)$ independent from 
the initial wealth $w$. The details and various interesting topics
can be found in a comprehensive review by Mania \& Tevzadze (2008)~\cite{Mania2}.
Similar qgBSDE arises in other economically important setups too, 
such as Power and  HARA (hyperbolic absolute risk aversion) utilities after appropriate change of variables.
We simply study (\ref{BSDE-util})  for a demonstrative purpose of the current approximation scheme.

It is important to notice that one cannot make use of Cole-Hopf transformation 
to convert the qgBSDE (\ref{BSDE-util}) to a solvable linear BSDE as long as $(\bar{H},\bar{\theta})$
depend on both of the $S$ and $\bar{Y}$. For example, if both of them depend only on $\bar{Y}$,
one can solve it analytically by following the arguments given by Zariphopoulou (2001)~\cite{Zaripho}.

As in Section~\ref{sec-Heston}, we perform the change of variables
\bea
&&X_t=\ln\left(\frac{S_t}{S_0}\right) \\
&&Y_t=\bar{Y}_t-1
\eea
and define 
\bea
\theta(s,X_s,Y_s):=\bar{\theta}(s,S_s,\bar{Y}_s)~.
\eea
The relevant forward SDEs are now given by
\bea
\label{X-util}
&&X_t=\int_0^t\left\{ \sigma(s)\sqrt{Y_s+1}\Bigl(dW_s+\theta(s,X_s,Y_s)ds\Bigr) 
-\frac{\sigma(s)^2}{2}(Y_s+1)ds \frac{\bigl.}{}\right\}\\
&&Y_t=\int_0^t\left\{\alpha(s)\sqrt{Y_s+1}\Bigl(dB_s+\rho(s)\theta(s,X_s,Y_s)ds\Bigr)
-\kappa(s)Y_s ds\frac{\bigl.}{}\right\}~.
\label{Y-util}
\eea
Simple redefinition of variables yields
\bea
&&V_t=\gamma H(X_T,Y_T)-\int_t^T Z_s dX_s-\int_t^T \Gamma_s dY_s -\int_t^T\left\{
\frac{1}{2}\Theta(s,X_s,Y_s) \right. \nn \\
&&\qquad \left. -\frac{\alpha(s)^2}{2}(1-\rho(s)^2)(1+Y_s)\Gamma_s^2
+\frac{\sigma(s)^2}{2}(1+Y_s)Z_s+\kappa(s)Y_s\Gamma_s\right\}ds
\label{V-util}
\eea
where $H(X_T,Y_T):=\bar{H}(S_T,\bar{Y}_T)$ and $\Theta(s,X_s,Y_s)=\theta(s,X_s,Y_s)^2$~.
The control variables are connected to those in (\ref{BSDE-util}) by
\be
\bar{Z}_s=Z_s\sigma(s)\sqrt{Y_s+1}, \quad \bar{\Gamma}_s=\Gamma_s\alpha(s)\sqrt{Y_s+1}~.
\ee
We assume the system of the forward and backward SDEs (\ref{X-util}), (\ref{Y-util}) and
(\ref{V-util}) has a well-posed solution in the reminder of the section.
Although it deviates from the main subject of the paper, 
it is interesting to notice that the above BSDE has a
simple exact solution in a special case.  The details
are give in Appendix~\ref{util-exact}.

\subsection{Polynomial Expansion}
In order to obtain the polynomial approximation for the system (\ref{X-util}), (\ref{Y-util})
and (\ref{V-util}), let us introduce $\ep$ and consider the perturbed BSDE:
\bea
&&V_t^\ep=\gamma H(\ep X_T, \ep Y_T) -\int_t^T Z_s^\ep dX_s -\int_t^T \Gamma_s^\ep dY_s
-\int_t^T \left\{\frac{\bigl.}{}
\frac{1}{2}\Theta(s,\ep X_s, \ep Y_s)  \right. \nn \\
&&\quad \left.-\frac{\alpha(s)^2}{2}(1-\rho(s)^2)(1+\ep Y_s)[\Gamma_s^\ep]^2+
\frac{\sigma(s)^2}{2}(1+\ep Y_s)Z_s^\ep +\ep \kappa(s)Y_s\Gamma_s^\ep \frac{\bigl.}{}\right\}ds
\label{epV-util}
\eea
and the associated expansion
\bea
\label{poly-V-util}
&&V_t^\ep=\sum_{n=0}^\infty \ep^n V_t^{[n]} \\
&&Z_t^\ep=\sum_{n=0}^\infty \ep^n Z_t^{[n]}, \qquad \Gamma_t^\ep=\sum_{n=0}^\infty \ep^n \Gamma_t^{[n]}~.
\label{poly-C-util}
\eea
The approximate solution of the original system is obtained by truncating the summation at a certain order $n$
and putting $(\ep=1)$ as explained in Section~\ref{sec-asympgeneral}.
For this model, we have the following result:

{\lemma{If it exists, the polynomial solution for the expansion in (\ref{poly-V-util}) and (\ref{poly-C-util})
is uniquely given  by
\bea
\label{Vn-util}
&&V_t^{[n]}=\sum_{m=0}^n \sum_{k=0}^m \frac{X_t^{m-k}Y_t^k}{(m-k)!k!}v^{[n]}_{m-k,k}(t) \\
\label{Zn-util}
&&Z_t^{[n]}=\sum_{m=1}^n \sum_{k=0}^{m-1}\frac{X_t^{m-k-1}Y_t^k}{(m-k-1)!k!}v^{[n]}_{m-k,k}(t)\\ 
&&\Gamma_t^{[n]}=\sum_{m=1}^n \sum_{k=1}^m \frac{X_t^{m-k}Y_t^{k-1}}{(m-k)!(k-1)!}v^{[n]}_{m-k,k}(t)
\label{Gn-util}
\eea
with the set of deterministic functions $v^{[n]}_{m-k,k}(t)$ of $(0\leq k\leq m\leq n)$ satisfying 
the following recursive system of linear ODEs
\bea
&&\dot{v}^{[n]}_{m-k,k}(t)=-\mbb{I}_{(m\leq n-1,1\leq k)}k \left(
\frac{\sigma_t^2}{2}v^{[n]}_{m-k+2,k-1}(t)+\rho_t\sigma_t\alpha_t v^{[n]}_{m-k+1,k}(t)
+\frac{\alpha_t^2}{2}v^{[n]}_{m-k,k+1}(t)\right)\nn \\
&&-\mbb{I}_{(m\leq n-2)}\left( \frac{\sigma_t^2}{2}v^{[n]}_{m-k+2,k}(t)
+\rho_t\sigma_t\alpha_t v^{[n]}_{m-k+1,k+1}(t)+\frac{\alpha_t^2}{2}v^{[n]}_{m-k,k+2}(t)\right)\nn \\
&&+\mbb{I}_{(m=n)}\frac{1}{2}\part_x^{n-k}\part_y^k \Theta(t,0,0)
+\mbb{I}_{(m\leq n-1)}\frac{\sigma_t^2}{2}v^{[n]}_{m-k+1,k}(t)\nn \\
&&+\mbb{I}_{(m\leq n-1,1\leq k)} k \left(\frac{\sigma_t^2}{2}v^{[n-1]}_{m-k+1,k-1}(t)+
\kappa_t v^{[n-1]}_{m-k,k}(t)\right)\nn \\
&&\hspace{-7mm}-\mbb{I}_{(m\leq n-2)}\sum_{l=1}^{n-1}\sum_{j=1\vee[l+2-n+m]}^{l\wedge [m+1]}
\sum_{p=1\vee[j-m+k]}^{j\wedge [k+1]} \frac{\alpha_t^2}{2}\xi_t^2
C_{(m-k,j-p)}C_{(k,p-1)}v^{[l]}_{j-p,p}(t)v^{[n-l]}_{m-k-j+p,k-p+2}(t)\nn \\
&&\hspace{-10mm}-\mbb{I}_{(1\leq m\leq n-2,1\leq k)}\sum_{l=1}^{n-2}
\sum_{j=1\vee[l+2-n+m]}^{l\wedge m} \sum_{p=1\vee[j-m+k]}^{j\wedge k}
\frac{\alpha_t^2}{2}\xi_t^2C_{(m-k,j-p)}C_{(k,p)}p v^{[l]}_{j-p,p}(t)
v^{[n-l-1]}_{m-k-j+p,k-p+1}(t)\nn \\
\eea
with $\xi_t^2:=(1-\rho(t)^2)$ and the terminal conditions $v^{[n]}_{n-k,k}(T)=\gamma \part_x^{n-k}\part_y^k H(0,0)$
with all the other components zero.
}\label{lemma-3}
}
\\\\
{\it Proof:} The proof is done in a similar way to  Lemma~\ref{lemma-1} and \ref{lemma-2}.
The details of the derivation are given in Appendix~\ref{ap-lemma-3}.

\subsection{Numerical Examples}
For numerical examples, we shall use 
\bea
&&\Theta(t,X_t,Y_t):=c_0 e^{-c_1 X_t} (Y_t+1)  \\
&&H(X_T,Y_T):=e^{-g_1 X_T} G(Y_T)
\eea
where $c_0, c_1, g_1$ are constants and $G(\cdot)$ a smooth function of $Y$.
Since the parameter of risk-averseness $\gamma$ appears only in a combination $\gamma H$,
the factor $e^{-g_1 X_T}$ can equivalently be interpreted as a $S_T$-dependent risk averseness. 

The problem analyzed in this section is intrinsically non-linear. Thus, we cannot use the density 
approximation and must directly approximate the terminal payoff  by a  smooth 
function. In practice, however, it should not be a prohibitive limitation.
Since the problem is non-linear, one has to consider the optimization 
in a portfolio level. Then, 
considering an appropriate hedging strategy based on a smooth approximate payoff function,
instead of treating it exactly,  should be reasonable.

We consider the next four choices of terminal liability (except $e^{-g_1 x}$ factor) 
in the numerical examples:
\bea
&&(1):~\sin\Bigl(y+\frac{\pi}{6}\Bigr) \\
&&(2):~\max\bigl(0,y\bigr) \\
&&(3):~\max\bigl(0,-y\bigr) \\
&&(4):~0.6-\max\bigl(0,0.2-y\bigr)
\eea
For $(2)$ to $(4)$, we have approximated it by a $5$th-order polynomial function determined
by a simple least-square method, and treat it as the {\it true} $G(y)$ in the evaluation.
Here, the shapes of the liability and the order of approximating polynomial function are chosen rather arbitrary.
In practice, one has to consider in a portfolio level and needs to choose a 
certain order of polynomial to recover its {\it overall} shape. The impact from adding another term would be 
easy to check directly. It is naturally expected, however, that the higher order terms plays only a minor role
otherwise it means that the firm is taking quite problematic positions and exposing it 
to the far-tail behavior of the underlying securities.

\begin{figure}[H]
\vspace{20mm}
\begin{center}	
\includegraphics[width=72mm]{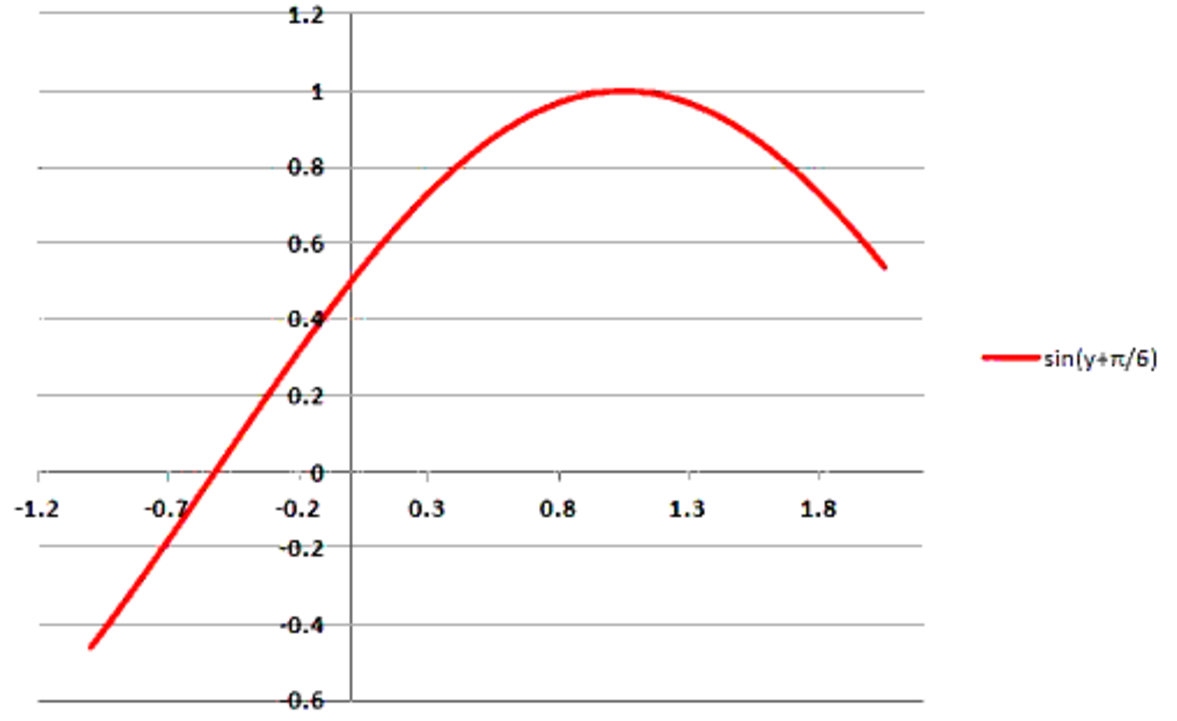}
\includegraphics[width=72mm]{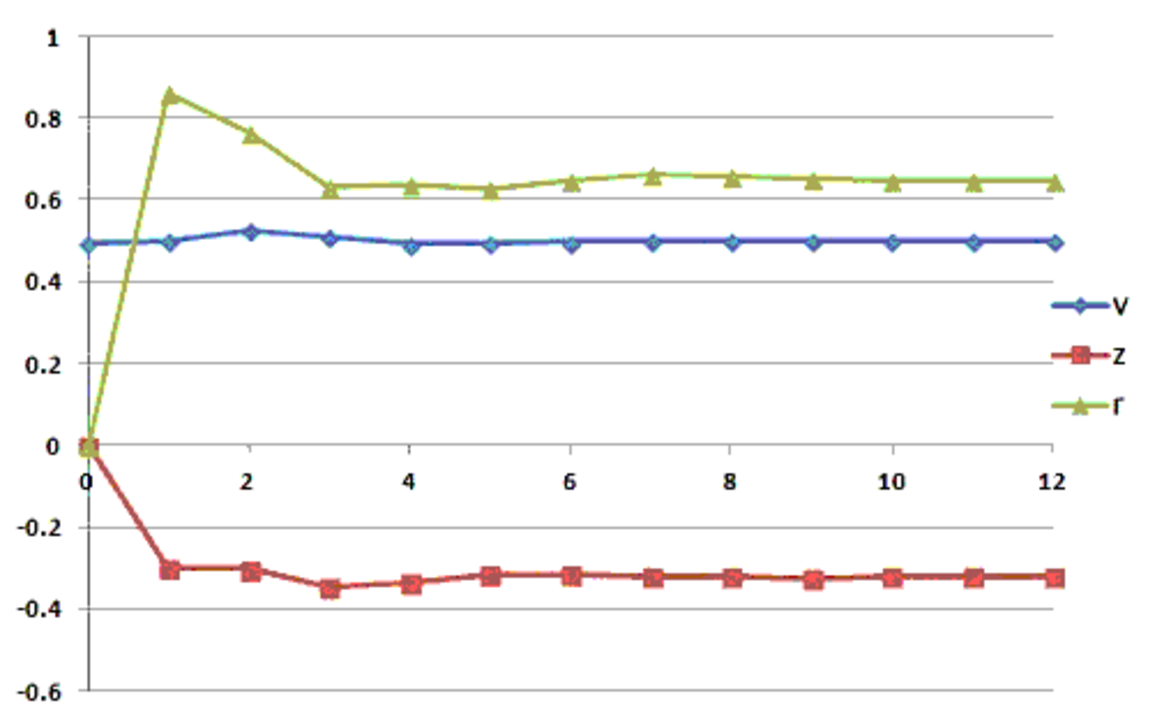}
\includegraphics[width=72mm]{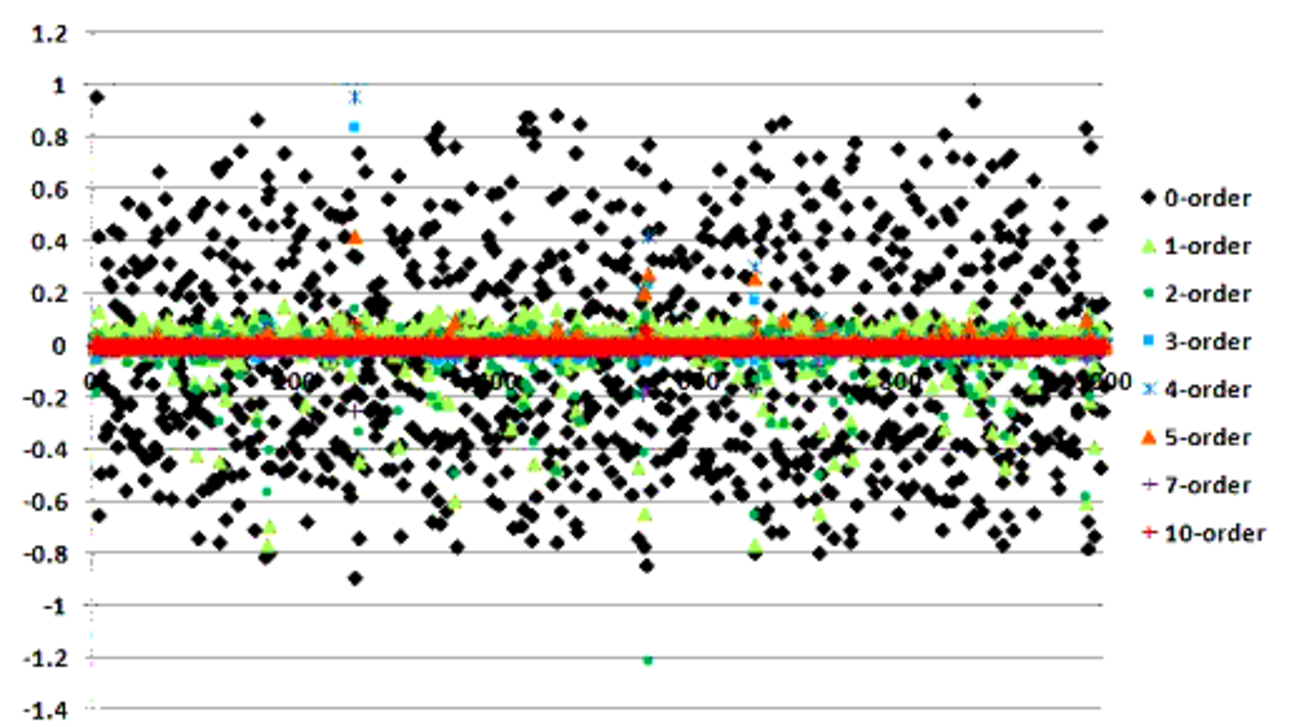}
\includegraphics[width=72mm]{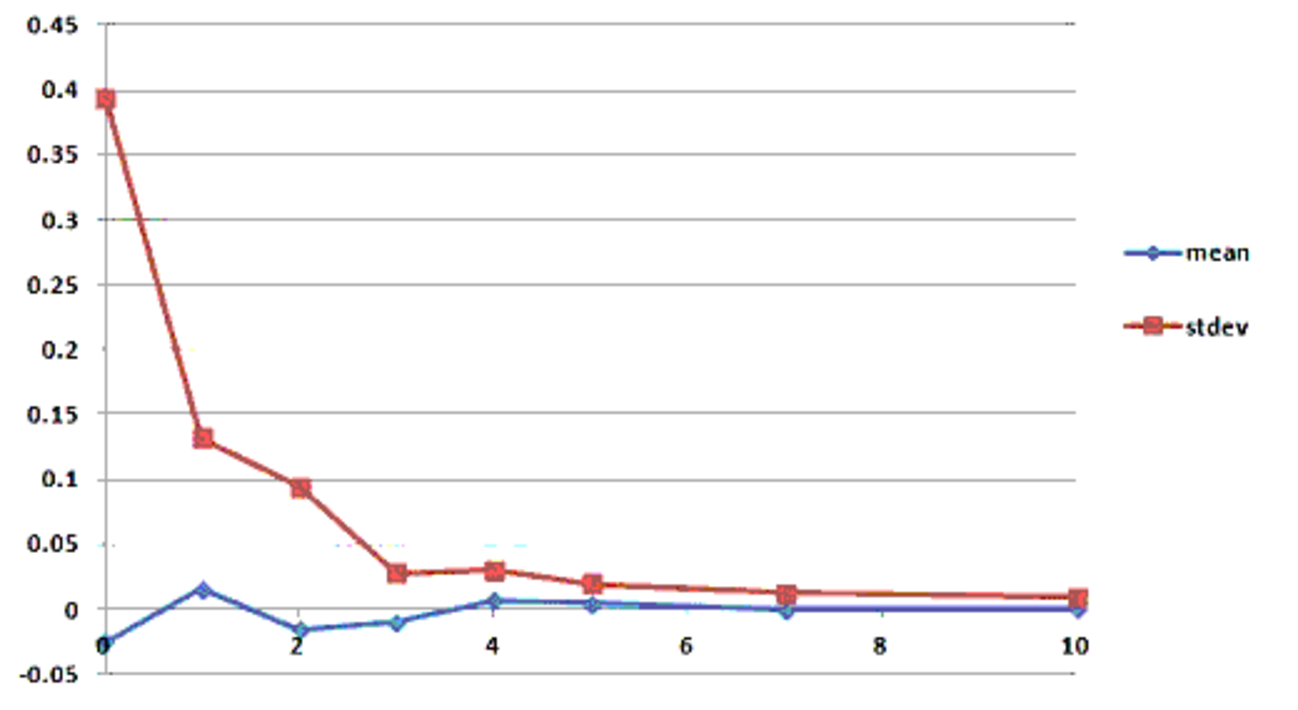}
\end{center}
\caption{$T=1, \sigma=0.2, \alpha=0.5, \rho=-0.7, \kappa=0.1, c_0=0.01$, 
$c_1=0.4, \gamma=1,g_1=0.6$. $G(y)=\sin(y+\pi/6)$.}
\label{Fig-Sin}

\begin{table}[H]
\footnotesize
\begin{tabular}{|c||c|c|c|c|c|c|c|c|}																			
\hline																			
     & $n=0$ & $n=1$ & $n=2$ & $n=3$ & $n=4$ & $n=5$ & $n=7$ & $n=10$ \\
\hline
mean & -0.026 & 0.016 & -0.016 & -0.011 & $6.3\times 10^{-3}$ & $3.9\times 10^{-3}$ & $-6.5\times 10^{-4}$ 
& $5.0\times 10^{-4}$ \\ 
\hline
stdev & 0.39 & 0.13 & 0.094 & 0.027 & 0.030 & 0.020 & 0.012 & $9.1\times 10^{-3}$ \\
\hline 
\end{tabular} 
\caption{
Mean and standard deviation of $[\gamma H_T-\wt{V}^{(n)}_T]$ for the setup in Figure~\ref{Fig-Sin}.}
\end{table}

\end{figure}

\begin{figure}[H]
\vspace{40mm}
\begin{center}	
\includegraphics[width=72mm]{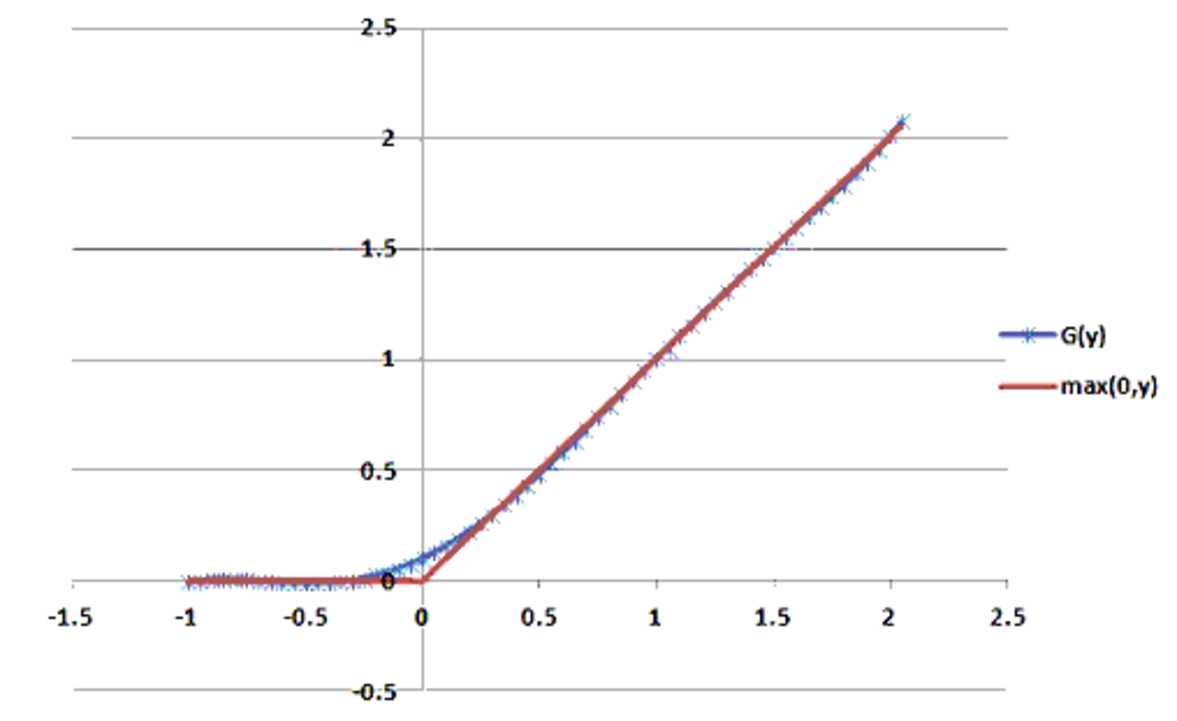}
\includegraphics[width=72mm]{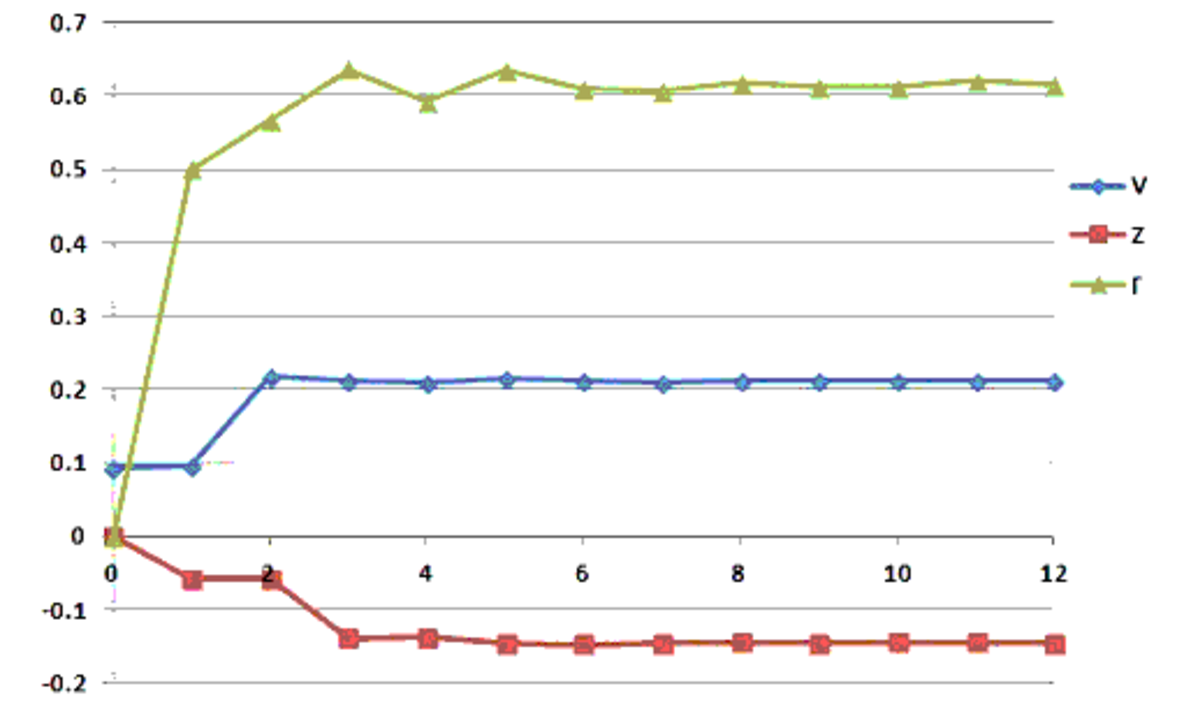}
\includegraphics[width=72mm]{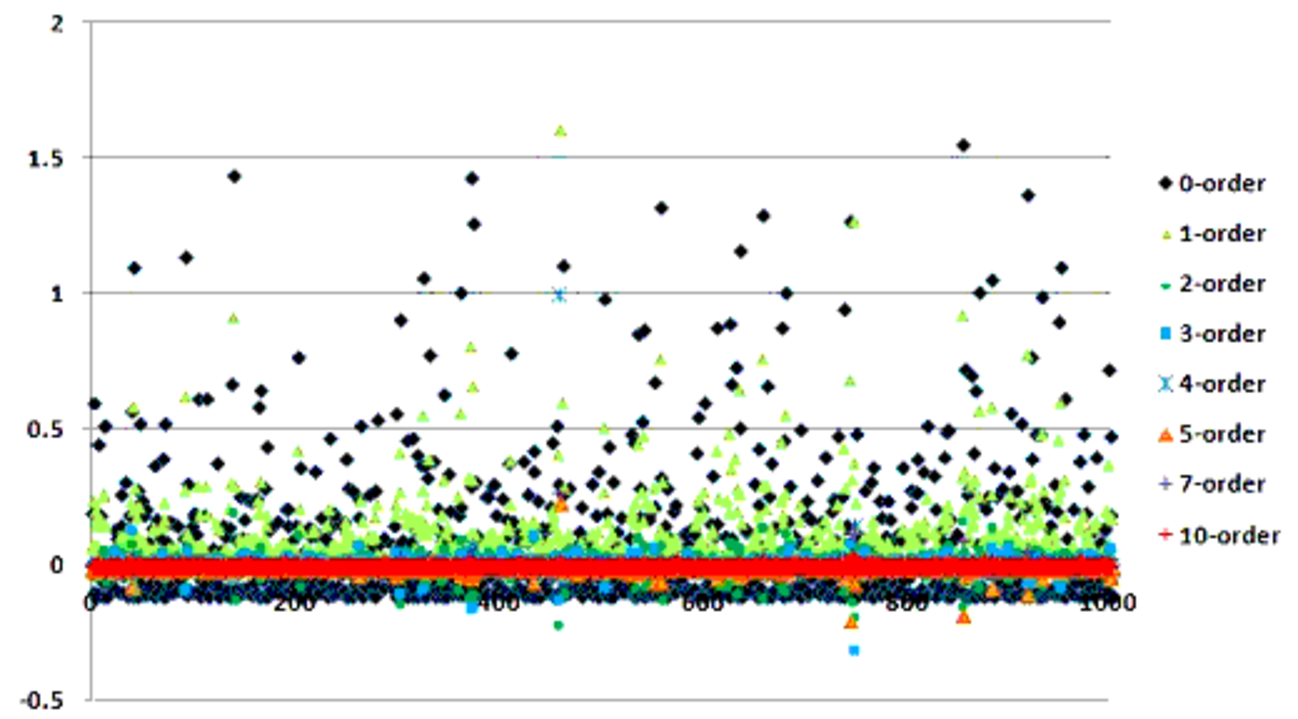}
\includegraphics[width=72mm]{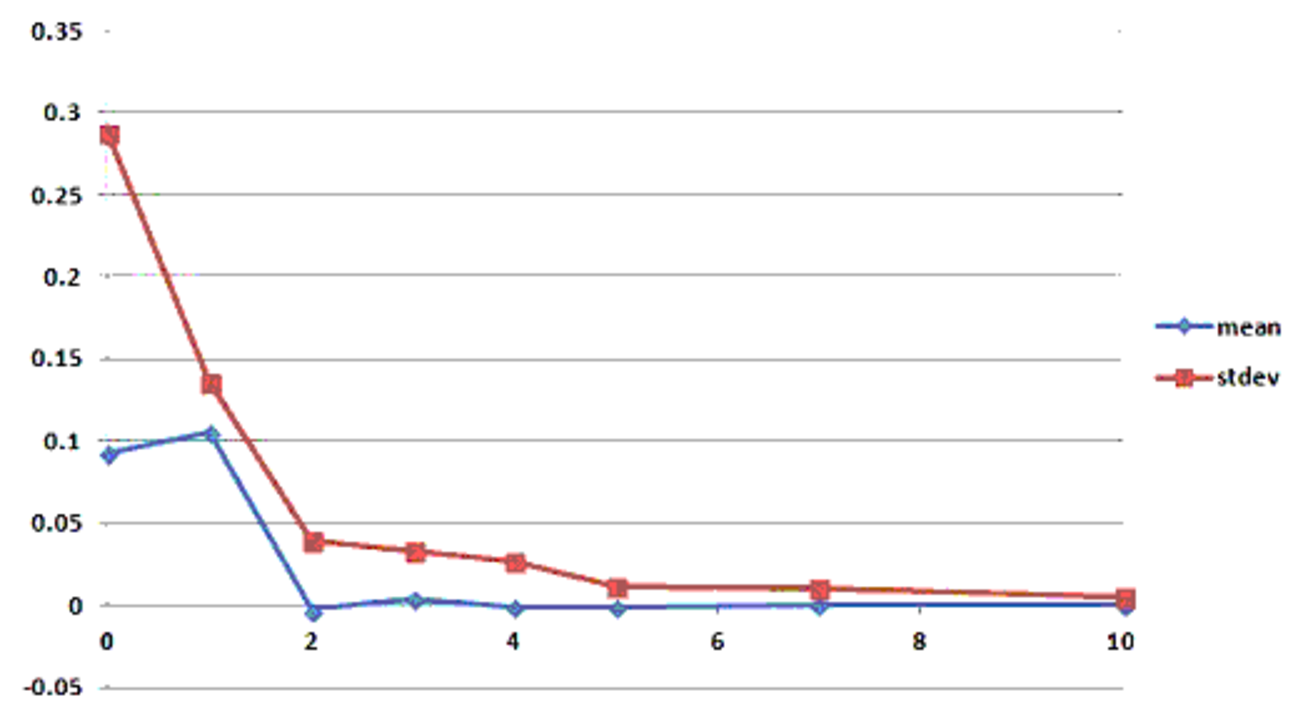}
\end{center}
\caption{
$T=1, \sigma=0.2, \alpha=0.4, \rho=-0.6, \kappa=0.1, c_0=0.01$,
$c_1=0.4, \gamma=1, g_1=0.6$. $G(y)$ is a $5$-th order polynomial approximating $\max(0,y)$.}
\label{Fig-Call}
\end{figure}

\begin{figure}[H]
\vspace{40mm}
\begin{center}	
\includegraphics[width=72mm]{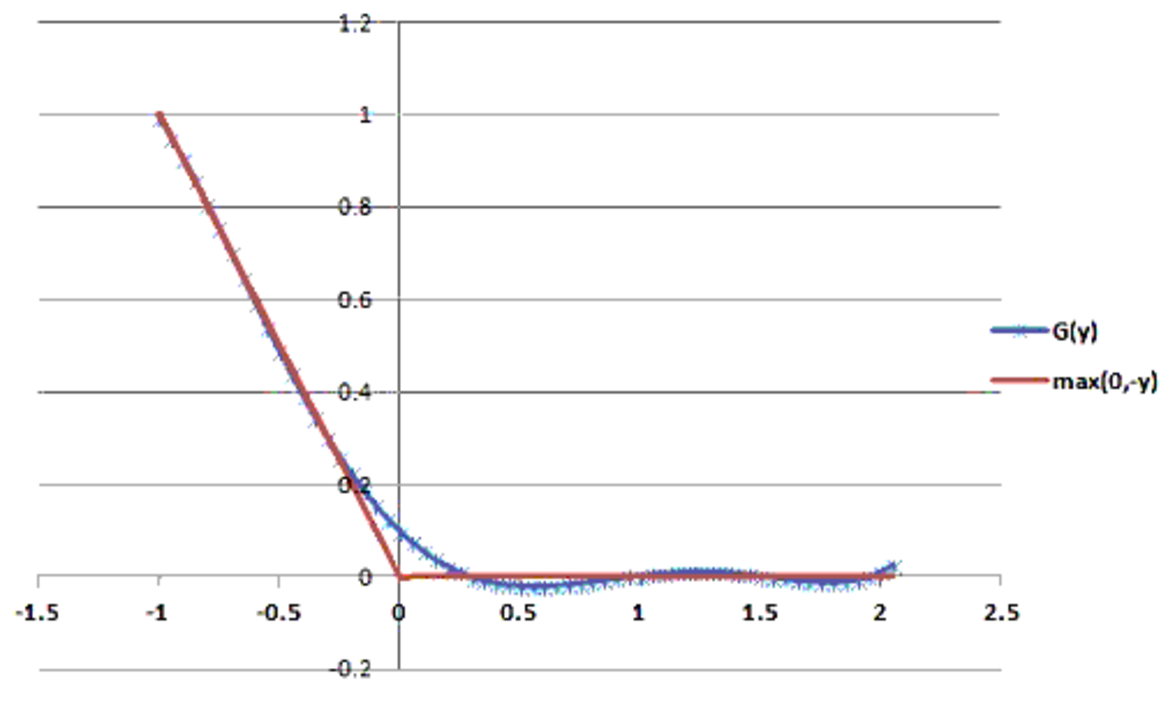}
\includegraphics[width=72mm]{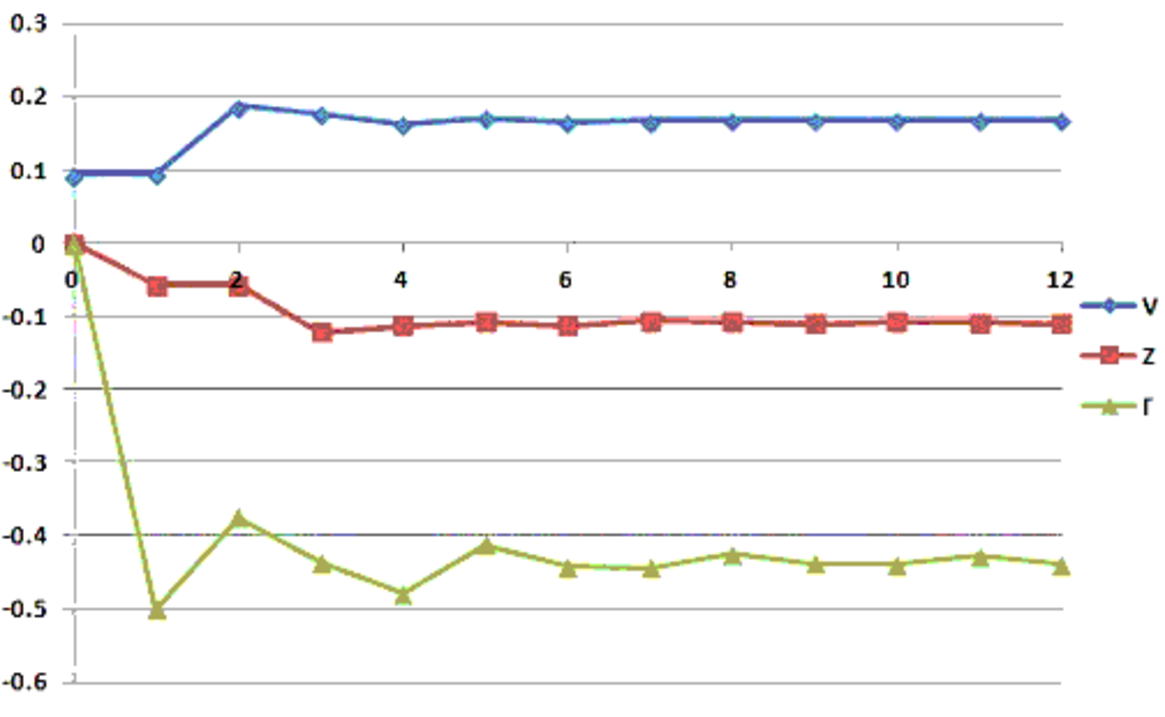}
\includegraphics[width=72mm]{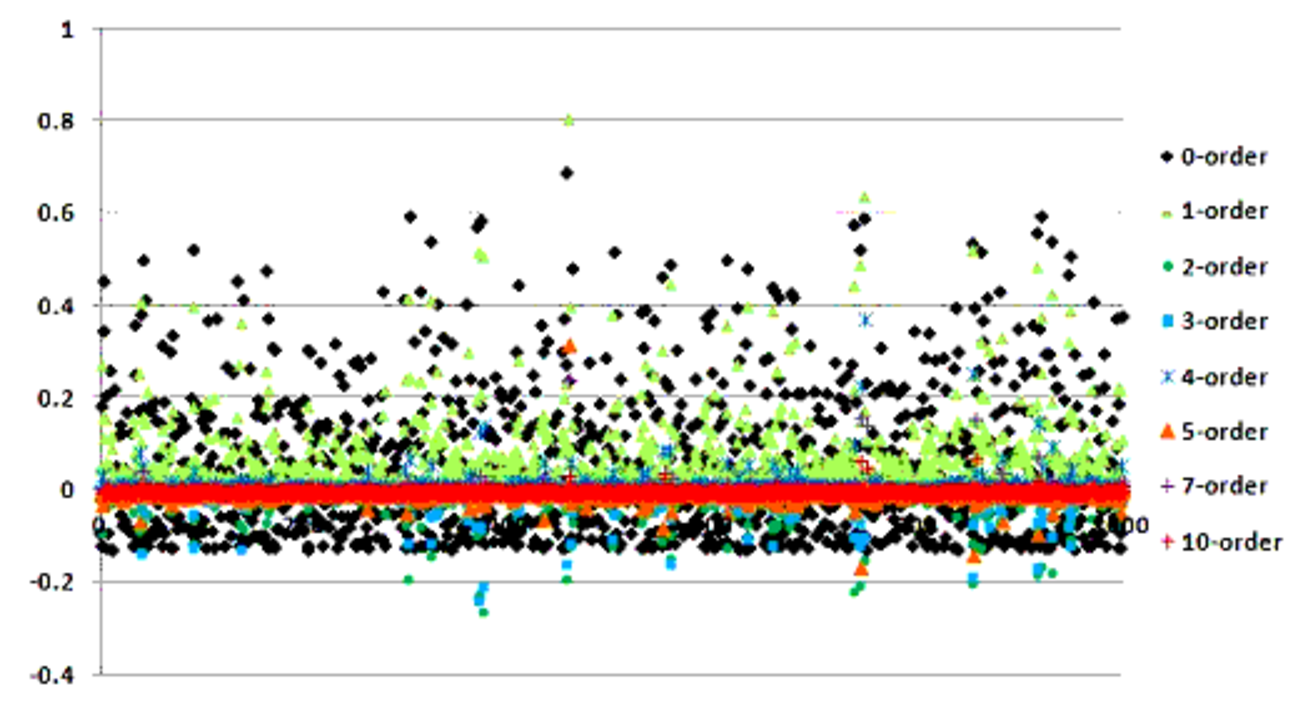}
\includegraphics[width=72mm]{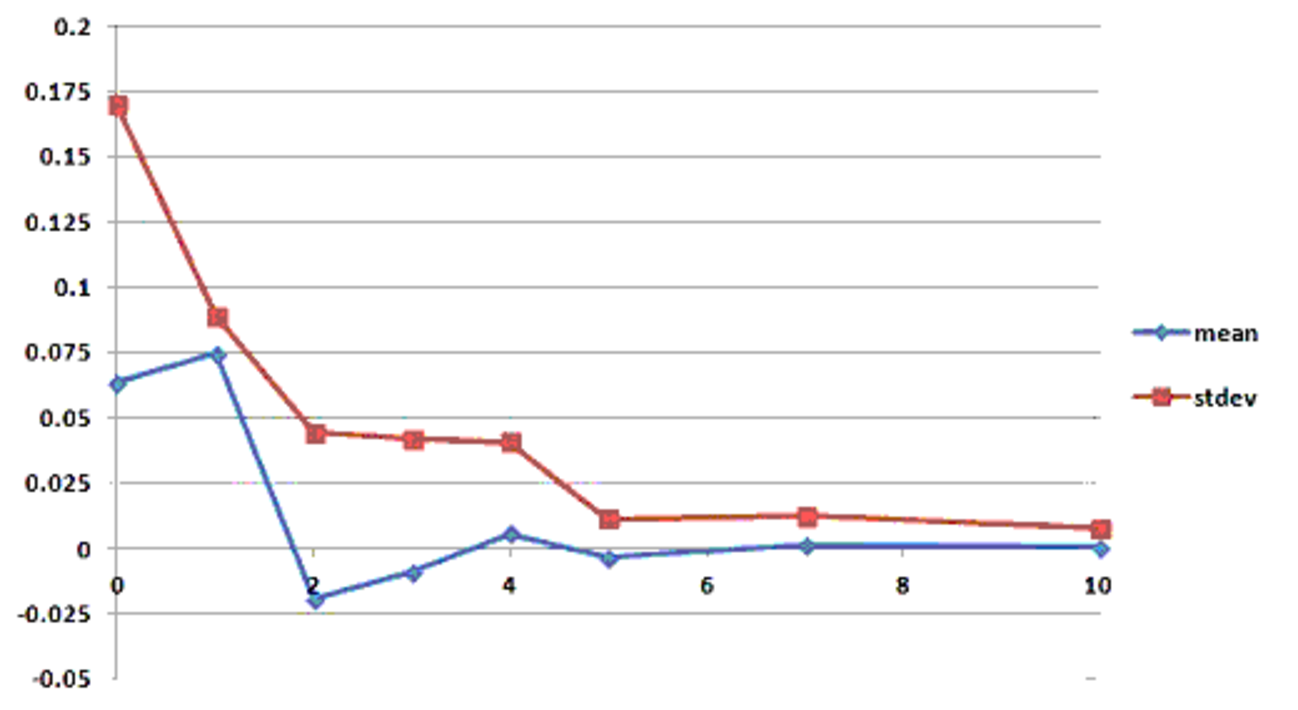}
\end{center}
\caption{
$T=1, \sigma=0.2, \alpha=0.4, \rho=-0.6, \kappa=0.1, c_0=0.01$, $c_1=0.4,\gamma=1, g_1=0.6$.
$G(y)$ is a $5$-th order polynomial approximating $\max(0,-y)$. }
\label{Fig-Put}
\end{figure}

\begin{figure}[H]
\vspace{40mm}
\begin{center}	
\includegraphics[width=72mm]{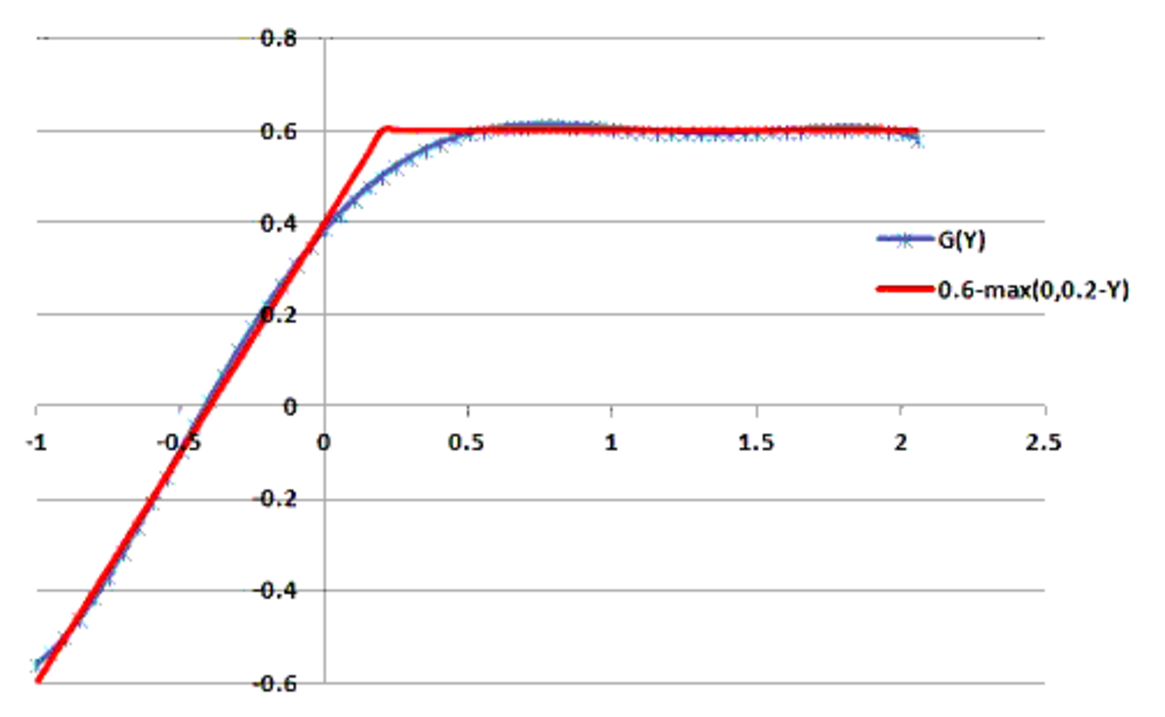}
\includegraphics[width=72mm]{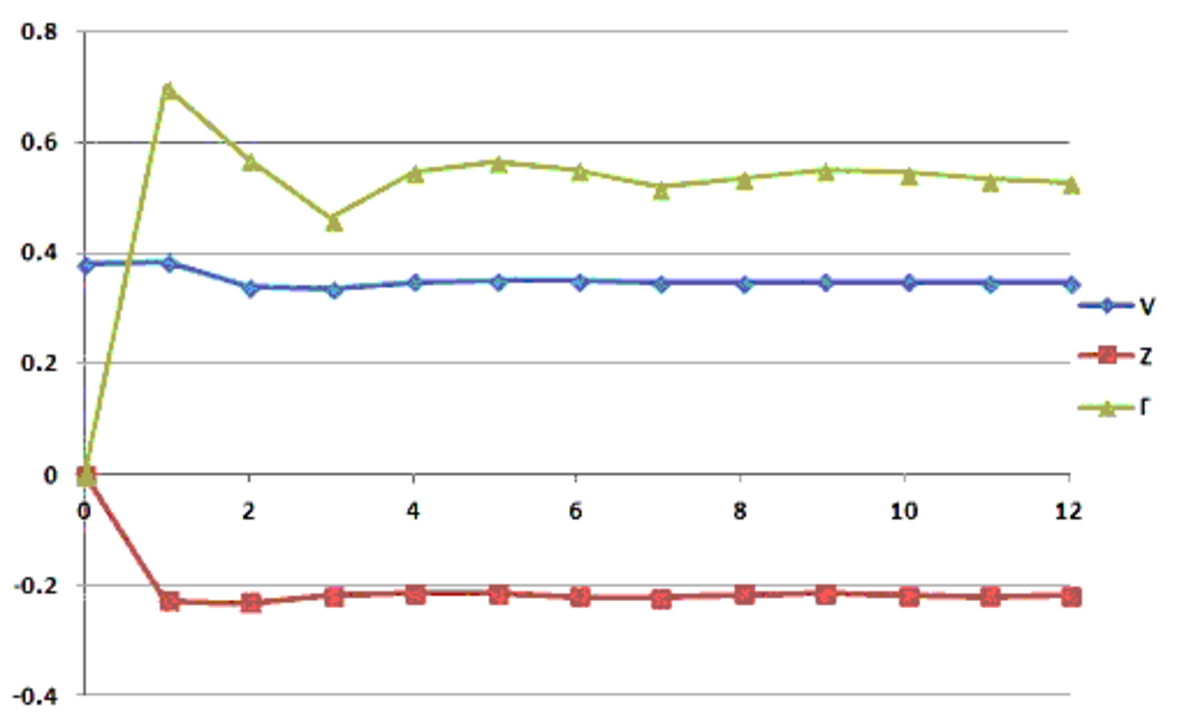}
\includegraphics[width=72mm]{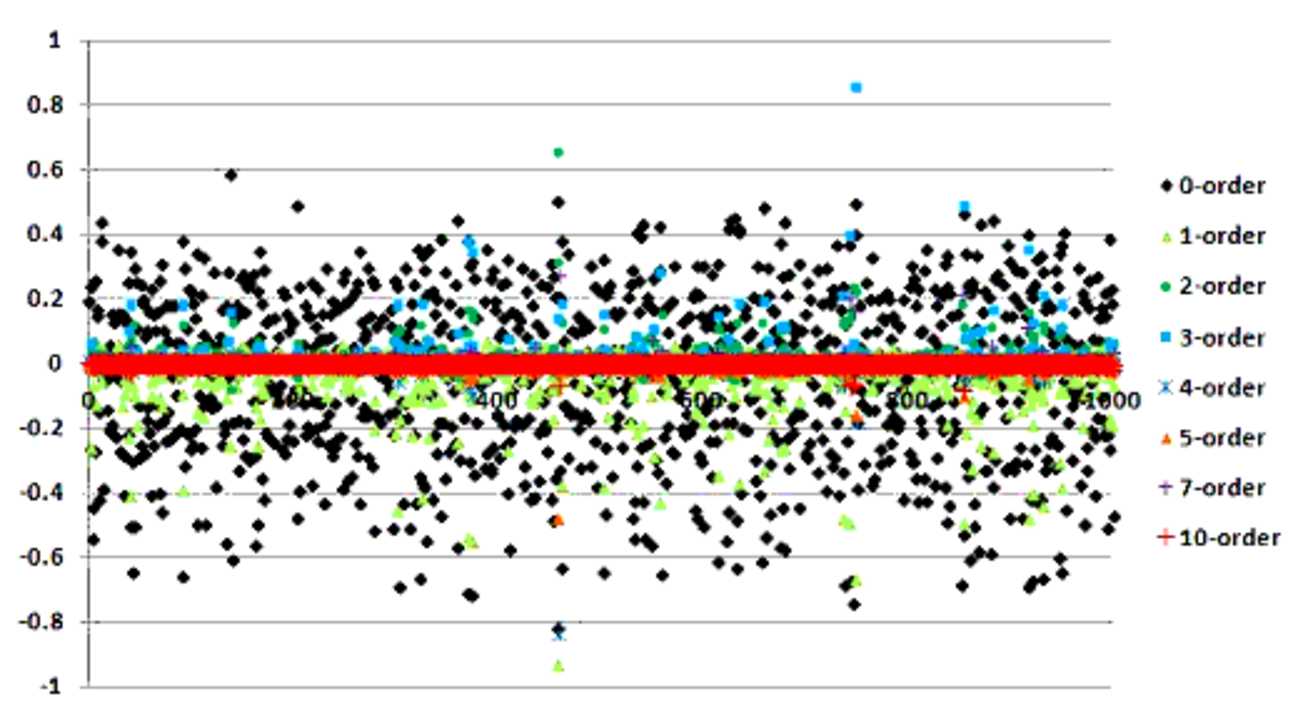}
\includegraphics[width=72mm]{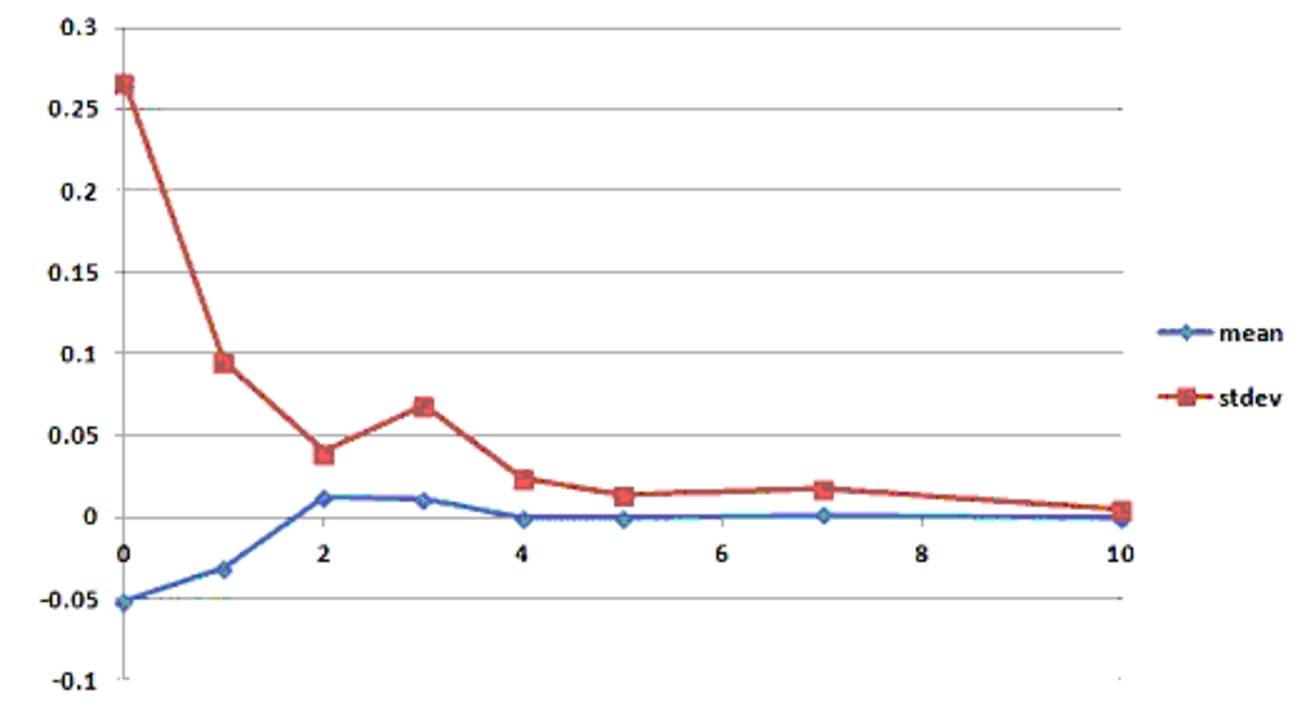}
\end{center}
\caption{
$T=1, \sigma=0.2, \alpha=0.4, \rho=-0.6, \kappa=0.1, c_0=0.01$, $c_1=0.4,\gamma=1, g_1=0.6$.
$G(y)$ is a $5$-th order polynomial approximating $[0.6-\max(0,0.2-y)]$.}
\label{Fig-SPut}
\end{figure}

Each of Figure~\ref{Fig-Sin} to \ref{Fig-SPut} consists of: 
1)Top left: a graph of $G(y)$, 2)Top right: a graph of the truncated value function and control variables $(V^{(n)}_0, Z^{(n)}_0, \Gamma^{(n)}_0)$ for each $n$ specified by the 
horizontal axis~\footnote{See, (\ref{Vn-truncated}) and (\ref{Cn-truncated}) for the definition
of truncated variables.}, 3)Bottom left: a scattered plot of $[\gamma H(X_T,Y_T)-\wt{V}^{(n)}_T]$ for each 
expansion order, 4) Bottom right: a graph of the means as well as the standard deviations of
 $[\gamma H(X_T,Y_T)-\wt{V}_T^{(n)}]$ for $(0\leq n\leq 10)$ with 100,000-path simulation, 
whose details are also given in a table associated with each example.
Note that the errors are measured relative to the smoothly modified terminal functions
in $(2)$ to $(4)$ cases.

From the definition of the truncated approximation, one can easily see that the mean of $[\gamma H(X_T,Y_T)-\wt{V}^{(n)}_T]$
is equivalent to the estimate of 
\bea
&&V_0^{(n)}-\mbb{E}\left[\gamma H(X_T,Y_T)-\int_0^T Z_t^{(n)}dX_t-\int_0^T \Gamma_t^{(n)}dY_t \right.\nn \\
&&\hspace{-10mm} \left. -\int_0^T \Bigl\{ \frac{1}{2}\Theta(t,X_t,Y_t)-
\frac{\alpha^2}{2}(1-\rho^2)(1+Y_t)[\Gamma_t^{(n)}]^2+\frac{\sigma^2}{2}(1+Y_t)Z_t^{(n)}+\kappa Y_t \Gamma_t^{(n)}
\Bigr\}dt \right]
\eea
by simulation. Its convergence to zero
gives one consistency test for the value function at the initial point.
The scattered plots of $[\gamma H(X_T,Y_T)-\wt{V}^{(n)}_T]$ and the corresponding 
standard deviations provide a much stronger test. They suggest that the truncated value functions and 
control variables give a good {\it path-wise} approximation for the original BSDE.
One can clearly observe that the deviations $[\gamma H(X_T,Y_T)-\wt{V}^{(n)}_T]$ at the maturity 
are strongly clustering around zero even for a relatively low expansion order $n\sim 3$. 
Of course, as one can imagine, 
the probability that the size of $\bigl(\gamma H(X_T,Y_T),X_T,Y_T\bigr)$ becomes (meaningfully) bigger than one
should be small enough in order to obtain a converging result.
This means that we need to adopt a proper ``scaling" for the wealth and the other parameters to make 
sure the chosen utility (or cost function) stays $\calo(1)$~\footnote{A similar scaling would 
be necessary for any risk-management in practice in the presence of non-linearities.}.

\subsubsection*{Remark}
For those who have checked the result of Lemma~3 by themselves, it must be clear that
deriving a closed form system of ODEs would be much harder in a realistic multi-asset
setup. In that sense, the above numerical result is quite encouraging by implying that one may
 get a reasonable approximation even by a lower order expansion, for example,  $n\sim 4$.
In this case, step-by-step derivation of the relevant ODEs following the 
instruction given in Section~\ref{sec-asympgeneral} can be done without much difficulty 
even for a more involved BSDE. Interesting practical applications are left for the 
future research.


\begin{table}[H]
\footnotesize
\begin{tabular}{|c||c|c|c|c|c|c|c|c|}																			
\hline																			
     & $n=0$ & $n=1$ & $n=2$ & $n=3$ & $n=4$ & $n=5$ & $n=7$ & $n=10$ \\
\hline
mean & 0.093 & 0.10  & $-2.8\times 10^{-3}$  & $3.7\times 10^{-3}$  & $-8.2\times 10^{-4}$  & $-1.3\times 10^{-3}$ 
& $2.9\times 10^{-4}$ & $1.3\times 10^{-4}$  \\ 
\hline
stdev & 0.29  & 0.14  & 0.040  & 0.030 & 0.026  & 0.011  &  0.011 & $5.3\times 10^{-3}$  \\
\hline 
\end{tabular}
\caption{Mean and standard deviation of $[\gamma H_T-\wt{V}^{(n)}_T]$ for the setup in Figure~\ref{Fig-Call}.}
\end{table}

\begin{table}[H]
\footnotesize
\begin{tabular}{|c||c|c|c|c|c|c|c|c|}																			
\hline																			
     & $n=0$ & $n=1$ & $n=2$ & $n=3$ & $n=4$ & $n=5$ & $n=7$ & $n=10$ \\
\hline
mean & 0.064 & 0.075  & -0.019  & $-8.6\times 10^{-3}$ & $5.8\times 10^{-3}$ & $-3.6\times 10^{-3}$ & $1.5\times 10^{-3}$  
& $4.4\times 10^{-4}$  \\ 
\hline
stdev & 0.17 &  0.089 & 0.044 & 0.042 & 0.041 & 0.011 & 0.012 & $8.0\times 10^{-3}$ \\
\hline 
\end{tabular}
\caption{Mean and standard deviation of $[\gamma H_T-\wt{V}^{(n)}_T]$ for the setup in Figure~\ref{Fig-Put}.}
\end{table}

\begin{table}[H]
\footnotesize
\begin{tabular}{|c||c|c|c|c|c|c|c|c|}																			
\hline																			
     & $n=0$ & $n=1$ & $n=2$ & $n=3$ & $n=4$ & $n=5$ & $n=7$ & $n=10$ \\
\hline
mean & -0.052 & -0.031  & 0.012  & 0.011 & $-9.2\times 10^{-4}$ & $-9.9\times 10^{-4}$ & $1.4\times 10^{-3}$  
& $-9.0\times 10^{-4}$  \\ 
\hline
stdev & 0.27 & 0.095 & 0.040 & 0.069 & 0.024 & 0.013 & 0.017&  $5.2\times 10^{-3}$ \\
\hline 
\end{tabular}
\caption{Mean and standard deviation of $[\gamma H_T-\wt{V}^{(n)}_T]$ for the setup in Figure~\ref{Fig-SPut}.}
\end{table}

\clearpage
\section{Conclusions}
In this paper,  a polynomial scheme of asymptotic expansion for BSDEs is proposed.
We have shown that the polynomial expansion is uniquely determined by the recursive
system of linear ODEs, which can be easily solved {\it one-by-one} by following 
the appropriate order of evaluation.
We have studied possible applications to the pricing of European contingent claims
as well as the exponential-utility optimization with terminal liability, each of 
which is provided several illustrative numerical examples.

A rigorous mathematical justification and more intensive numerical studies with 
realistic models are left for the future works.
For example, a class of multi-factor Heston model proposed by Col et al. (2013)~\cite{Multi-Heston}
has a nice structure of dependence to which the current scheme can be applied. 
Studying the BSDEs associated with the control problem 
with defaultable securities, such as 
those give by Pham (2010)~\cite{Pham-prog}, looks interesting, too.

\appendix
\section{Proof of  Lemma~\ref{lemma-2}}
\label{ap-lemma-2}
We proceed as in the Heston's model. Based on the dynamics (\ref{SABR-X}) and (\ref{SABR-Y}),
the forward dynamics of the hypothesized polynomial solution is given by
\bea
&&dV_t^{[n]}=\sum_{m=0}^n \sum_{k=0}^m \frac{X_t^{m-k}Y_t^k}{(m-k)!k!}\left\{ \frac{\bigl.}{\bigr.}
\dot{v}^{[n]}_{m-k,k}(t) \right.\nn \\
&&\quad+\mbb{I}_{(2\leq k)}k(k-1)\left(
\frac{\sigma_t^2}{2}v^{[n]}_{m-k+2,k-2}(t)+\rho_t\sigma_t\alpha_t v^{[n]}_{m-k+1,k-1}(t)
+\frac{\alpha_t^2}{2}v^{[n]}_{m-k,k}(t)\right)\nn\\ 
&&\quad+\mbb{I}_{(m\leq n-1,1\leq k)}k\Bigl(
\sigma_t^2 v^{[n]}_{m-k+2,k-1}(t)+2\rho_t\sigma_t\alpha_t v^{[n]}_{m-k+1,k}(t)+
\alpha_t^2 v^{[n]}_{m-k,k+1}(t)\Bigr)\nn \\
&&\quad\left.+\mbb{I}_{(m\leq n-2)}\left(\frac{\sigma_t^2}{2}v^{[n]}_{m-k+2,k}(t)
+\rho_t \sigma_t \alpha_t v^{[n]}_{m-k+1,k+1}(t)+
\frac{\alpha_t^2}{2}v^{[n]}_{m-k,k+2}(t)\right) \frac{\bigl.}{\bigr.} \right\}dt\nn  \\
&&\quad +\sum_{m=1}^n \sum_{k=0}^{m-1} v^{[n]}_{m-k,k}(t)\frac{X_t^{m-k-1}Y_t^k}{(m-k-1)!k!}dX_t \nn \\
&&\quad +\sum_{m=1}^n \sum_{k=1}^{m} v^{[n]}_{m-k,k}(t)
\frac{X_t^{m-k}Y_t^{k-1}}{(m-k)!(k-1)!}dY_t 
\label{SABR-Vn-SDE}
\eea
which implies the control variables given in (\ref{SABR-Ztn}) and (\ref{SABR-Gtn}).

On the other hand, the $n$-th order part of the BSDE (\ref{SABR-epV}) is
\bea
&&V_t^{[n]}=\frac{X_T^n}{n!}\part_x^n H(0)-\int_t^T Z_s^{[n]}dX_s
-\int_t^T  \Gamma_s^{[n]} dY_s
\nn \\
&&\quad-\int_t^T \left\{ \frac{\beta}{2}\sigma(s)^2\left(
\sum_{l=0}^{n-1}\frac{\part_x^l b(0)}{l!} X_s^l Z_s^{[n-l]}+
2\sum_{l=0}^{n-2}\frac{\part_x^l b(0)}{l!}X_s^l Y_s Z_s^{[n-l-1]} \right. \right.  \nn \\
&&\qquad\left. \left.+\sum_{l=0}^{n-3}\frac{\part_x^l b(0)}{l!}X_s^l Y_s^2 Z_s^{[n-l-3]}\right)
+\kappa_s Y_s \Gamma_s^{[n-1]}\right\}ds
\label{SABR-Vn-BSDE}
\eea
Substituting the control variables by those in (\ref{SABR-Ztn}) and (\ref{SABR-Gtn}),
one obtains
\bea
&&V_t^{[n]}=\frac{X_T^n}{n!}\part_x^n H(0)-\int_t^T Z_s^{[n]}dX_s
-\int_t^T  \Gamma_s^{[n]} dY_s -\sum_{m=0}^n \sum_{k=0}^m \int_t^T 
\frac{X_s^{m-k}Y_s^k}{(m-k)!k!} \nn \\
&&\times \left\{ \mbb{I}_{(m\leq n-1,1\leq k)}k \kappa_s v^{[n-1]}_{m-k,k}(s)
+\mbb{I}_{(m\leq n-1)}\frac{\beta}{2}\sigma_s^2 \sum_{l=0}^{m-k}
C_{(m-k,l)}\part^l_x b(0)
\right. \nn \\
&&\quad \left. \times\left(
v^{[n-l]}_{m-k-l+1,k}(s)+\mbb{I}_{(1\leq k)}2k~v^{[n-l-1]}_{m-k-l+1,k-1}(s)
+\mbb{I}_{(2\leq k)}k(k-1)v^{[n-l-2]}_{m-k-l+1,k-2}(s)
\right)\frac{\bigl.}{\bigr.} \right\} ds~.\nn \\
\eea
By comparing the coefficients in the drift term, one obtains the linear ODEs 
as (\ref{SABR-ODE}). If the forward SDE (\ref{SABR-Vn-SDE}) is well-defined
with the solution of (\ref{SABR-ODE}), then it is clear that it gives 
one solution for the BSDE (\ref{SABR-Vn-BSDE}). The uniqueness of the 
polynomial solution is clear due to the linearity of the ODEs.

\section{Proof of Lemma~\ref{lemma-3}}
\label{ap-lemma-3}
The dynamics of (\ref{X-util}) and (\ref{Y-util}) gives 
the forward SDEs of the assumed polynomial (\ref{Vn-util}) as
\bea
&&dV_t^{[n]}=\sum_{m=0}^n \sum_{k=0}^m\frac{X_t^{m-k}Y_t^k}{(m-k)!k!}\left\{\dot{v}^{[n]}_{m-k,k}(t) 
\frac{\bigl.}{\bigr.}
\right. \nn \\
&&\quad+\mbb{I}_{(m\leq n-1,1\leq k)} k\left(
\frac{\sigma_t^2}{2}v^{[n]}_{m-k+2,k-1}(t)+\rho_t\sigma_t\alpha_t v^{[n]}_{m-k+1,k}(t)+
\frac{\alpha_t^2}{2}v^{[n]}_{m-k,k+1}(t)\right) \nn\\
&&\quad\left.+\mbb{I}_{(m\leq n-2)}\left(
\frac{\sigma_t^2}{2}v^{[n]}_{m-k+2,k}(t)+\rho_t\sigma_t\alpha_t v^{[n]}_{m-k+1,k+1}(t)+
\frac{\alpha_t^2}{2} v^{[n]}_{m-k,k+2}(t)\right)\frac{\bigl.}{\bigr.} \right\}dt\nn\\
&&\quad+\sum_{m=1}^n \sum_{k=0}^{m-1}v^{[n]}_{m-k,k}(t)\frac{X_t^{m-k-1}Y_t^k}{(m-k-1)!k!}dX_t\nn \\
&&\quad+\sum_{m=1}^n \sum_{k=1}^{m}v^{[n]}_{m-k,k}(t)\frac{X_t^{m-k}Y_t^{k-1}}{(m-k)!(k-1)!}dY_t 
\label{Vn-util-SDE}
\eea
which then implies the control variables as in (\ref{Zn-util}) and (\ref{Gn-util}).
On the other hand, the $n$-order BSDE of (\ref{epV-util}) is
\bea
&&V_t^{[n]}=\sum_{k=0}^n \frac{X_T^{n-k}Y_T^k}{(n-k)!k!}\gamma \part_x^{n-k}\part_y^k H(0,0)-
\int_t^T Z_s^{[n]}dX_s-\int_t^T \Gamma_s^{[n]}dY_s \nn \\
&&-\int_t^T \left\{ \sum_{k=0}^n \frac{X_s^{n-k}Y_s^k}{(n-k)!k!}\frac{1}{2}\part_x^{n-k}\part_y^k
\Theta(s,0,0)+\kappa_s Y_s \Gamma_s^{[n-1]}
\right. \nn \\
&&\qquad -\frac{\alpha(s)^2}{2}(1-\rho(s)^2) \left(
\sum_{l=1}^{n-1}\Gamma_s^{[l]}\Gamma_s^{[n-l]}+Y_s \sum_{l=1}^{n-2}\Gamma_s^{[l]}\Gamma_s^{[n-l-1]}
\right) \nn \\
&&\qquad\left.  +\frac{\sigma(s)^2}{2}\Bigl(Z_s^{[n]}+Y_s Z_s^{[n-1]}\Bigr) \frac{\bigl.}{\bigr.} \right\}ds~.
\label{Vn-BSDE-util}
\eea
Substituting (\ref{Zn-util}) and (\ref{Gn-util}) into the above expression and reordering the summation 
yield
\bea
&&V_t^{[n]}=\sum_{k=0}^n \frac{X_T^{n-k}Y_T^k}{(n-k)!k!}\gamma \part_x^{n-k}\part_y^k H(0,0)-
\int_t^T Z_s^{[n]}dX_s-\int_t^T \Gamma_s^{[n]}dY_s \nn \\
&&\hspace{-5mm}-\sum_{m=0}^n \sum_{k=0}^m \int_t^T \frac{X_s^{m-k}Y_s^k}{(m-k)!k!}\left\{ 
\mbb{I}_{(m=n)}\frac{1}{2}\part_x^{n-k}\part_y^k \Theta(s,0,0)+
\mbb{I}_{(m\leq n-1)}\frac{\sigma_s^2}{2}v^{[n]}_{m-k+1,k}(s)  \frac{\bigl.}{}
\right. \nn \\
&&\quad +\mbb{I}_{(m\leq n-1,1\leq k)}k\Bigl(\frac{\sigma_s^2}{2}v^{[n-1]}_{m-k+1,k-1}(s)+\kappa_s v^{[n-1]}_{m-k,k}(s)
\Bigr)\nn \\
&&\hspace{-5mm}-\frac{\alpha_s^2}{2}\xi_s^2 \mbb{I}_{(m\leq n-2)}\sum_{l=1}^{n-1}
\sum_{j=1\vee[l+2-n+m]}^{l\wedge[m+1]} \sum_{p=1\vee [j-m+k]}^{j\wedge [k+1]}
C_{(m-k,j-p)}C_{(k,p-1)}v^{[l]}_{j-p,p}(s)v^{[n-l]}_{m-k-j+p,k-p+2}(s)\nn \\
&&\hspace{-10mm}\left.-\frac{\alpha_s^2}{2}\xi_s^2 \mbb{I}_{(1\leq m\leq n-2, 1\leq k)}\sum_{l=1}^{n-2}
\sum_{j=1\vee [l+2-n+m]}^{l\wedge m} \sum_{p=1\vee[j-m+k]}^{j\wedge k}
C_{(m-k,j-p)}C_{(k,p)}p v^{[l]}_{j-p,p}(s)v^{[n-l-1]}_{m-k-j+p,k-p+1}(s)\right\}ds\nn\\
\label{Vn-util-BSDE}
\eea
By comparing the drift terms of (\ref{Vn-util-SDE}) and (\ref{Vn-util-BSDE}),
one obtains the system of linear ODEs as in Lemma~\ref{lemma-3}.
It is clear that if the forward SDE (\ref{Vn-util-SDE}) with the solution of the ODEs
is well-defined, it at least gives one solution for the BSDE of the $n$-th order (\ref{Vn-BSDE-util}).
Due to the linearity of the ODEs, the solution should be unique within the assumed polynomial form.

\section{An exact solution for (\ref{V-util})}
\label{util-exact}
Suppose that both of the $H(x,y)$ and $\Theta(t,x,y)$
are linear functions of $(x,y)$:
\bea
&&H(x,y)=h_x x+h_y y+h_0 \\
&&\Theta(t,x,y)=\Theta_x(t)x+\Theta_y(t) y+ \Theta_0(t)
\eea
where $(h_x,h_y,h_0)$ are constants and $(\Theta_x(t), \Theta_y(t),\Theta_0(t))_t$
are some deterministic functions of time.
Then, it is almost immediate to notice that a linear value function and
the deterministic control variables can provide an exact solution:
\bea
&&V_t=v_x(t)X_t+v_y(t) Y_t+v_0(t)\\
&&Z_t=v_x(t)\\
&& \Gamma_t= v_y(t)
\eea
where $(v_x(t),v_y(t),v_0(t))$ are deterministic functions of time.

The ODEs which fixes $v_x,v_y,v_0$ can be obtained quite similarly to
the discussed approximation scheme.  On the one hand, the dynamics of the proposed solution 
becomes
\bea
dV_t&=&\Bigl(\dot{v}_x(t)X_t+\dot{v}_y(t) Y_t+\dot{v}_0(t)\Bigr)dt\nn \\
&&+v_x(t) dX_t+v_y(t) dY_t~.
\eea
On the other hand, by inserting the assumed form of control variables to (\ref{V-util}),
one obtains
\bea
&&V_t=\gamma\Bigl[h_x X_T+h_y Y_T+h_0\Bigr]-\int_t^T v_x(s) dX_s -\int_t^T v_y(s) dY_s \nn \\
&&\quad-\int_t^T \left\{
\frac{1}{2}\Bigl(\Theta_x(s)X_s+\Theta_y(s)Y_s+\Theta_0(s)\Bigr)+\kappa(s)v_y(s)Y_s \right. \\
&&\qquad \left.+\frac{1}{2}\Bigl(\sigma(s)^2 v_x(s)-\alpha(s)^2(1-\rho(s)^2)v_y(s)^2\Bigr)
(1+Y_s)\right\}ds~.
\eea

Thus, the the system of ODEs including a Riccati-type for $v_y$
\bea
&&\dot{v}_x(t)=\frac{1}{2}\Theta_x(t) \\
&&\dot{v}_y(t)=\frac{1}{2}\Bigl(\sigma(t)^2 v_x(t)-\alpha(t)^2(1-\rho(s)^2)v_y(t)^2\Bigr)
+\kappa(t)v_y(t)+\frac{1}{2}\Theta_y(t)\\
&&\dot{v}_0(t)=\frac{1}{2}\Bigl(\sigma(t)^2 v_x(t)-\alpha(t)^2(1-\rho(s)^2)v_y(t)^2\Bigr)+\frac{1}{2}\Theta_0(t)
\eea
with the terminal conditions
\be
v_x(T)=\gamma h_x, \quad v_y(T)=\gamma h_y, \quad v_0(T)=\gamma h_0
\ee
gives an exact solution if $v_y$ (and hence the others) has a finite solution 
for the relevant time interval $t\in[0,T]$.

\newpage
\section*{Acknowledgement}
This research is partially supported by Center for Advanced Research in Finance (CARF).
The author is grateful to professor Takahashi for many helpful discussions 
and encouragements. The author also thanks anonymous referees whose comments
significantly clarifies the presentation of the material.



\end{document}